\documentclass{ws-procs975x65}

\makeatletter
\newdimen\normalarrayskip              
\newdimen\minarrayskip                 
\normalarrayskip\baselineskip
\minarrayskip\jot
\newif\ifold             \oldtrue            \def\new{\oldfalse}
\def\arraymode{\ifold\relax\else\displaystyle\fi} 
\def\@arrayskip{\ifold\baselineskip\z@\lineskip\z@
     \else
     \baselineskip\minarrayskip\lineskip2\minarrayskip\fi}
\def\@arrayclassz{\ifcase \@lastchclass \@acolampacol \or
\@ampacol \or \or \or \@addamp \or
   \@acolampacol \or \@firstampfalse \@acol \fi
\edef\@preamble{\@preamble
  \ifcase \@chnum
     \hfil$\relax\arraymode\@sharp$\hfil
     \or $\relax\arraymode\@sharp$\hfil
     \or \hfil$\relax\arraymode\@sharp$\fi}}
\def\@array[#1]#2{\setbox\@arstrutbox=\hbox{\vrule
     height\arraystretch \ht\strutbox
     depth\arraystretch \dp\strutbox
     width\z@}\@mkpream{#2}\edef\@preamble{\halign \noexpand\@halignto
\bgroup \tabskip\z@ \@arstrut \@preamble \tabskip\z@ \cr}%
\let\@startpbox\@@startpbox \let\@endpbox\@@endpbox
  \if #1t\vtop \else \if#1b\vbox \else \vcenter \fi\fi
  \bgroup \let\par\relax
  \let\@sharp##\let\protect\relax
  \@arrayskip\@preamble}
\makeatother

\newcommand{\beq}{\begin{eqnarray}}
\newcommand{\eeq}{\end{eqnarray}}

\newcommand{\G}{\Gamma}

\renewcommand{\theequation}{\thesection.\arabic{equation}}

\def\appendix#1{\addtocounter{section}{1}\setcounter{equation}{0}
\renewcommand{\thesection}{\Alph{section}}
\section*{Appendix\thesection\protect\indent \parbox[t]{11.715cm} {#1}}
\addcontentsline{toc}{section}{Appendix \thesection\ \ \ #1} }
\newcommand{\complex}{{\bb C}} 
\newcommand{\zed}{{\bb Z}} 
\newcommand{\real}{{\bb R}} 
\newcommand{\NO}{\,\mbox{$\circ\atop\circ$}\,} 
\def\slash{{\!\!\!/\,}} 

\def\hil{{\cal H}}

\font\mybb=msbm10 at 12pt
\def\bb#1{\hbox{\mybb#1}}

\font\mybbs=msbm10 at 9pt
\def\bbs#1{\hbox{\mybbs#1}}
\newcommand{\nd}[1]{/\hspace{-0.6em} #1}
\def\nn{\nonumber}
\newcommand{\tr}[1]{\:{\rm tr}\,#1}

\def\e{{\,\rm e}\,}

\newcommand{\non}{\nonumber \\ \nopagebreak}
\newcommand{\nopg}{\\ \nopagebreak}

\newcommand{\modul}{{\bb M}} 
\newcommand{\moduls}{{\bbs M}} 
\newcommand{\semiplus}{{\supset\!\!\!\!\!\!\!+~}} 

\def\be{\begin{equation}}
\def\ee{\end{equation}}
\def\bea{\begin{eqnarray}}
\def\eea{\end{eqnarray}}
\def\bd{\begin{displaymath}}
\def\ed{\end{displaymath}}

\def\dd{{\rm d}}

\def\ii{{\,{\rm i}\,}}

\def\scT{{\sf T}}

\def\sc\Phi{{\sf \Phi}}
\def\scV{{\sf V}}
\def\scz{{\sf z}}
\def\scw{{\sf w}}
\def\scC{{\sf C}}
\def\scD{{\sf D}}

\def\scx{{\sf x}}

\def\cal{\mathcal }
\def\deriv{{\cal D}}

\begin{document}

\title{Logarithmic Conformal Field Theories and Strings in Changing Backgrounds}

\author{N.~E.MAVROMATOS}

\address{King's College London, Department of Physics, Strand WC2R 2LS, U.K.\\
E-mail: Nikolaos.Mavromatos@kcl.ac.uk}


\maketitle

\abstracts{I review a particular class 
of physical applications of Logarithmic Conformal 
Field Theory in strings propagating in changing (not necessarily 
conformal) backgrounds,
namely D-brane recoil in flat or time-dependent cosmological backgrounds. 
The role of recoil logarithmic vertex operators as non-conformal deformations
requiring in some cases Liouville dressing is pointed out. 
It is also argued that, although in the case of non-supersymmetric
recoil deformations the representation of target 
time as a Liouville zero mode may lead to 
non-linear quantum mechanics for stringy defects, such non-linearities
disappear (or, at least, are strongly suppressed) 
after world-sheet supersymmetrization. A possible 
link is therefore suggested
between (world-sheet) 
supersymmetry and linearity of quantum mechanics in 
this framework.}

\vspace{0.5cm}
\tableofcontents
\newpage

\section{Introduction}\label{sec:intro}

In this article, as a tribute to Ian Kogan, I would like to review
some work that I have done partly with him, in connection with 
some physical applications
of Logarithmic Conformal Field Theory (LCFT) to strings propagating in 
changing backgrounds. Such a situation is encountered, for instance,
when a macroscopic number of closed strings hits a D-particle, embedded in a
$d$-dimensional space time, and forces the D-particle to recoil via 
an impulse~\cite{kmw,ms1}. Equivalently, pairs of logarithmic operators 
may occur in some (nearly conformal) cosmological backgrounds of string theory,
such as late times Robertson-Walker (RW) 
cosmology~\cite{grav}. In the first case, 
the logarithmic pair consists of 
the velocity and the position of the non-relativistic recoiling brane defect,
while in the second example it is the cosmic velocity and acceleration
that enter in a logarithmic fashion.

In the context of non-supersymmetric 
D-particle recoil an interesting situation arises.
The target time dynamics of the recoiling defects can be described in terms 
of the (irreversible) flow of 
a renormalization-group (RG) scale on the world-sheet of the underlying 
$\sigma$-model describing the stringy excitations of the 
recoiling D-particle. This leads to non-linear terms in the 
associated ``evolution equation'' based on the identification of target time
with such a RG scale~\cite{emn,kogan}. The latter is nothing other than the zero mode of the 
associated Liouville $\sigma$-model field 
required for restoration of the conformal
invariance, which was broken by the recoil/impulse (non marginal) 
deformations. Upon going to the supersymmetric case, however, which is 
realized via 
appropriately supersymmetrized recoil operators, such non-linearities
disappear (or at least are strongly suppressed), and the associated 
evolution dynamics for the D-particles 
is that of a linear Schr\"odinger-like quantum mechanics.
This may have some interesting implications in linking supersymmetry
(of some form) to linearity of quantum mechanics. 
The key result in our analysis was that 
world-sheet leading ultraviolet (UV) divergences in an appropriate 
Zamolodchikov metric of the recoil operators, which in the 
non-supersymmetric (bosonic) string case lead to diffusion like terms
in the quantum evolution, and hence to non-linearities, 
cancel out in the supersymmetric
case, 
thereby leading to ordinary Schr\"odinger evolution
under the identification of the world-sheet  RG scale with target time.  
This latter result provides a highly non-trivial consistency check 
of the above identification, at least in this specific context.

Logarithmic conformal field theories~\cite{gurarie,lcftfurther} have
been attracting a lot of attention in recent years 
because of their diverse range of
applications, from condensed matter models of disorder~\cite{ckt,disorder} to
applications involving gravitational dressing of two-dimensional field
theories~\cite{liouv}, a general analysis of target space symmetries
in string theory~\cite{km}, D-brane recoil~\cite{kmw,ms1}, $AdS$
backgrounds in string theory and also M-theory~\cite{logads},
as well as $PP$-wave backgrounds in string theory\cite{sfetsos}
(see~\cite{LCFTRev} for reviews and more exhaustive lists of
references). They lie on the border between conformally invariant and
general renormalizable field theories in two dimensions. A logarithmic
conformal field theory is characterized by the property that its correlation
functions differ from the standard conformal field theoretic
ones by terms which contain logarithmic branch cuts. Nevertheless, it
is a limiting case of an ordinary conformal field theory which is
still compatible with conformal invariance and which can
still be classified to a certain extent by means of conformal data.

The current understanding of logarithmic conformal field theories
lacks the depth and generality that characterizes the conventional
conformally invariant field theories. Most of the analyses so far
pertain to specific models, and usually to those involving free field
realizations. Nevertheless, some general properties of logarithmic
conformal field theories are now very well understood. For example,
an important deviation from standard conformal field theory is the
non-diagonalizable spectrum of the Virasoro Hamiltonian operator
$L_0$, which connects vectors in a Jordan cell of a certain size. This
implies that the logarithmic operators of the theory, whose
correlation functions exhibit logarithmic scaling violations, come in
pairs, and they appear in the spectrum of a conformal field theory
when two primary operators become degenerate. It would be most
desirable to develop methods that would
classify and analyze the origin of logarithmic singularities in these
models in as general a way as possible, and in particular beyond
the free field prescriptions. Some modest steps in this direction have
been undertaken recently using different approaches. For instance, an
algebraic approach is advocated in~\cite{GabKausch1,Gab1} and used to
classify the logarithmic triplet theory as well as certain
non-unitary, fractional level Wess-Zumino-Witten (WZW) models. The
characteristic features of logarithmic conformal field theories are
described within this setting in terms of the representation theory of
the Virasoro algebra. An alternative approach to the construction of
logarithmic conformal field theories starting from conformally invariant
ones is proposed in~\cite{fuchs}. In this setting, logarithmic behavior arises
in extended models obtained by appropriately deforming the fields, including
the energy-momentum tensor, in the chiral algebra of an ordinary
conformal field theory.

{}From whatever point of view one wishes to look at logarithmic
conformal field theory, an important issue concerns the nature of the
extensions of these models to include worldsheet supersymmetry. In many
applications, most notably in string theory, supersymmetry plays a
crucial role in ensuring the overall stability of the target space theory.
A partial purpose of this review is to analyze in some detail the general
characteristics of the $N=1$ supersymmetric extension of logarithmic
conformal field theory. These models were introduced
in~\cite{CKLT}--\cite{MavSz}, 
where some features of the
Neveu-Schwarz~(NS) sector of the superconformal algebra were
described. In ref. \cite{MSsuper}
we extended and elaborated on these
studies, and further incorporated the Ramond~(R) sector of the
theory. In addition to unveiling some general features of logarithmic
superconformal field theories, in this article 
we shall also study in some detail how these
novel structures emerge in the super D-particle recoil 
problem~\cite{MSsuper} and connect it with the 
above mentioned problem of the linearity of the
emerging quantum mechanics of D-particles upon the identification
of the Liouville mode with target time.

The structure of the article is as follows: in section 2 we 
discuss the propagation of strings in recoiling D-brane
backgrounds embedded in both flat and (late times) 
Robertson-Walker cosmological
space times. We point out the emergence of a Logarithmic 
Conformal Field Theory as a result of the D-brane recoil.
Although our recoil formalism is general, however, for definiteness,
we restrict ourselves to the case of D-particles (D0-branes).
An interesting consequence of the LCFT is the 
possibility of the 
identification of target time with a world-sheet Renormalization Group (RG) 
scale (Liouville zero mode). 
In section 3 we review some consequences of this identification
in bosonic D-particles, in particular 
the emergence (due to recoil) of diffusion-like terms in the 
probability distribution for the position of the D-particles, and hence
non-linearities in the associated temporal evolution of 
their wavefunctionals. In section 4 we give a 
general description of N=1 superconformal logarithmic algebras
on the world sheet, which is used in section 5 to discuss
the recoil/impulse 
problem of supersymmetric D particles under their scattering 
from a (macroscopic) number of closed string states. 
It is shown that, as a result of special properties of the
world-sheet supersymmetric algebras involved, the 
non-linearities of the bosonic case, 
associated with diffusion like terms in the 
probability distribution, disappear (or, at least, are strongly suppressed),
thereby restoring the linear quantum mechanical Schr\"odinger evolution
of the recoiling super D-particle. This provides a non-trivial consistency
check of the r\^ole of time as a Liouville field in superstring theory.
Our conclusions are presented in section 6. 

\section{Strings in Changing Backgrounds and Logarithmic Conformal 
Field Theory}

\subsection{Logarithmic Conformal Field Theories}

The Virasoro algebra of a two-dimensional conformal field theory is generated
by the worldsheet energy-momentum tensor $T(z)$ with the operator product
expansion
\beq
T(z)\,T(w)=\frac{c/2}{(z-w)^4}+\frac2{(z-w)^2}\,T(w)+\frac1{z-w}\,\partial_w
T(w)+\dots \ ,
\label{TOPE2}\eeq
where $c$ is the central charge of the theory, and an ellipsis always denotes
terms in the operator product expansion which are regular as $z\to w$. For a
closed surface these relations are accompanied by their anti-holomorphic
counterparts, while for an open surface the coordinates $z,w$ are real-valued
and parametrize the boundary of the worldsheet. In the following we will be
concerned with the latter case corresponding to open strings and so will not
write any formulas for the anti-holomorphic sector. We shall always set the
worldsheet infrared scale to unity to simplify the formulas which follow.

The simplest logarithmic conformal field theory is characterized by a pair of
operators $C$ and $D$ which become degenerate and span a $2\times2$ Jordan cell
of the Virasoro operators. The two operators then form a logarithmic pair and
their operator product expansion with the energy-momentum tensor involves a
non-trivial mixing~\cite{gurarie,lcftfurther}
\bea
T(z)\,C(w)&=&\frac\Delta{(z-w)^2}\,C(w)+\frac1{z-w}\,\partial_wC(w)+\dots
 \ , \nn\\T(z)\,D(w)&=&\frac\Delta{(z-w)^2}\,D(w)+\frac1{(z-w)^2}\,C(w)
+\frac1{z-w}\,\partial_wD(w)+\dots \ ,
\label{TCD}\eea
where $\Delta$ is the conformal dimension of the operators determined by the
leading logarithmic terms in the conformal blocks of the theory, and an
appropriate normalization of the $D$ operator has been chosen. Because of
(\ref{TCD}), a conformal transformation $z\mapsto w(z)$ mixes the logarithmic
pair as
\beq
\begin{pmatrix} C(z)\cr D(z)\cr \end{pmatrix}=
\left(\frac{\partial w}{\partial z}
\right)^{\begin{pmatrix}\Delta&0\cr1&\Delta\cr \end{pmatrix}}\,
\begin{pmatrix} C(w)\cr D(w)\cr \end{pmatrix} \ ,
\label{conftransf}\eeq
from which it follows that their two-point functions are given
by~\cite{gurarie,lcftfurther}
\bea
\Bigl\langle C(z)\,C(w)\Bigr\rangle&=&0 \ , \nn\\
\Bigl\langle C(z)\,D(w)\Bigr\rangle&=&\frac\xi{(z-w)^{2\Delta}} \ , \nn\\
\Bigl\langle D(z)\,D(w)\Bigr\rangle&=&\frac1{(z-w)^{2\Delta}}\,
\Bigl(-2\xi\ln(z-w)+d\Bigr) \ ,
\label{CD2pt}\eea
where the constant $\xi$ is fixed by the leading logarithmic divergence of the
conformal blocks of the theory and the integration constant $d$ can be changed
by the field redefinition $D\mapsto D+({\rm const.})\,C$. The vanishing of the
$CC$ correlator in (\ref{CD2pt}) is equivalent to the absence of double or
higher logarithmic divergences. From these properties it is evident that the
operator $C$ behaves similarly to an ordinary primary field of scaling
dimension $\Delta$, while the properties of the $D$ operator follow from the
formal identification $D=\partial C/\partial\Delta$.

\subsection{Impulse Operators for Moving D-Branes}

The bosonic part of the vertex
operator describing the motion of the super D-brane is given by~\cite{ms1,kmw}
\bea
V_{\rm D}^{\rm bos}&=&\exp\left(-\frac1{2\pi\alpha'}\,\int\limits_\Sigma
d^2\sigma~\eta^{\alpha\beta}\,\partial_\alpha\left[Y_i\Bigl(x^0(\sigma)\Bigr)
\,\partial_\beta
x^i(\sigma)\right]\right)\nn\\&=&\exp\left(-\frac1{2\pi\alpha'}\,
\int\limits_0^1d\tau~Y_i\Bigl(x^0(\tau)\Bigr)\,\partial^{~}_{\!\perp} x^i(
\tau)\right) \ ,
\label{vertexD}\eea
where $\alpha'$ is the string slope,
$\partial_\alpha=\partial/\partial\sigma^\alpha$, and
$Y_i(x^0)=\delta_{ij}\,Y^j(x^0)$ describes the trajectory of the D0-brane as it
moves in spacetime.

The recoil of a heavy D-brane due to the scattering of closed string states may
be described in an impulse approximation by inserting appropriate factors of
the usual Heaviside function $\Theta(x^0)$ into (\ref{vertexD}). This describes
a non-relativistic 0-brane which begins moving at time $x^0=0$ from the initial
position $y_i$ with a constant velocity $u_i$. The appropriate trajectory is
given by the operator~\cite{kmw}
\beq
Y_i(x^0)=y_i\,C_{\epsilon}(x^0)+u_i\,D_{\epsilon}(x^0) \ ,
\label{Yirecoil}\eeq
where we have introduced the operators
\beq
C_\epsilon(x^0)=\alpha'\,\epsilon\,\Theta_\epsilon(x^0) \ , ~~
D_\epsilon(x^0)=x^0\,\Theta_\epsilon(x^0) \ ,
\label{CDops}\eeq
with $\Theta_\epsilon(x^0)$ the regulated step function which is defined by the
Fourier integral transformation
\beq
\Theta_\epsilon(x^0)=\frac1{2\pi i}\,\int\limits_{-\infty}^\infty\frac{d\omega}
{\omega-i\epsilon}~\e^{i\omega x^0} \ .
\label{ThetaFourier}\eeq
This integral representation is needed to make the Heaviside function
well-defined as an {\it operator}. In the limit $\epsilon\to0^+$, it reduces
via the residue theorem to the usual step function. The operator
$C_\epsilon(x^0)$ is required in (\ref{Yirecoil}) by scale invariance. Note
that the center of mass coordinate $y_i$ appears with a factor of
$\epsilon\to0^+$, so that the first operator in (\ref{Yirecoil}) represents a
small uncertainty in the initial position of the D-brane induced by stringy
effects~\cite{kmw}. The pair of fields (\ref{CDops}) are interpreted as
functions of the coordinate $z$ on the upper complex half-plane, which is
identified with the boundary variable $\tau$ in (\ref{vertexD}). This
interpretation is possible because the boundary vertex operator (\ref{vertexD})
is a total derivative and so can be thought of as a bulk deformation of the
underlying free bosonic conformal $\sigma$-model on $\Sigma$ (in the conformal
gauge). The impulse operator (\ref{vertexD},\ref{Yirecoil}) then describes the
appropriate change of state of the D-brane background because it has
non-vanishing matrix elements between different string states. It can be
thought of as generating the action of the Poincar\'e group on the 0-brane,
with $y_i$ parametrizing translations and $u_i$ parametrizing boosts in the
transverse directions.

By using the representation (\ref{ThetaFourier}) and the fact that the tachyon
vertex operator $\e^{i\omega x^0}$ has conformal dimension $\alpha'\omega^2/2$,
it can be shown~\cite{kmw} that the operators (\ref{CDops}) form a degenerate
pair which generate a logarithmic conformal algebra (\ref{TCD}) with conformal
dimension $\Delta=\Delta_\epsilon$, where
\beq
\Delta_\epsilon=-\frac{\alpha'\epsilon^2}2 \ .
\label{Deltavarep}\eeq
The total dimension of the impulse operator (\ref{vertexD},\ref{Yirecoil}) is
$h_\epsilon=1+\Delta_\epsilon$, and so for $\epsilon\neq0$ it describes a
relevant deformation of the underlying worldsheet conformal $\sigma$-model. The
existence of such a deformation implies that the resulting string theory is
slightly non-critical and leads to the change of state of the D-brane
background.

The two-point functions of the operators (\ref{CDops}) can be computed
explicitly to be~\cite{kmw}
\bea
\Bigl\langle C_\epsilon(z)\,C_\epsilon(w)\Bigr\rangle&=&\frac1{4\pi}\,
\sqrt{\frac{(\alpha')^3}{\epsilon^2\ln\Lambda}}\,\left[\frac{\sqrt\pi}2~
{}^{~}_1F^{~}_1\Bigl(\mbox{$\frac12$}\,,\,\mbox{$\frac12$}\,;4\epsilon^2\alpha'
\ln(z-w)\Bigr)\right.\nn\\&&-\left.2\,\sqrt{\epsilon^2\alpha'\ln(z-w)}~
{}^{~}_1F^{~}_1\Bigl(1\,,\,\mbox{$\frac32$}\,;4\epsilon^2\alpha'\ln(z-w)\Bigr)
\right] \ , \nn\\\Bigl\langle C_\epsilon(z)\,D_\epsilon(w)\Bigr\rangle&=&
\frac1{4\pi\epsilon^3}\,\sqrt{\frac1{2\alpha'\ln\Lambda}}\,\left[
\frac{\sqrt\pi}8~{}^{~}_1F^{~}_1\Bigl(\mbox{$\frac12$}\,,\,-\mbox{$\frac12$}\,;
4\epsilon^2\alpha'\ln(z-w)\Bigr)\right.\nn\\&&+\left.\frac{16}3\,\Bigl(
\epsilon^2\alpha'\ln(z-w)\Bigr)^{3/2}~{}^{~}_1F^{~}_1\Bigl(2\,,\,
\mbox{$\frac52$}\,;4\epsilon^2\alpha'\ln(z-w)\Bigr)\right] \ , \nn\\
\Bigl\langle D_\epsilon(z)\,D_\epsilon(w)\Bigr\rangle&=&
\frac1{\epsilon^2\alpha'}
\,\Bigl\langle C_\epsilon(z)\,D_\epsilon(w)\Bigr\rangle \ ,
\label{CDepsilon2pt}\eea
where $\Lambda\to0$ is the worldsheet ultraviolet cutoff which arises from the
short-distance propagator
\beq
\lim_{z\to w}\,\Bigl\langle x^0(z)\,x^0(w)\Bigr\rangle=-2\alpha'\ln\Lambda \ .
\label{shortdistprop}\eeq
Here we have used the standard bulk Green's function in the upper half-plane,
as the effects of worldsheet boundaries will not be relevant for the ensuing
analysis.\footnote{\baselineskip=12pt Boundary effects in logarithmic conformal
field theories have been analyzed in \cite{ms1,kw}.} 
This is again justified by
the bulk form of the vertex operator (\ref{vertexD}), and indeed it can be
shown that using the full expression for the propagator on the disc does not
alter any results~\cite{ms1}. It is then straightforward to see~\cite{kmw} that
in the correlated limit $\epsilon,\Lambda\to0^+$, with
\beq
\frac1{\epsilon^2}=-2\alpha'\ln\Lambda \ ,
\label{epsilonLambdarel}\eeq
the correlators (\ref{CDepsilon2pt}) reduce at order $\epsilon^2$ to the
canonical two-point correlation functions (\ref{CD2pt}) of the logarithmic
conformal algebra, with conformal dimension (\ref{Deltavarep}) and the
normalization constants
\beq
\xi=\frac{\pi^{3/2}\,\alpha'}2 \ , ~~ d=d_\epsilon
=\frac{\pi^{3/2}}{2\epsilon^2} \ .
\label{aconstsrecoil}\eeq
Note that the singular behavior of the constant $d_\epsilon$ in
(\ref{aconstsrecoil}) is not harmful, because it can be removed by considering
instead the connected correlation functions of the theory~\cite{kmw}.

\subsection{Target Space Formalism}

Let us now describe the target space properties of the logarithmic
(super)conformal algebra that we have derived. A worldsheet finite-size scaling
\beq
\Lambda~\longmapsto~\Lambda'=\Lambda~\e^{-t/\sqrt{\alpha'}}
\label{Lambdascale}\eeq
induces from (\ref{epsilonLambdarel}) a transformation of the target space
regularization parameter,
\beq
\epsilon~\longmapsto~\epsilon'=\epsilon+\epsilon^3\,t\,\sqrt{\alpha'}
+O(\epsilon^5) \ .
\label{epsilonscale}\eeq
By using (\ref{aconstsrecoil}) and the ensuing scale dependence of the
correlation functions (\ref{CD2pt}) we may then infer the transformation rules
\beq
C_{\epsilon'}=C_\epsilon \ , ~~ D_{\epsilon'}=D_\epsilon-\frac
t{\sqrt{\alpha'}}\,C_\epsilon
\label{CDscale}\eeq
to order $\epsilon^2$. It follows that, in order to maintain conformal
invariance, the $\sigma$-model coupling constants in (\ref{vertexDsusy}) must
transform as $y_i\mapsto y_i+(t/\sqrt{\alpha'}\,)\,u_i$, $u_i\mapsto u_i$, and
thus a worldsheet scale transformation leads to a Galilean boost of the D-brane
in target space. This provides a non-trivial indication that a 
world-sheet RG scale can be identified with the target time. 
In fact we shall discuss important consequences of this in section 3.

\subsection{Recoiling D-particles in Robertson-Walker backgrounds}

Above we discussed recoil in flat target space times. 
Placing D-branes in curved space times is not understood well at
present. The main problem originates from the lack of knowledge of the
complete dynamics of such solitonic objects. One would hope that such
a knowledge would allow a proper study of the back reaction of such
objects onto the surrounding space time geometry (distortion), and
eventually a consistent discussion of their dynamics in curved
spacetimes.  Some modest steps towards an incorporation of curved
space time effects in D-brane dynamics have been taken in the recent
literature from a number of authors~\cite{curved}.  These works are
dealing directly with world volume effects of D-branes and in some
cases string dualities are used in order to discuss the effects of
space time curvature.

A different approach has been adopted in~\cite{km},\cite{ms1},\cite{kmw},
\cite{recoil},
in which we have attempted to approach some aspects of the problem
from a world sheet view point, which is probably suitable for a study
of the effects of the (string) excitations of the heavy brane.  We
have concentrated mainly on heavy D-particles, embedded in a {\it
flat} target background space time.  We have discussed the
instantaneous action (impulse) of a `force' on a heavy
D-particle. The impulse may be viewed either as a consequence of
`trapping' of a \emph{macrosopic number} of closed string states on the
defect, and their eventual splitting into pairs of open strings, or,
in a different context, as the result of a more general phenomenon
associated with the \emph{sudden} appearance of such defects.  Our
world sheet approach is a valid approximation only if one looks at
times \emph{long after} the event.  Such impulse approximations usually
characterize classical phenomena. In our picture we view the whole
process as a \emph{semi-classical} phenomenon, due to the fact that the
process involves open string \emph{recoil} excitations of the heavy
D-particle, which are \emph{quantum} in nature.  It is this point of
view that we shall adopt in the present article.

Such an approach should be distinguished from the problem of studying
single-string scattering of a D-particle with closed string states in
flat space times~\cite{paban}.  We have shown
in~\cite{km},\cite{ms1},\cite{recoil} that for a D-particle embedded in a
$d$-dimensional \emph{flat Minkowski} space time such an impulse
action is described by a world-sheet $\sigma$-model deformed by
appropriate `recoil' operators, which obey a logarithmic conformal
algebra~\cite{lcftfurther}. The appearance of such algebras, which lie on the
border line between conformal field theories and general
renormalizable field theories in the two-dimensional world sheet, but
can still be classified by conformal data, is associated with the fact
that an impulse action (recoil) describes a \emph{change} of the
string/D-particle background, and as such it cannot be described by
conformal symmetry all along.  The \emph{transition} between the two
asymptotic states of the system before and (long) after the event is
precisely described by deforming the associated $\sigma$-model by
operators which \emph{spoil} the conformal symmetry.

In this section we shall extend~\cite{grav} the flat space time results
of~\cite{km,kmw,recoil}, reviewed above, 
to the physically relevant case of a
Robertson-Walker (RW) cosmological background space time.  As well
known in string $\sigma$-model perturbation theory, Robertson-Walker
space times are not solutions of conformal invariance conditions of
the $\sigma$-model, in the sense of having $\sigma$-model
$\beta$-function different from zero.  This would affect in general
the two point correlators (c.f.\ below) $\langle X^\mu (z) X^\nu (w)
\rangle$ which is modified from the standard $G^{\mu\nu}\ln |z-w|^2$ form by the inclusion of $\beta^{\mu\nu}$-dependent
terms. Nevertheless, in the particular case of (large) cosmological
times we are interested in here, which describe well the present era
of the Universe, such terms are subleading, given that $\beta^{\mu\nu}
\propto R^{\mu\nu} \sim 1/t^2$, and thus can be safely neglected.  In
this sense, discussing recoil in such (almost conformal) backgrounds
is a physically interesting and non-trivial exercise in conformal
field theory, which we would like to pursue here.

Although, our results do not depend on the target space dimension,
however, for definiteness we shall concentrate on the case of a
D-particle embedded in a four-dimensional RW spacetime.  It must be
stressed that we shall not attempt here to present a complete
discussion of the associated space time curvature effects, which - as
mentioned earlier - is a very difficult task, still unresolved.
Nevertheless, by concentrating on times much larger than the moment of
impulse on the D-particle defect, one may ignore such effects to a
satisfactory approximation.  As we shall see, our analysis produces
results which look reasonable and are of sufficient interest to
initiate further research.

The vertex operators which describe the impulse in curved RW
backgrounds obey a suitably extended (higher-order) logarithmic
algebra.  The algebra is valid at, and in the neighborhood of, a
non-trivial infrared fixed point of the world-sheet Renormalization
Group.  For a RW spacetime of scale factor of the form $t^p$, where
$t$ is the target time, and $p > 1$ in the horizon case, the algebra
is actually a set of logarithmic algebras up to order $[2p]$, which
are classified by the appropriate higher-order Jordan
blocks~\cite{gurarie,lcftfurther}.

As in the flat case, which is obtained as a special limit of this more
general case, the recoil deformations are relevant operators from a
world-sheet Renormalization-Group viewpoint.  One distinguishes two
cases. In the first, the initial RW spacetime does not possess
cosmological horizons.  In this case it is shown that the limit to the
conformal world-sheet non-trivial (infrared) fixed point can be taken
smoothly without problems and one has a standard logarithmic algebra.
On the other hand, in the case where the initial spacetime has {\it
cosmological horizons}, such a limit is plagued by world-sheet
divergences.  These should be carefully subtracted in order to allow
for a smooth approach to the fixed point, leading to {\it
higher-order} logarithmic algebras.  We find this an interesting
result which requires further study. The presentation 
below follows that in \cite{grav}, where we refer the reader
for more details.

\subsubsection{Geodesic paths and recoil} 

Let us consider a D-particle, located (for convenience) at the
origin of the spatial coordinates of a four-dimensional space time,
which at a time $t_0$ experiences an impulse.  In a $\sigma$-model
framework, the trajectory of the D-particle $y^i(t)$, $i$ a spatial
index, is described by inserting the following vertex operator
\begin{equation} \label{path}
V = \int _{\partial \Sigma} G_{ij}y^j(t)\partial_n X^i
\end{equation}
where $G_{ij}$ denotes the spatial components of the metric, $\partial
\Sigma$ denotes the world-sheet boundary, $\partial _n$ is a normal
world-sheet derivative, $X^i$ are $\sigma$-model fields obeying
Dirichlet boundary conditions on the world sheet, and $t$ is a
$\sigma$-model field obeying Neumann boundary conditions on the world
sheet, whose zero mode is the target time.

This is the basic vertex deformation which we assume to describe the
motion of a D-particle in a curved geometry to leading order at least,
where spacetime back reaction and curvature effects are assumed weak.
Such vertex deformations may be viewed as a generalization of the
flat-target-space case~\cite{dparticle}.

Perhaps a formally more desirable approach towards the construction of
the complete vertex operator would be to start from a T-dual (Neumann)
picture, where the deformation~(\ref{path}) should correspond to a
proper Wilson loop operator of an appropriate gauge vector field. Such
loop operators are by construction independent of the background
geometry. One can then pass onto the Dirichlet picture by a T-duality
transformation viewed as a canonical transformation from a
$\sigma$-model viewpoint~\cite{otto}. In principle, such a procedure
would yield a complete form of the vertex operator in the Dirichlet
picture, describing the path of a D-particle in a curved geometry.
Unfortunately, such a procedure is not free from ambiguities at a
quantum level~\cite{otto}, which are still unresolved for general
curved backgrounds.  Therefore, for our purposes here, we shall
consider the problem of writing a complete form for the
operator~(\ref{path}) in a RW spacetime background in the Dirichlet
picture as an open issue. Nevertheless, for RW backgrounds at large
times, ignoring curvature effects proves to be a satisfactory
approximation, and in such a case one may consider the vertex
operator~(\ref{path}) as a sufficient description for the physical
vertex operator of a D-particle.  As we shall show below, the results
of such analyses appear reasonable and interesting enough to encourage
further studies along this direction.

For times long after the event, the trajectory $y^i(t)$ will be that
of free motion in the curved space time under consideration. In the
flat space time case, this trajectory was a straight
line~\cite{dparticle,kmw,ms1}, and in the more general case here it
will be simply the associated \emph{geodesic}.  Let us now determine
its form, which will be essential in what follows.

The space time assumes the form:
\begin{equation}\label{rwmetric}
ds^2 = -dt^2 + a(t)^2 (dX^i)^2  
\end{equation}
where $a(t)$ is the RW scale factor. We shall work with 
expanding RW space times with scale factors 
\begin{equation}  
a(t) =a_0 t^p\,, \qquad p \in R^+ 
\end{equation} 
The geodesic equations in this case read:
\begin{eqnarray}
{\ddot t} + pt^{2p-1}({\dot y}^i)^2 &=& 0 
\nonumber \\
{\ddot y} + 2\frac{p}{t}({\dot y}^i) {\dot t}&=& 0
\end{eqnarray}
where the dot denotes differentiation with respect to the proper time
$\tau$ of the D-particle.

With initial conditions $y^i(t_0)=0$, and $dy^i/dt (t_0) \equiv v^i$,
one easily finds that, for long times $t \gg t_0$ after the event, the
solution acquires the form:
\begin{equation}\label{pathexpre} 
y^i(t) =\frac{v^i}{1-2p}\left(t^{1-2p}t_0^{2p} - t_0 \right) + {\cal
O}\left(t^{1-4p}\right), \qquad t \gg t_0
\end{equation} 

To leading order in $t$, therefore, the appropriate vertex operator
(\ref{path}), describing the recoil of the D-particle, is:
\begin{equation}\label{path2} 
V=\int _{\partial \Sigma} a_0^2 \frac{v^i}{1-2p}\Theta
(t-t_0)\left(tt_0^{2p}-t_0t^{2p}\right) \partial_n X^i
\end{equation} 
where $\Theta (t-t_0)$ is the Heaviside step function, expressing an
instantaneous action (\emph{impulse}) on the D-particle at
$t=t_0$~\cite{kmw,recoil}.  As we shall see later on, such deformed
$\sigma$-models may be viewed as providing rather generic mathematical
prototypes for models involving phase transitions at early stages of
the Universe, leading effectively to time-varying speed of light.  In
the context of the present work, therefore, we shall be rather vague
as far as the precise physical significance of the
operator~(\ref{path2}) is concerned, and merely exploit the
consequences of such deformations for the expansion of the RW
spacetime after time $t_0$, from both a mathematical and physical
viewpoint.

\begin{figure}[ht]
\centerline{\epsfxsize=3.9in\epsfbox{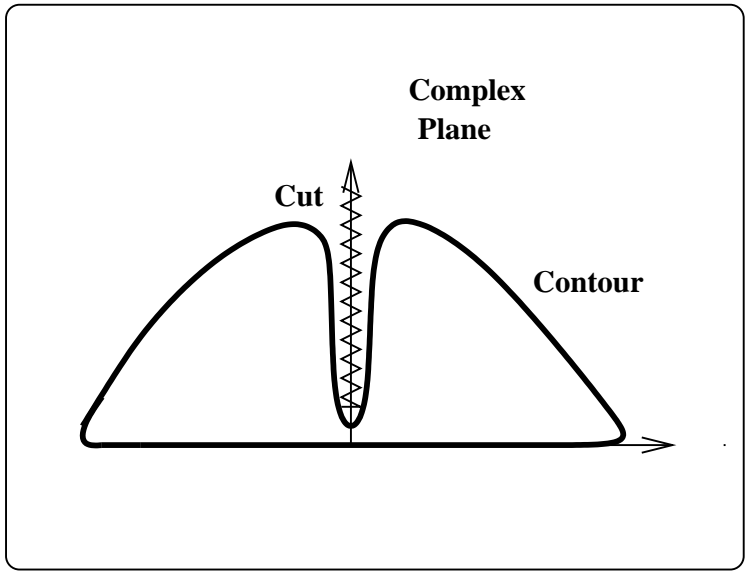}}   
\caption{Contour of integration in the
complex plane to define the recoil operators ${\cal D}^{(q)}$, by
proper treatment of the associated cuts.\label{cut}}
\end{figure}

In~\cite{kmw}, we have studied the case $p=0$, $a_0=1$, where the
operators assumed the form $t\Theta_\epsilon (t)$ to leading order in
$t$, where $\Theta_\epsilon (t)$ is the regulated form of the step
function, given by~\cite{kmw}:
\begin{equation}\label{rep} 
\Theta_\epsilon =-i\int_{-\infty}^{+\infty} \frac{d\omega}{2\pi}
\frac{1}{\omega -i\epsilon} e^{i\omega\,t}\,, \qquad \epsilon
\rightarrow 0^+
\end{equation}
As discussed in that reference, this operator forms a logarithmic
pair~\cite{lcftfurther} with $\epsilon \Theta_\epsilon (t)$, expressing
physically fluctuations in the initial position of the D-particle.

In the current case, one may expand the integrand of~(\ref{path2}) in
a Taylor series in powers of $(t-t_0)$, which implies the presence of
a series of operators, of the form $(t-t_0)^q\Theta_\epsilon (t-t_0)$,
where $q $ takes on the values $2p, 2p-1, \ldots $, i.e.\ it is not an
integer in general.  In a direct generalization of the Fourier
integral representation (\ref{rep}), we write in this case:
\begin{eqnarray}\label{opd}  
{\cal D}^{(q)} \equiv v_i (t-t_0)^q \Theta_\epsilon
(t-t_0)\partial_n X^i &=&v_i~N_q~\int _{-\infty}^{+\infty}d\omega
\frac{1}{(\omega -i\epsilon)^{q+1}}~e^{i\omega(t-t_0)}\partial_nX^i~,
\nonumber \\
N_q \equiv \frac{i^q}{\Gamma (-q)(1-e^{-i2\pi q})}&=&
\frac{(-i)^{q+1}\Gamma (q+1)}{2\pi}~,
\end{eqnarray} 
where we have incorporated the velocity coupling $v_i$ in the
definition of the $\sigma$-model deformation, and we have defined the
integral along the contour of figure~\ref{cut}, having chosen the cut to be
from $+i\epsilon$ to $+i\infty$.

\subsubsection{Extended logarithmic world-sheet algebra of recoil in RW 
backgrounds}

Following the flat space time analysis of~\cite{kmw}, we now proceed
to discuss the conformal structure of the recoil operators in RW
backgrounds.  We shall do so by acting on the operator ${\cal D^{(q)}}$~(\ref{opd}) with the world-sheet energy momentum tensor
operator $T_{zz} \equiv T$ (in a standard notation).  Due to the form
of the background space time~(\ref{rwmetric}), the stress tensor $T$
assumes the form
\begin{equation}\label{stress}  
2T =-(\partial t)^2 + a^2(t) (\partial X^i)^2
\end{equation}
where, from now on, $\partial \equiv \partial _z$, unless otherwise
stated.  One can then obtain the relevant operator-product expansions
(OPE) of $T$ with the operators ${\cal D}^{(q)}$. For convenience in
what follows we shall consider the action of each of the two terms
in~(\ref{stress}) on the operators ${\cal D}^{(q)}$ separately. For
the first (time $t$-dependent part), one has, as $z \to w$:
\begin{eqnarray} 
-\frac{1}{2}(\partial t(z))^2\cdot{\cal D}^{(q)}(w) &=&
\frac{v_i}{(z-w)^2}\left[ N_q\int _{-\infty}^{+\infty}d\omega
  \frac{\omega^2/2}{(\omega -i\epsilon)^{q+1}}~e^{i\omega~t(w)}\right]
\partial_nX^i =
\nonumber \\
&=&\frac{1}{(z-w)^2}\left[-\frac{\epsilon^2}{2}{\cal D}^{(q)} +
q\epsilon {\cal D}^{(q-1)} + \frac{q(q-1)}{2}{\cal D}^{(q-2)} \right]
\nonumber \\\end{eqnarray} 
The above formul\ae\, were derived for asymptotically large time $t$,
assuming the two-point correlators
\begin{equation}\label{correl}
\langle X^\mu(z) X^\nu(w) \rangle = 2G^{\mu\nu}\ln |z-w|^2 +
\cdots\,,
\end{equation}
where the $\cdots $ denote terms with negative powers of $t$, related
to space-time curvature, which are subleading in the limit $t \to
\infty$.

At this point, we stress again that Robertson-Walker space times are
not solutions of conformal invariance conditions of the
$\sigma$-model, as having $\beta$-functions different from zero.
Non-zero $\beta$ functions affect in general the two point correlators
(\ref{correl}) by $\beta^{\mu\nu}$-dependent terms. In the particular
case of (large) cosmological times, however, which describe well the
present era of the Universe we are interested in here, such terms are
subleading, given that $\beta^{\mu\nu} \propto R^{\mu\nu} \sim 1/t^2$,
and thus can be safely neglected in the limit $t \to \infty$.

For the spatial part of~(\ref{stress}) we consider the OPE $a(t(z))^2
(\partial X^i(z))^2 {\cal D}^{(q)}(w)$ as $z \to w$. Again, for
convenience we shall do the time and space contractions separately:
\begin{eqnarray}
t^{2p}(z)\cdot {\cal D}^{(q)}(w) &=& \int d\omega 
\tilde{\cal{D}}^{(q)}(\omega) t^{2p}(z)\cdot e^{i\omega (t(w)-t_0)} = 
\nonumber \\
&=& \int _0^{\infty} \frac{d\nu}{\Gamma (-2p)} \nu^{-1-2p} \int
d\omega \tilde{\cal{D}}^{(q)}(\omega) e^{-\nu t(z)}\cdot e^{i\omega
(t(w)-t_0)}
\end{eqnarray} 
Using the OPE $e^{-\nu t(z)} \cdot e^{-i\omega (t(w)-t_0)} \sim
|z-w|^{i\nu\omega}e^{-\nu t(z) - i\omega (t(z)-t_0) + {\cal O}(z-w)}$
one obtains (as $z \sim w$):
\begin{eqnarray} 
t(z)^{2p}\cdot{\cal D}^{(q)}(w)&=&
\int _0^\infty \frac{d\nu}{\Gamma (-2p)}\nu^{-1-2p}e^{-\nu t(z)}
{\cal D}^{(q)} (t-t_0 -\nu\ln |z-w|) = 
\\
&=& t^{2p}~\int _0^\infty \frac{d\nu}{\Gamma (-2p)}\nu^{-1-2p}e^{-\nu}
{\cal D}^{(q)} (t-t_0 -\frac{\nu}{t}\ln |z-w|) = 
\nonumber \\ 
&=& t^{2p}\left[{\cal D}^{(q)}(t-t_0) -\frac{1}{t}\ln |z-w|
\frac{\Gamma (1-2p)}{\Gamma (-2p)}\frac{d}{dt}{\cal D}^{(q)}(t-t_0)
+ {\cal O}(t-t_0)^{q-2}\right].
\nonumber 
\end{eqnarray}
We now observe that $\frac{d{\cal D}^{(q)}}{dt}=q{\cal D}^{(q-1)} -
\epsilon~{\cal D}^{(q)} $, where both terms have vacuum expectation
values of the same order in $\epsilon$, as we shall see below, and
hence both should be kept in our perturbative expansion.
 
Expanding the various terms around $t_0$, 
$$
t^{s}=(t-t_0)^{s} + s~t_0(t-t_0)^{s-1}+
\frac{t_0^{2}}{2}(s)(s-1)(t-t_0)^{s-2}+{\cal O}([t-t_0]^{s-3})\,,
$$ 
one has:
\begin{eqnarray}  
 t^{2p}(z)\cdot{\cal D}^{(q)}(w) &=& {\cal D}^{(2p+q)}(t-t_0)+ 
\left(2p~t_0 - 2p~\epsilon~\ln |z-w|\right)~{\cal D}^{(2p+q-1)}+ 
\nonumber \\
&& {}+ \left(\frac{t_0^{2}}{2}2p(2p-1)+\!\left[2pq + 
(2p-4p^2)\epsilon~t_0\right]\! \ln |z-w|\!\right)
{\cal D}^{(2p+q-2)}(t-t_0)\!+ 
\nonumber \\
&& {}+ {\cal O}\left([t-t_0]^{2p+q-3}\right)
\end{eqnarray}
where it is worthy of mentioning that inside the subleading terms
there are higher logarithms of the form $\ln ^n|z-w|$, where 
$n =2,3,4, \ldots$

We now come to the OPE between the spatial parts. In view of
(\ref{correl}), upon expressing $\partial _z$ in normal $\partial_n$
and tangential parts, and imposing Dirichlet boundary conditions on
the world-sheet boundary where the operators live on, we observe that
such operator products take the form:
\begin{equation}
(\partial X^j(z))^2 \cdot \partial_n X^i(w) \sim
G^{ii}\frac{1}{(z-w)^2}\partial_n X^i \sim \frac{t^{-2p}}{(z-w)^2}
\partial_n X^i\,, \qquad 
(\hbox{no sum over}~i)
\end{equation} 
Performing the last contraction with the $t^{-2p}$, following the
previous general formul\ae\, and collecting appropriate terms, one
obtains:
\begin{eqnarray}\label{elalg}
T(z)\cdot{\cal D}^{(q)}[(t-t_0)(w)] 
&=&\frac{1-\frac{\epsilon^2}{2}}{(z-w)^2}{\cal D}^{(q)} [(t-t_0)(w)] +
\frac{q\epsilon}{(z-w)^2}{\cal D}^{q-1}[(t-t_0)(w)] + 
\nonumber\\
&& {}+\! \frac{\frac{q(q-1)}{2} -2p^{2}\ln |z-w| -2p^{2}\epsilon^2\ln
  ^2|z-w|}{(z-w)^2}{\cal D}^{(q-2)} [(t-t_0)(w)]\! + 
\nonumber\\
&& {}+{\cal O}([t-t_0]^{q-3}) 
\end{eqnarray} 
where again inside the subleading terms there are higher logarithms.

We next notice that, as a consistency check of the formalism, one can
calculate the OPE~(\ref{elalg}) in case one considers matrix elements
between \emph{on-shell} physical states.  In the context of
$\sigma$-models, we are working with, the physical state condition
implies the constraint of the vanishing of the world-sheet
stress-energy tensor $2T=-(\partial t)^2 + a(t)^2(\partial X^i)^2 =0$.
This condition allows $(\partial X^i)^2 $ to be expressed in terms of
$(\partial t)^2$, which is consistent even at a correlation function
level in the case of very target times $t \gg t_0$, since in that
case, the correlator $\langle X^i t \rangle$ is subleading, as
mentioned previously.  Implementing this, it can be then seen that the
OPE between the spatial parts of $T$ and ${\cal D}^{(q)}$ is:
\begin{eqnarray} 
\lefteqn{a^2(t)(\partial X^i)^2 \cdot {\cal D}^{(q)} =}\ &&
\nonumber\\&&
= t^{-2p}(\partial t)^2\cdot \biggl\{ {\cal D}^{(2p+q)}(t-t_0)+
\left(2p t_0 - 2p \epsilon \ln 
|z-w|\right) {\cal D}^{(2p+q-1)}+ 
\nonumber \\ && 
\hphantom{=t^{-2p}(\partial t)^2\cdot \biggl\{}
+\left(\frac{t_0^{2}}{2}2p(2p-1)+\left[2pq +
  \left(2p-4p^2\right)\epsilon t_0\right] \ln |z-w|\!\right) {\cal
  D}^{(2p+q-2)}(t-t_0)+ 
\nonumber \\ && 
\hphantom{=t^{-2p}(\partial t)^2\cdot \biggl\{}
+{\cal O}\left([t-t_0]^{2p+q-3}\right)\biggr\}.
\end{eqnarray} 
Performing the appropriate contractions, and adding to this result the
OPE of the temporal part of $T$ with ${\cal D}^{(q)}$, i.e.\ the
quantity $-\frac{\epsilon^2}{2}{\cal D}^{(q)} + q\epsilon {\cal
D}^{(q-1)} + \frac{1}{2}q(q-1){\cal D}^{(q-2)}$, we obtain:
\begin{eqnarray}
T\cdot {\cal D}^{(q)}|_{\rm on-shell}&=&\left(-2p\epsilon
-pt_0\epsilon^2 + p\epsilon^2\ln \left(\frac aL\right)\right){\cal
  D}^{(q-1)} +
\nonumber \\
&& {}+ \biggl\{t_0^2\epsilon^2 2p(2p+1) - 3\epsilon^2p(2p+q)\ln
\left(\frac aL \right) 
- 2\epsilon^3\left(p+p^2\right)t_0 \ln \left( \frac aL \right) -
\nonumber \\&&
\hphantom{+ \biggl\{}
-2p^2\epsilon^4\ln ^2\left(\frac aL\right)+  
\epsilon (2p+q)2p t_0 - \left(4p^2 + 4pq - 2p\right)\biggr\}
{\cal D}^{(q-2)} + 
\nonumber\\&&
{}+{\cal O}\left([t-t_0]^{q-3}\right)
\label{onshell}
\end{eqnarray}
From the above we observe that the on-shell operators become marginal
as they should, given that an on-shell theory ought to be conformal.
Moreover, and more important, the world-sheet divergences
\emph{disappear} upon imposing the condition
\begin{equation}\label{xi0}  
\epsilon^{2}\ln \left(\frac La \right)^2 = \xi_0 = {\rm
constant~independent~of}~\epsilon,~a,~L
\end{equation}
where $L$ ($a$) is the world-sheet (ultraviolet) infrared cut-off on
the world sheet.  As we shall discuss later on, this condition will be
of importance for the closure of the logarithmic algebra, which
characterizes the fixed point~\cite{kmw}. Hence, the conformal
invariance is preserved by the on-shell states, any dependence from it
being associated with \emph{off-shell} states.

We next notice that, in the context of the RW metric~(\ref{rwmetric}),
there are two cases of expanding universes, one corresponding to $0< p
\le 1$, and the other to $p > 1$.  Whenever $p \le 1$ (which notably
incorporates the cases of both radiation and matter dominated
Universes) there is \emph{no horizon}, given that the latter is given
by:
\begin{equation}\label{horizon}
 \delta (t) = a(t)\int_{t_0}^\infty \frac{dt'}{a(t')}
\end{equation} 
In this case the relevant value for $q$ is $q=2p \le 2$.  On the other
hand, for the case $p > 1$, i.e.\ $q > 2$ there is a non-trivial
cosmological \emph{horizon}, which as we shall see requires special
treatment from a conformal symmetry viewpoint.

We commence with the no-horizon case, $1 < q \le 2$.  We first notice
that the linear in $t$ term in~(\ref{path2}) leads to the conventional
logarithmic algebra, discussed in~\cite{kmw}, corresponding to a pair
of impulse (`recoil') operators $C,D$. The main point of our
discussion below is a study of the $t^{2p}$ terms in~(\ref{path2}),
and their connection to other logarithmic algebras.  Indeed, we
observe that a logarithmic algebra~\cite{lcftfurther,km,kmw} can be
obtained for these terms of the operators, if we define ${\cal D}
\equiv {\cal D}^{(q)}$ and ${\cal C} \equiv q\epsilon {\cal
D}^{(q-1)}$.  In this case we have the following OPE with $T$:
\begin{eqnarray}\label{tdope}
(z-w)^2~T\cdot {\cal D}&=&\left(1-\frac{\epsilon^2}{2}\right){\cal D} + {\cal
    C}\,, 
\nonumber \\
(z-w)^2~T\cdot {\cal C}&=&\left(1-\frac{\epsilon^2}{2}\right){\cal C} + {\cal
    O}\left([t-t_0]^{q-2}\right), 
\end{eqnarray}
where throughout this work we ignore terms with negative powers in
$t-t_0$ (e.g.\ of order $q-2$ and higher), for large $t\gg t_0$.
Notice that in the case $q < 1$ (i.e.\ $p < 1/2$) the ${\cal C}$
operator defined above is absent.

In the second case $p > 1$ one faces the problem of having
cosmological horizons (cf.~(\ref{horizon})), which recently has
attracted considerable attention in view of the impossibility of
defining a consistent scattering $S$-matrix for asymptotic
states~\cite{challenge,emnsmatrix}.  In this case the operator ${\cal
D}^{(q-2)}$ is \emph{not subleading} and one has an \emph{extended
(higher-order) logarithmic algebra} defined by~(\ref{elalg}).  It is
interesting to remark that now the logarithmic world-sheet terms in
the coefficient of the ${\cal D}^{(q-2)}$ operator imply that the
limit $z \to w$ is plagued by ultraviolet world-sheet divergences, and
hence the world-sheet conformal invariance is spoiled. This
necessitates Liouville dressing, in order to restore the conformal
symmetry~\cite{ddk}.  Such a dressing implies the presence of an extra
space-time dimension given by the Liouville mode.  The signature
depends on the signature of the central charge deficit.  We shall not
deal with this procedure further in this article, the reason being
that the RW background is itself \emph{not conformal}.

We now turn to a study of the correlators of the various ${\cal
D}^{(q)}$ operators, which will complete the study of the associated
logarithmic algebras, in analogy with the flat target-space case
of~\cite{kmw}.  From the algebra~(\ref{elalg}) we observe that we need
to evaluate correlators between ${\cal D}^{(q)}, {\cal
D}^{(q-n)}$, $n=0,1,2, \dots$  We shall evaluate correlators $\langle
\dots \rangle $ with respect to the free world-sheet action, since we
work to leading order in the (weak) coupling $v_i$.  For convenience
below we shall restrict ourselves only to the time-dependent part of
the operators ${\cal D}$. The incorporation of the $\partial_n X^i$ is
trivial, and will be implied in what follows.  With these in mind one
has:
\begin{equation}\label{qqn}  
\langle {\cal D}^{(q)}(z) {\cal D}^{(q-n)}(w) \rangle = N_q N_{q-n}
\int \int_{-\infty}^{+\infty} \frac{d \omega d\omega '}{ (\omega -
i\epsilon)^{q+1}(\omega' - i\epsilon)^{q-n+1}} \langle e^{-i\omega
t(z)}~e^{-i\omega' t(w)} \rangle
\end{equation}
where $\epsilon \to 0^+$.  As already mentioned, we work to leading
order in time $t \gg \infty$, and hence we can we apply the
formula~(\ref{correl}) for two-point correlators of the $X^\mu$ fields
to write\footnote{Here we use simplified propagators on the boundary,
with the latter represented by a straight line; this means that the
arguments of the logarithms are real~\cite{kmw}.  To be precise, one
should use the full expression for the propagator on the disc, along
the lines of~\cite{ms1}. As shown there, and can be checked here as
well, the results are unaffected.}
\begin{eqnarray}
\langle e^{-i\omega t(z)}~e^{-i\omega' t(w)}\rangle  
&=&e^{-\frac{\omega^2}{2}\langle t(z) t(z)\rangle -
\frac{\omega^{'2}}{2}\langle t(w) t(w)\rangle - 
\omega \omega'\langle t(z) t(w)\rangle} = 
\nonumber \\
 &=& e^{-(\omega+\omega')^2\ln (L/a)^2 + 2\omega\omega'\ln (|z-w|/a)^2}\,,
\end{eqnarray} 
where we took into account that ${\rm Lim}_{z\to w}~\langle
t(z)t(w)\rangle = -2\ln (a/L)^2$. Given that $\ln (L/a)$ is very
large, one can approximate 
$$
e^{-(\omega+\omega')^2\ln (L/a)^2} \simeq
\frac{\sqrt{\pi}}{\sqrt{\ln (L/a)^2}}\delta(\omega+\omega')\,.
$$  
Thus we obtain:
\begin{eqnarray} 
\langle {\cal D}^{(q)}(z) {\cal D}^{(q-n)}(w) \rangle 
&=& (-1)^{-q+n-1} N_q N_{q-n}{\cal J}_n^{(q)}\,,
\nonumber\\
{\cal J}_n^{(q)} &\equiv&
\sqrt{\frac{\pi}{\alpha}}\int_{-\infty}^{+\infty}
\frac{d\omega e^{-\omega^2 \lambda} (\omega+i\epsilon)^{n}}{(\omega^2
+ \epsilon^2)^{q+1}}
\end{eqnarray}
where $\lambda \equiv 2\ln (|z-w|/a)^2$, and $\alpha \equiv\ln
(L/a)^2$.

Below, for definiteness, we shall be interested in the case $2<q<3$,
in which the relevant correlators are given by $n=0,1,2$.  One has:
\begin{eqnarray}\label{defj} 
{\cal J}_{0}^{(q)}&=&\sqrt{\frac{\pi}{\alpha}} \epsilon^{-2q-1} f_q
\left(\epsilon^2 \lambda\right)\,; 
\nonumber \\
f_q(\xi) &=&\sqrt{\pi}\frac{\Gamma(\frac{1}{2}+q)}{\Gamma (1 + q)}
F\left(\frac{1}{2}, \frac{1}{2}-q ;\xi\right) 
+ \xi^{\frac{1}{2}+q} \Gamma\left(-\frac{1}{2}-q\right) F\left(1+q,
\frac{3}{2}+q ;\xi\right)  
\nonumber \\
{\cal J}_{1}^{(q)} &=& i\epsilon{\cal J}_{0}^{(q)} \,, 
\nonumber \\
{\cal J}_{2}^{(q)}&=&-2\epsilon^2 {\cal J}_0^{(q)} + {\cal
  J}_0^{(q-1)}=-\frac{\partial}{\partial \lambda}{\cal
  J}_0^{(q)}-\epsilon^2{\cal J}_0^{(q)} 
\end{eqnarray} 
where $F(a,b;z)=1 + \frac{a}{b}\frac{z}{1!}+
\frac{a(a+1)}{b(b+1)}\frac{z^2}{2!} + \cdots $ is the degenerate
(confluent) hypergeometric function.  Thus, the form of the algebra
away from the fixed point (`\emph{off-shell form}'), i.e.\ for
$\epsilon^2 \ne 0$,~is:
\begin{eqnarray}\label{offshellalg}   
\langle {\cal D}^{(q)}(z) {\cal D}^{(q)}(0) \rangle &=&
\nonumber\\
\lefteqn{\hspace{-1.5cm} ={\tilde N}_q^2
\sqrt{\frac{\pi}{\xi_0}} \biggl(f_q(2\xi_0)
\left(\frac{\alpha}{\xi_0}\right)^q +
2  f'_q(2\xi_0)\left(\frac{\alpha}{\xi_0}\right)^{q-1} \ln 
\left(\left|\frac zL \right|^2\right) +}
\nonumber\\
\lefteqn{\hspace{.7cm} {}
+ \frac{1}{2} f''_q(2\xi_0)\left(\frac{\alpha}{\xi_0}\right)^{q-2} 
4\ln ^2\left(\left|\frac zL \right|^2\right)
+ {\cal O}\left(\alpha^{q-3}\right)\biggr) \,,} 
\nonumber \\
\epsilon q \langle {\cal D}^{(q)}(z) {\cal D}^{(q-1)}(0) \rangle &=&
\nonumber\\
\lefteqn{ \hspace{-3cm}={\tilde N}_q^2\sqrt{\frac{\pi}{\xi_0}}
  \biggl(f_q(2\xi_0) 
\left(\frac{\alpha}{\xi_0}\right)^{q-1} 
+ 2 f'_q(2\xi_0)\left(\frac{\alpha}{\xi_0}\right)^{q-2} 
\ln \left(\left|\frac zL \right|^2\right) + {\cal
  O}\left(\alpha^{q-3}\right)\biggr) \,, } 
\nonumber \\
\epsilon^2 q^2 \langle {\cal D}^{(q-1)}(z) {\cal D}^{(q-1)}(0)\rangle
&=& {\tilde N}_q^2 \sqrt{\frac{\pi}{\xi_0}} f_{q-1}(2\xi_0)
\left(\frac{\alpha}{\xi_0}\right)^{q-2} + {\cal O}\left(\alpha^{q-3}\right), 
\nonumber \\
\epsilon^2 q (q-1) \langle {\cal D}^{(q)}(z) {\cal D}^{(q-2)}(0)\rangle
&=& -{\tilde N}_q^2 \sqrt{\frac{\pi}{\xi_0}}  
\left(f_q(2\xi_0) + f'_q(2\xi_0)\right)
\left(\frac{\alpha}{\xi_0}\right)^{q-2} 
+ {\cal O}\left(\alpha^{q-3}\right) , 
\nonumber \\
\epsilon^3 q^2(q-1) \langle {\cal D}^{(q-1)}(z) {\cal D}^{(q-2)}(0)\rangle
&=& {\cal O}\left(\alpha^{q-3}\right) , 
\nonumber \\
\epsilon^4 q^2(q-1)^2 \langle {\cal D}^{(q-2)}(z) {\cal D}^{(q-2)}(0)\rangle
&=& {\cal O}\left(\alpha^{q-4}\right) 
\end{eqnarray} 
where ${\tilde N}_q = \frac{\Gamma (1+q)}{2\pi}$, and $\xi_0$ has been
defined in~(\ref{xi0}).

Notice that the above algebra is plagued by world-sheet ultraviolet
divergences as $\epsilon^2 \to 0^+$, thereby making the approach to
the fixed (conformal) point subtle.  As becomes obvious
from~(\ref{xi0}), the non-trivial fixed point $\epsilon \to 0^+$
corresponds to $L/a \to +\infty$, i.e.\ it is an infrared world-sheet
fixed point.  In order to understand the approach to the infrared
fixed point, it is important to make a few remarks first, motivated by
physical considerations.

From the integral expression of the regularized Heaviside
function~\cite{kmw} (\ref{rep}) it becomes obvious that a scale
$1/\epsilon$ for the target time is introduced. This, together with
the fact that the scale $\epsilon$ is connected~(\ref{xi0}) to the
world renormalization-group scales $L/a$, implies naturally the
introduction of a `renormalized' $\sigma$-model coupling/velocity
$v_{R,i}({1}/{\epsilon}) $ at the scale ${1}/{\epsilon}$:
\begin{equation} 
v_{R,i}\left(\frac{1}{\epsilon}\right) \sim
\left(\frac{1}{\epsilon}\right)^{q-1}
\label{velrenorm}
\end{equation}
for a trajectory $y_i (t) \sim t^q$.  This normalization would imply
the following rescaling of the operators
\begin{equation}
{\cal D}^{(q-n)} \rightarrow \epsilon^{q-1}{\cal D}^{(q-n)}
\end{equation} 
As a consequence, the factors $\epsilon^{2(1-q)}$
in~(\ref{defj}),~(\ref{offshellalg}) are removed.  In the context of
the world-sheet field theory this renormalization can be interpreted
as a subtraction of the ultraviolet divergences by the addition of
appropriate counterterms in the $\sigma$ model.

The approach to the infrared fixed point $\epsilon \to 0^+$ can now be
made by looking at the \emph{connected} two point correlators between
the operators ${\cal D}^{(q)}$ defined by
\begin{equation}\label{connected}  
\langle {\cal A}{\cal B} \rangle_c = \langle {\cal A}{\cal B} \rangle
- \langle {\cal A} \rangle~\langle {\cal B} \rangle\,,
\end{equation}
where the one-point functions are given by:
\begin{eqnarray} 
\langle {\cal D}^{(s)}\rangle &=& N_s\int 
\frac{d\omega}{(\omega -i\epsilon)^{s+1}}\langle e^{i\omega t} \rangle 
=N_s\int \frac{d\omega}{(\omega -i\epsilon)^{s+1}} e^{-\omega^2\alpha}
= {\tilde N}_s \epsilon^{-s} h_s\left(\epsilon^2\alpha\right), 
\nonumber \\
h_s(x) 
&=& -\frac{x^{s/2}}{2}\left(\frac{4\pi}{\Gamma (\frac{1+s}{2})} 
\sqrt{\pi} F\left(1+\frac{s}{2} , \frac{3}{2} , x\right) -
\frac{2\pi}{\Gamma (1 +  
\frac{s}{2})} F\left(\frac{1 + s}{2} , \frac{1}{2} , x\right)\right).
\end{eqnarray}  
 
For the two-point function of the ${\cal D}^{(q)}$ operator the result
is:
\begin{equation} 
 \langle {\cal D}^{(q)}(z){\cal D}^{(q)}(0) \rangle_c = {\tilde N}_q
\epsilon^{-2}\left(\frac{\sqrt{\pi}}{\xi_0} f_q\left(2\xi_0+2\epsilon^2\ln
\left|\frac zL \right| ^2\right) - h^2_q(\xi_0)\right).
\end{equation}

Expanding in powers of $\epsilon$, we obtain
\begin{eqnarray} \label{dd}
\langle {\cal D}^{(q)}(z) {\cal D}^{(q)}(0) \rangle_c &=&  
{\tilde N}_q \epsilon^{-2}\left(\frac{\sqrt{\pi}}{\sqrt{\xi_0}}
f_q(2\xi_0) - h^2_q(\xi_0)\right) + 
{\tilde N}^2_q~\frac{\sqrt{\pi}}{\sqrt{\xi_0}}~f'_q(2\xi_0)2
\ln \left|\frac zL \right|^2 + 
\nonumber \\&&
{}+ \epsilon^2 {\tilde N}_q^2 \sqrt{\frac{\pi}{\xi_0}}
\frac{1}{2}f''_q(2\xi_0)4\ln ^2\left|\frac zL \right|^2 + \cdots
\end{eqnarray} 
where $\cdots $ denote terms that vanish as $\epsilon \to 0^+$.

\begin{figure}[ht]
\centerline{\epsfxsize=3.9in\epsfbox{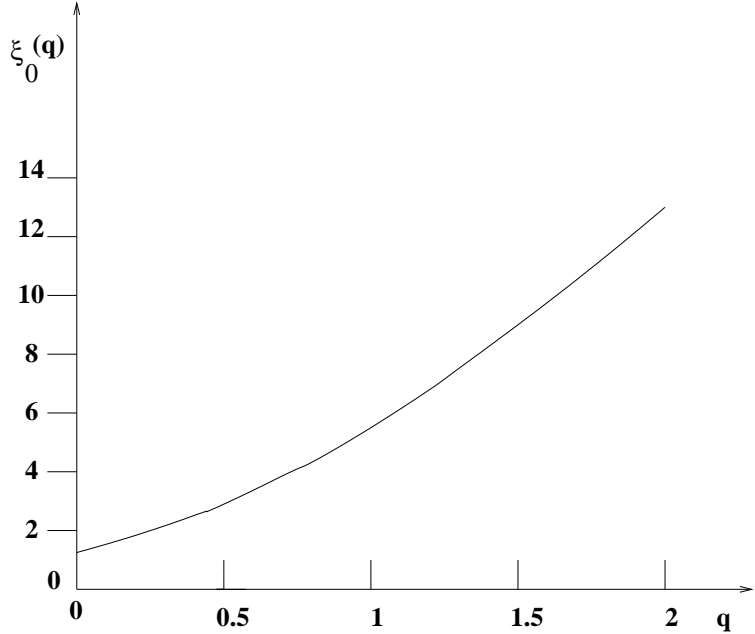}}  
\caption{Graphic solution of the equation
$\frac{\sqrt{\pi}}{\sqrt{\xi_0}} f_q(2\xi_0) -
h^2_q(\xi_0)=0$.\label{sol}}
\end{figure}

To avoid the divergences coming from the $\epsilon^{-2}$ factors, the
following condition must be satisfied: there must be a solution
$\xi_0=\xi_0(q)$ of the equation: 
$$
{\cal H}(\xi_0) \equiv \frac{\sqrt{\pi}}{\sqrt{\xi_0}} f_q(2\xi_0) -
h^2_q(\xi_0)=0\,.
$$ 
The existence of such a solution can be verified numerically (see
figure~\ref{sol}). Analytically this can be confirmed by looking at
the asymptotic behaviour of the function ${\cal H}(x)$ as $x \to
\infty$, which yields a negative value: 
$$
{\cal H}(x\to \infty) \sim -\frac{\pi^3x^{2q}e^{2x}}{\Gamma^2
  (\frac{1+q}{2}) \Gamma^2(1 + \frac{q}{2})} < 0\,.
$$ 
This behavior comes entirely from the term
$h_q^2(x)$, given that $f_q(x\to \infty) \to 0^+$.

As we shall show below, for various values of $q$, near the fixed
point $\epsilon \to 0^+$, one can construct higher order logarithmic
algebras, whose highest power is determined by the dominant terms in
the operator algebra of correlators~(\ref{offshellalg}),
(\ref{tdope}).  To this end, we first remark that in the above
analysis we have dealt with a small but otherwise arbitrary parameter
$\epsilon$, which allows us to keep as many powers as required
by~(\ref{offshellalg}) in conjunction with the value of $q$. The value
of $\epsilon$ determines the distance from the fixed point.

For $1< q <2$, there are only two dominant operators as the time $t
\to \infty$, ${\cal D}$,${\cal C}$.  In this case one obtains a
conventional logarithmic conformal algebra of two-point functions near
the fixed point:
\begin{eqnarray}\label{logalg}
\langle {\cal D}^{(q)}(z) {\cal D}(0)^{(q)} 
\rangle_c &=& \langle {\cal D}(z) {\cal D}(0) \rangle_c  
\sim {\tilde N}^2_q \frac{\sqrt{\pi}}{\sqrt{\xi_0}} f'_q(2\xi_0)2
\ln \left|\frac zL \right|^2, 
\nonumber \\
\epsilon q \langle {\cal D}^{(q-1)}(z) {\cal D}^{(q)}(0) \rangle_c &=& 
\langle {\cal C}(z) {\cal D}(0) \rangle_c \sim {\tilde N}^2_q 
\left(h^2_q(\xi_0) - h_{q-1} h_q (\xi_0)\right) , 
\end{eqnarray} 
and all the other correlators are subleading as $t \to \infty$.

Therefore, the \emph{on shell algebra} is of the conventional
\emph{logarithmic form}~\cite{lcftfurther}, between a pair of operators, and
hence, ${\cal D}^{(q-2)}$ and subsequent operators, which owe their
existence to the non-trivial RW metric, do not modify the two-point
correlators of the standard logarithmic algebra of `recoil'
(impulse)~\cite{kmw}.\footnote{We note at this stage that, in our case
of non-trivial cosmological RW spacetimes, the pairs of operators
${\cal D},{\cal C}$ do not represent velocity and position as in the
flat space time case of ref.~\cite{kmw}, but rather velocity and
acceleration.  This implies that, under a finite-size scaling of the
world sheet, the induced transformations of these operators do not
form a representation of the Galilean transformations of the
flat-space-time case.}

Next, we consider the case where $2< q < 3$. In this case, from
(\ref{offshellalg}) we observe that there are now three operators
which dominate in the limit $t \to \infty$, ${\cal D}$, ${\cal C}$ and
${\cal B} = \epsilon^2~q~(q-1) {\cal D}^{(q-2)}$, whose form is
implied from~(\ref{tdope}), in analogy with ${\cal C}$. The
corresponding algebra of correlators consists of parts forming a
conventional logarithmic algebra, and parts forming a second-order
logarithmic algebra, the latter being obtained from terms of order
$\epsilon^2$ in the appropriate two-point connected correlators
(cf.~(\ref{dd}) etc.), which are denoted by a superscript $\langle
\cdots \rangle_c^{(2)}$:
\begin{eqnarray}\label{higherord}  
 \langle {\cal D}(z) {\cal D}(0) \rangle_c^{(2)} &=& 
{\tilde N}_q^2 \sqrt{\frac{\pi}{\xi_0}}
\frac{1}{2}f''_q(2\xi_0)4\ln ^2\left|\frac zL \right|^2 , 
\nonumber \\
 \langle {\cal C}(z) {\cal D}(0) \rangle_c^{(2)}&=& 
{\tilde N}_q^2 \sqrt{\frac{\pi}{\xi_0}}
2f'_q(2\xi_0)\ln \left|\frac zL \right|^2 , 
\nonumber \\
 \langle {\cal C}(z) {\cal C}(0) \rangle_c^{(2)} &=& 
{\tilde N}_q^2 \sqrt{\frac{\pi}{\xi_0}}
f_{q-1}(2\xi_0) \,, 
\nonumber \\
 \langle {\cal B}(z) {\cal D}(0) \rangle_c^{(2)} &=& 
-{\tilde N}_q^2 \sqrt{\frac{\pi}{\xi_0}}
\left(f_{q}(2\xi_0) + f'_q(2\xi_0)\right) , 
\nonumber \\
 \langle {\cal C}(z) {\cal B}(0) \rangle_c^{(2)} &=& 
\langle {\cal B}(z) {\cal B}(0) \rangle_c^{(2)} = 0
\end{eqnarray} 
where the last two correlators are of order $\epsilon^4$ and
$\epsilon^6$ respectively, that is of higher order than the
$\epsilon^2$ terms, and hence they are viewed as zero to the order we
are working here.

An important relation in logarithmic conformal field theories is a
``formal derivative'' relation with respect to the anomalous dimension
$\Delta$, between the logarithmic set of operators~\cite{gezel}. In
this respect, we mention that in the case of logarithmic algebras of
order $[q]$ we encounter here one has:
\begin{eqnarray} 
 \frac{\partial {\cal C}}{\partial \Delta}&=&q{\cal D} 
+ \frac{{\cal C}}{2\Delta} \,,
\nonumber \\
 \frac{\partial^2 {\cal B}}{\partial \Delta^2}&=&q(q-1){\cal D} 
+ 3\frac{{\cal C}}{2\Delta} \,,  
\nonumber \\ 
  \cdots &&
\label{derivation}
\end{eqnarray} 
where $\Delta = -{\epsilon^2}/{2}$, ${\cal C}$, ${\cal D}$ and ${\cal
B}$ have been defined previously (c.f.~(\ref{tdope})), and the $\dots
$ denote similar relations for higher-order logarithmic algebras than
the ones examined in detail above, whose pattern can already be
inferred easily.  The first terms on the right-hand-side of these
relations would be exactly the derivative relation of a standard
logarithmic conformal field theory of order $[q]$~\cite{gezel}.
However in the recoil case one encounters singular $1/\sqrt{-\Delta}
\sim 1/\epsilon$ terms due to the specific form of the operator ${\cal
C}$.  Such singular terms characterize also the corresponding
derivative relations in the flat-space recoil case~\cite{kmw,ms1}. 
It is worthy of stressing, though, that such
singularities seem to characterize only the formal derivative
relations and not the logarithmic O.P.E.'s or \emph{the connected}
correlators, as we have seen in detail above.

In general, if one considers $q > 3$ one arrives at higher order
logarithmic algebras~\cite{lcftfurther}, with the highest power given by the
integer value of $q$, $[q]$. This is an interesting feature of the
recoil-induced motion of D-particles in RW backgrounds with scale
factors $\sim t^{p}$, $p >1$, corresponding to cosmological horizons
and accelerating Universes. In such a case the order of the
logarithmic algebra is given by $[2p]$.  It is interesting to remark
that radiation and matter (dust) dominated RW Universes would imply
simple logarithmic algebras.
 
We now notice that, under a world-sheet finite-size scaling,
\begin{equation} 
L \to L' = L e^{{\cal T}\,{\cal K}(q)} \,, \qquad 
\epsilon^{-2} \to (\epsilon')^{-2} = \epsilon^{-2} + {\cal T}
\end{equation}
with ${\cal K}(q)$ a function of $q$ determined by (\ref{logalg}), the
operators ${\cal C}, {\cal D}, \dots $, and consequently the
target-time $t$, transform in a non trivial way. In particular, for
$t$ one has:
\begin{equation}\label{timeshift2}  
\left(\frac{\epsilon'}{\epsilon}\right)^{q-1}{\cal Z} ({\cal T})^q
t({\cal T})^q = t^q + q \epsilon {\cal T} t^{q-1} + {\cal
O}\left(\epsilon^2\right)
\end{equation}
where ${\cal Z}({\cal T})$ is a wave function renormalization of the
world-sheet field $t(z)$, which can be chosen in a natural way so that
$(\epsilon' / \epsilon)^{q-1}{\cal Z}\,({\cal T})^q
=1$.  This implies
\begin{eqnarray}\label{timeshift}  
 t({\cal T})^q &=& (t + \epsilon {\cal T})^q + {\cal
 O}\left(\epsilon^2\right) ,
\nonumber \\
 t({\cal T}) &=& t + \epsilon {\cal T} + {\cal O}\left(\epsilon^2\right),
\end{eqnarray}
i.e.\ that a shift in the target time is represented as
$\epsilon\,{\cal T}$.  Of course, at the fixed point, $\epsilon =0$,
the field $t(z)$ does not run, as expected.

\subsubsection{Vertex operator for the path and associated spacetime geometry}

In this subsection we shall discuss the implications of the
world-sheet deformation~(\ref{path}) for the spacetime geometry. In
particular, we shall show that its r\^ole is to preserve the Dirichlet
boundary conditions on the $X^i$ by changing coordinate system, which
is encoded in an induced change in the space time geometry $G_{ij}$.
The final coordinates, then, are coordinates in the rest frame of the
recoiling particle, which naturally explains the preservation of the
Dirichlet boundary condition.

To this end, we first rewrite the world-sheet boundary vertex
operator~(\ref{path}) as a bulk operator:
\begin{eqnarray} \label{bulkop}
 V &=& \int _{\partial \Sigma} G_{ij}y^j(t)\partial_n X^i =
\int_{\Sigma} \partial_\alpha \left(y_i(t)\partial^\alpha X^i\right)
\nonumber \\
 &=& \int _{\Sigma} \left({\dot y}_i(t)\partial_\alpha t \partial^\alpha X^i 
+ y_i \partial^2 X^i \right)
\end{eqnarray}
where the dot denotes derivative with respect to the target time $t$,
and $\alpha$ is a world-sheet index. Notice that it is the covariant
vector $y_i$ which appears in the formula, which incorporates the
metric $G_{ij}$, $y_i=G_{ij}y^j$.

To determine the background geometry, which the string is moving in,
it is sufficient to use the classical motion of the string, described
by the world-sheet equations:
\begin{eqnarray}\label{sem} 
\partial^2 X^i + {\Gamma^i}_{\mu\nu}\partial_\alpha X^\mu
\partial^\alpha X^\nu =0\,,
\end{eqnarray}
where $\mu, \nu$ are space time indices, $\alpha=1,2$ is a world-sheet
index, $\partial^2$ is the laplacian on the world sheet, and $i$ is a
target spatial index.

The relevant Christoffel symbol in our RW background case, is
${\Gamma^i}_{ti}$, and thus the operator~(\ref{bulkop}) becomes:
\begin{eqnarray} 
\int _{\Sigma} \left({\dot y}_i -
2y_i(t){\Gamma^i}_{ti}\right)\partial_\alpha t \partial^\alpha X^i
\end{eqnarray} 
from which we read an induced non-diagonal component for the space
time metric
\begin{eqnarray}\label{indmetr} 
2G_{0i}= {\dot y}_i - 2y_i(t){\Gamma^i}_{ti}
\end{eqnarray}
In the RW background~(\ref{rwmetric}) the path $y_i(t)$ is
described~(\ref{pathexpre}) by (notice again we work with covariant
vector $y_i$):
\begin{eqnarray} 
y_i(t) = \frac{v_ia_0^2}{1-2p}\left(t t_0^{2p}-t_0 t^{2p}\right)
\end{eqnarray}
which gives $2G_{0i}=a^2(t_0) v_i$, yielding for the metric line
element:
\begin{equation}\label{fixedpoint}
ds^2=-dt^2 + v_ia^2(t_0)dtdX^i + a^2(t)(dX^i)^2\,, 
\qquad {\rm for} \qquad t > t_0
\end{equation} 
As expected, this spacetime has precisely the form corresponding to a
Galilean-boosted frame (the D-particle's rest frame), with the boost
occurring suddenly at time $t=t_0$.

This can be understood in a general fashion by first noting that
(\ref{indmetr}) can be written in a general covariant form as:
\begin{eqnarray}
2G_{0i}=\nabla_t y_i~~(= \nabla_t y_i + \nabla_i t)
\end{eqnarray} 
which is the general coordinate transformation associated with $y_i$
from a passive (Lie derivative) point of view.

In general, given the boundary condition $\partial_n t=0$, one can
write the operator~(\ref{path}), in a covariant form by expressing it
as a world-sheet bulk operator:
\begin{equation}
V= \int _{\partial \Sigma} y_\mu \partial_n X^\mu = \int _{\Sigma}
\partial_\alpha \left(y_\mu \partial^\alpha X^\mu \right) =
\int_{\Sigma} \nabla_{\mu} y_\nu \partial_\alpha X^\mu \partial^\alpha
X^\nu
\end{equation}
where in the last step, we have used again the string equations of
motion~(\ref{sem}). From this expression, one then derives the induced
change in the metric
\begin{equation}\label{lie} 
2 \delta G_{\mu\nu} = \nabla_\mu y_\nu + \nabla_\nu y_\mu  
\end{equation}
which is the familiar expression of the Lie derivative under the
coordinate transformation associated with $y_\mu$.

In all the above expressions we have taken the limit $\epsilon \to 0$,
which corresponds to considering the ratio of world-sheet cut-offs
$a/L \to 0$, implying that one approaches the infrared fixed point in
a Wilsonian sense.  As noted previously, in the context of the
logarithmic conformal analysis of the path $y^i(t)$, we have seen that
this limit can be reached without problems only in the case $p \le 1$,
which corresponds to the absence of cosmological horizons.  On the
other hand, the case of non-trivial horizons, $p > 1$, implies
ultraviolet divergences, which prevent one from taking this limit in a
way consistent with conformal invariance of the underlying $\sigma$
model. In such a case, the operators are relevant, with finite
anomalous dimensions $-\epsilon^2/2$.  One way to deal with such
relevant operators is by Liouville dressing~\cite{ddk,recoil} which
would in principle restore the conformal symmetry at the cost of
implying an extra target-space-time dimension.  However in our case,
such a restoration would not solve the full problem, since as we
mentioned above we have neglected in our approach terms proportional
to the graviton $\beta$-functions.

\section{Time as a RG Scale and Non-Linear Dynamics of Bosonic D-particles }

\subsection{General remarks} 

In \cite{msnl} 
we formulated an effective Schr\"odinger wave equation describing the quantum
dynamics of a system of D0-branes by applying the Wilson renormalization group
equation to the worldsheet partition function of a deformed $\sigma$-model
describing the system, which includes the quantum recoil due to the exchange of
string states between the individual D-particles. We arrived at an effective
Fokker-Planck equation for the probability density with diffusion coefficient
determined by the total kinetic energy of the recoiling system.
We used Galilean invariance of the system
to show that there are three possible solutions of the associated non-linear
Schr\"odinger equation depending
on the strength of the open string interactions among the D-particles. When
the open string energies are small compared to the total kinetic energy of the
system, the solutions are governed by freely-propagating solitary waves. When
the string coupling constant reaches a dynamically determined critical value,
the system is described by minimal uncertainty wavepackets which describe the
smearing of the D-particle coordinates due to the distortion of the surrounding
spacetime from the string interactions. For strong string interactions, bound
state solutions exist with effective mass determined by an energy-dependent
shift of the static BPS mass of the D0-branes.

The effective worldvolume dynamics of a single D$p$-brane coupled to a
worldvolume gauge field and to background supergravity fields is described by
the action \cite{polchinski}
\bea
&& I_{{\rm D}p}={\cal T}_p\int
d^{p+1}\sigma~\e^{-\phi}\,\sqrt{-\det_{\alpha,\beta}\left[G_{\alpha\beta}
+B_{\alpha\beta}+2\pi\alpha'F_{\alpha\beta}\right]}+ \nonumber \\
&& {\cal T}_p\int
d^{p+1}\sigma~\left[C\wedge\e^{2\pi\alpha'F+B}\wedge{\cal G}\right]_{p+1}
\label{IDp}\eea
The first term in (\ref{IDp}) is the Dirac-Born-Infeld action with ${\cal T}_p$
the $p$-brane tension, $\alpha'$ the string Regge slope, $\phi$ the dilaton
field, $F=dA$ the worldvolume field strength, and $G$ and $B$ the pull-backs of
the target space metric and Neveu-Schwarz two-form fields, respectively, to the
D$p$-brane worldvolume. It is a generalization of the geometric volume of the
brane trajectory. The second term is the Wess-Zumino action (restricted to its
$p+1$-form component) with $C$ the pullback of the sum over all electric and
magnetic Ramond-Ramond (RR) form potentials and $\cal G$ a geometrical factor
accounting for the possible non-trivial curvature of the tangent and normal
bundles to the $p$-brane worldvolume. It describes the coupling of the
D$p$-brane to the supergravity RR $p+1$-form fields as well as to the
topological charge of the worldvolume gauge field and to the worldvolume
gravitational connections. The fermionic completion of
the action (\ref{IDp}), compatible with spacetime supersymmetry and worldvolume
$\kappa$-symmetry, has been described in \cite{susybi}. For a recent review of
the Born-Infeld action and its various extensions in superstring theory, see
\cite{birev}.

While the generalization of the Wess-Zumino Lagrangian to multiple D$p$-branes
is obvious (one simply traces over the worldvolume gauge group in the
fundamental representation), the complete form of the non-abelian Born-Infeld
action is not known. In \cite{tseytlin} it was proposed that the background
independent terms can be derived using $T$-duality from a 9-brane action
obtained from the corresponding abelian version by symmetrizing all gauge
group traces in the vector representation \cite{birev}. A
direct calculation of the leading terms in a weak supergravity background has
been calculated using Matrix Theory methods in \cite{van}. Based on the Type I
formulation, i.e. by viewing a D-particle in the Neumann picture and imposing
$T$-duality as a functional canonical transformation in the string path
integral \cite{dorn}, the effective moduli space Lagrangian was derived in
\cite{ms1} and shown to coincide (to leading orders in a velocity expansion)
with the non-abelian Born-Infeld action of \cite{tseytlin}. In the following we
will use this moduli space approach to D-brane dynamics to describe some
properties of the multiple D-brane wavefunction.

The novel aspect of the approach of \cite{ms1} is that the moduli space
dynamics induces an effective target space geometry for the D-branes which
contains information about the short-distance spacetime structure probed by
multiple D-particles. Based on this feature, string-modified spacetime and
phase space uncertainty relations can be derived and thereby represent a proper
quantization of the noncommutative spacetime seen by low-energy D-particle
probes\cite{ms1}. The crucial property of the derivation is the incorporation
of proper recoil operators for the D-branes and the short open string
excitations connecting them. The smearing of the spacetime coordinates $y_i^a$
(in general $i=1,\dots,9-p$ label the transverse coordinates of the D$p$-brane
and $a=1,\dots,N$ the component branes of the multiple D-brane configuration)
of a given D-particle as a result from its open string interactions with other
branes can be seen directly from the formula for the variance
\be
\left(\Delta
y_i^a\right)^2\equiv\left[\left(Y_i-Y_i^{aa}\,I_N\right)^2\right]^{aa}
=\sum_{b\neq a}|Y_i^{ab}|^2
\label{Deltayia}\ee
where $Y_i^{ab}$ are the $u(N)$-valued positions of the D-particles ($a=b$) and
of the open strings connecting branes $a$ and $b$ ($a\neq b$), and $I_N$ is the
$N\times N$ identity matrix. The recoil operators give a relevant deformation
of the conformal field theory describing free open strings, and thus lead to
non-trivial renormalization group flows on the moduli space of coupling
constants. The moduli space dynamics is thereby governed by the Zamolodchikov
metric and the associated $C$-theorem. Physically, the recoil operators
describe the appropriate change of quantum state of the D-brane background
after the emission or absorption of open or closed strings. They are a
necessary ingredient in the description of multiple D-brane dynamics, in which
coincident branes interact with each other via the exchange of open string
states. The quantum uncertainties derived in \cite{ms1} were found to
exhibit quantum decoherence effects through their dependence on the recoil
energies of the system of D-particles. This suggests that the appropriate
quantum dynamics of D0-branes should be described by some sort of stochastic
string field theory involving a Fokker-Planck Hamiltonian.

As in \cite{birev,tseytlin}, the derivation in \cite{ms1} assumes constant
background supergravity fields. However, another important ingredient missing
in the moduli space description are the appropriate residual fermionic terms
from the supersymmetry of the initial static D-brane configuration. While the
recoil of the D-branes breaks supersymmetry, it is necessary to include these
terms to have a complete description of the stability of the D-particle bound
state. As shown in \cite{kallosh}, the energy of the bound states of D-branes
and strings is determined by the central charge of the corresponding spacetime
supersymmetry algebra. Nonetheless, the bosonic formalism that we display below
can be exploited to a large extent to describe at least heuristically the
quantum phase structure of the multiple D-particle system and, in particular,
determine the mass and stability conditions of the candidate bound state. One
reason that this approach is expected to yield reliable results is that we view
the system of D-branes and strings as a quantum mechanical system (rather than
a quantum field theoretical system as might be the case from the fact that
$T$-duality is used to effectively integrate over the transverse coordinates of
the branes), with the D-brane recoil constituting an excitation of this system.
The recoiling system of D-branes and strings can be viewed as an excited state
of a supersymmetric (static) vacuum configuration.
The breaking of target space supersymmetry by the excited state of the
system may thereby constitute a symmetry obstruction situation in the spirit
of \cite{symob}. According to the symmetry
obstruction hypothesis, the ground state of a system of
(static) strings and D-branes is a BPS state,
but the excited (recoiling) states do not respect the supersymmetry
due to quantum diffusion and other effects.
Phenomenologically, the supersymmetry
breaking induced by the excited system of recoiling D-particles
will distort the spacetime surrounding them
and may result in a decohering spacetime foam,
on which low energy (point-like) excitations live. This motivates the study
of non-supersymmetric D-branes recoiling under the exchange of strings.
Such quantum mechanical systems exhibit diffusion
and may be viewed as non-equilibrium (open) quantum systems,
with the non-equilibrium state being related naturally to the picture
of viewing the recoiling D-brane system as an excited state of some
(non-perturbative) supersymmetric D-brane vacuum configuration.

The main relationship we shall exploit in obtaining the quantum dynamics of
multiple D-particle systems is that between the Dirichlet
partition function in
the background of Type II string fields and the semi-classical (Euclidean)
wavefunctional $\Psi[Y^i]$ of a D$p$-brane.
This relation is usually expressed as \cite{hlp6,emn2}
\be
{\cal Z}=\int DY^i~\Psi[Y^i]
\label{Dppartfn}\ee
The wavefunction $\Psi[Y^i]$ is expressed in terms of the generating functional
which sums up all one-particle irreducible connected worldsheet diagrams whose
boundaries are mapped onto the D-brane worldvolume. Integration over the
worldvolume gauge field is implicit in $\Psi$ to ensure Type II winding number
conservation. Dirichlet string perturbation theory yields
\be
\Psi[Y^i]=\exp\sum_{h=1}^\infty\e^{(h-2)\phi}\,{\cal S}_h[Y^i]
\label{Dpert}\ee
where ${\cal S}_h$ denotes the amplitude with $h$ holes, in which an implicit
sum over handles is assumed. However, as we will discuss in the following,
the identification (\ref{Dppartfn}) is {\it not} the only one
consistent with the approach to D-brane dynamics advocated in
\cite{ms1}, and one may instead identify the worldsheet Dirichlet partition
function, summed over all genera, with the probability
distribution corresponding to the wavefunction $\Psi$.
Using this identification, the Wilson
renormalization group equation has been proposed as a defining principle for
obtaining string field equations of motion, including the appropriate
Fischler-Susskind mechanism for the contributions from higher genera
\cite{hlp6}. When applied to Dirichlet string theory, we shall find that the
consistent D-brane equation of motion follows from the renormalization group
equation.

More precisely, within the framework of a perturbative logarithmic conformal
field theory approach to multiple D-brane dynamics \cite{ms1}, we will show
that the intricate quantum dynamics of a system of interacting 
non-supersymmetric (bosonic) D-particles is
described by a non-linear Schr\"odinger wave equation. The corresponding
probability density is of the Fokker-Planck type, with quantum diffusion
coefficient $\cal D$ given by the square of the modulus of the recoil velocity
matrix of the bound state system of 
non-supersymmetric (bosonic) D-particles and strings:
\be
{\cal D}=c_G\,\sqrt{\alpha'}\,\sum_{i=1}^9\tr\,|\bar U^i|^2
\label{diffintro}\ee
where $c_G$ is a numerical constant and
$\bar U_{ab}^i$ is the (renormalized) constant velocity matrix of a system of
$N$ D-particles arising due to the D-particle recoil from the scattering of
string states. This phenomenon is in fact characteristic of Liouville string
theory, on which the above approach is based. Since the D-particle interactions
distort their surrounding spacetime, these non-linear structures may be thought
of as describing short-distance quantum gravitational properties of the D-brane
spacetime. Non-linear equations of motion for string field theories have been
derived in other contexts in \cite{cst}. From this nonlinear Schr\"odinger
dynamics we shall describe a multitude of classes of solutions, using Galilean
invariance of the D-brane dynamics which is a consequence of the corresponding
logarithmic conformal algebra. We will show that bound state solutions do
indeed exist for string couplings $g_s$ larger than a dynamically determined
critical value. The effective bound state mass is likewise determined as an
energetically induced shift of the static, BPS mass of the D0-branes.
In fact, we
shall find that there are essentially three different phases of the quantum
dynamics in string coupling constant space. Below the critical string coupling
the multiple D-brane wavefunction is described by solitary waves, in agreement
with the description of free D-branes as string theoretic solitons, while at
the critical coupling the quantum dynamics is described by coherent Gaussian
wavepackets which determine the appropriate quantum smearing of the multiple
D-particle spacetime. These results are shown to be in agreement with the
previous results concerning the structure of quantum spacetime \cite{ms1}.

We close this subsection by summarizing some of the generic guidelines that we
used in \cite{msnl}, and shall review below, 
for constructing a wavefunctional for the system of
bosonic D-branes. We will use a field theoretic approach by identifying the
Hartle-Hawking
wavefunction
\be
\Psi_0\simeq\e^{-S_E}
\label{hhwaveintro}\ee
where $S_E$ is the effective Euclidean action. We shall discuss the extension
to string theory and highlight the advantages and disadvantages of using this
identification. We shall also identify the probability density with the genus
expansion of an appropriate worldsheet $\sigma$-model:
\be
{\cal P}=\Psi_0^\dagger\Psi_0=\sum_{\rm genera}\int Dx~\e^{-S_\sigma[x]}
\label{probintro}\ee
The arguments in favour of this identification will be reality, and the
occurrence of statistical probability distribution factors which appear in the
wormhole parameters after resummation of (\ref{probintro}) over pinched genera.
We stress, however, that this turns out to be a feature of the 
bosonic D-particle case. Upon supersymmetrization, the leading (ultraviolet)
world-sheet modular divergences associated with such degenerate 
two-dimensional surfaces disappear, thereby making the summation over
genera a quite complicated technical issue not completely 
resolved to date. As we shall discuss in section 5, 
this will also have important
physical consequences for the linearity of the 
associated quantum dynamics of the super D-particles.

For the moment, we remark 
the Wilson-Polchinski worldsheet renormalization group flow, coming from the
sum over genera as in (\ref{probintro}), yields a Fokker-Planck diffusion
equation for the bosonic D-particle case 
\be
\partial_t{\cal P}={\cal D}\,\nabla^2{\cal P}-\nabla\cdot{\cal J}
\label{FPintro}\ee
where $\cal D$ is the diffusion operator defined in (\ref{diffintro}) in terms
of (renormalized) recoil velocity matrices, and $\cal J$ is the associated
probability current density. The equation (\ref{FPintro}) will follow from the
gradient flow property of the $\sigma$-model $\beta$-functions, which is also
necessary for the Helmholtz conditions or equivalently for canonical
quantization of the string moduli space.

The knowledge of the Fokker-Planck equation (\ref{FPintro}) alone does {\it
not} lead to an unambiguous construction of the wavefunction $\Psi$. There are
ambiguities associated with non-linear $\Psi$-dependent phase transformations
of the wavefunction:
\bea
\Psi&\mapsto&\e^{i{\cal N}_{\gamma,\lambda}(\Psi)}\,\Psi\nonumber\\{\cal
N}_{\gamma,\lambda}(\Psi)&=&\gamma\log|\Psi|+\lambda\,{\rm
arg}\,\Psi+\theta\Bigl(\{Y_i^{ab}\},t\Bigr)
\label{phaseintro}\eea
where $t$ is the Liouville zero mode.
Furthermore, $\Psi$ is then necessarily determined by a non-linear wave
equation if a diffusion coefficient $\cal D$ is present, as will be the case in
what follows. The non-linear Schr\"odinger equation has the form
\be
i\hbar\,\partial_t\Psi={\cal H}_0\Psi+\frac{i\hbar}2\,{\cal D}\,
\frac{\nabla^2{\cal P}}{\cal P}\,\Psi
\label{nlseintro}\ee
where ${\cal P}=\Psi^\dagger\Psi$ is the probability density. This is a
Galilean-invariant but time-reversal violating equation, exactly as expected
from previous considerations of non-relativistic D-brane dynamics and Liouville
string theory. Eq. (\ref{nlseintro}) will be the proposal in the following for
the non-linear quantum dynamics of matrix bosonic 
D-branes (this was noted in passing
in \cite{emn10}).

\subsection{Quantum Mechanics on Moduli Space}

In \cite{ms1} it was shown how a description of non-abelian D-particle
dynamics, based on canonical quantization of a $\sigma$-model moduli space
induced by the worldsheet genus expansion (i.e. the quantum string theory),
yields quantum fluctuations of the string soliton collective coordinates and
hence a microscopic derivation of spacetime uncertainty relations, as seen by
short distance D-particle probes. In the following we will proceed to construct
a wavefunction for the system of D0-branes which encodes the pertinent quantum
dynamics. To start, in this section we shall clarify certain facts about
wavefunctionals in non-critical string theories in general, completing the
discussion put forward in \cite{emn2}.

\subsubsection{Liouville-dressed Renormalization Group Flows}

Consider quite generally a non-critical string $\sigma$-model, defined as a
deformation of a conformal field theory $S_*$ with coupling constants
$\{g^I\}$. The worldsheet action is
\be
S_\sigma[x;\{g^I\}]=S_*[x]+\int\limits_\Sigma d^2z~g^IV_I[x]
\label{sigma1}\ee
where $V_I$ are the deformation vertex operators and an implicit sum over
repeated upper and lower indices is always understood. We assume that the
deformation is relevant, so that the worldsheet theory must be dressed by
two-dimensional quantum gravity in order to restore conformal invariance in the
quantum string theory. The corresponding Liouville-dressed renormalized
couplings $\{\lambda^I\}$ satisfy the renormalization group equations
\be
\ddot\lambda^I+Q\dot\lambda^I=-\beta^I(\lambda)
\label{beta2}\ee
where the dots denote differentiation with respect to the worldsheet zero mode
of the Liouville field. Here $Q$ is the square root of the running central
charge
deficit on moduli space and
\be
\beta^I(\lambda)=h^I\lambda^I+c^I_{~JK}\lambda^J\lambda^K+\dots
\label{betadef3}\ee
are the flat worldsheet $\beta$-functions, expressed in terms of
Liouville-dressed coupling constants. In (\ref{betadef3}), $h^I$ are the
conformal dimensions and $c^I_{~JK}$ the operator product expansion
coefficients of the vertex operators $V_I$. The minus sign in (\ref{beta2})
owes to the fact that we confine our attention here to the case of central
charge $c>25$ (corresponding to supercritical bosonic or fermionic strings).

Upon interpreting the Liouville zero mode as the target space time evolution
parameter, eq. (\ref{beta2}) is reminiscent of the equation of motion for the
inflaton field $\phi$ in inflationary cosmological models \cite{l5,lm3}. In the
present case of course one has a collection of fields $\{g^I\}$, but the
analogy is nevertheless precise. The role of the Hubble constant $H$ is played
by the central charge deficit $Q$. The precise correspondence actually follows
from the gradient flow property of the string $\sigma$-model $\beta$-functions
for flat worldsheets:
\be
\beta^I=G^{IJ}\,\frac\partial{\partial g^J}C
\label{gflow}\ee
where $C=Q^2$ is the Zamolodchikov $C$-function which is associated with the
generating functional for one-particle irreducible correlation functions
\cite{mm3b}, and $G^{IJ}$ is the matrix inverse of the Zamolodchikov metric
\be
G_{IJ}=2|z|^4\left\langle V_I(z,\bar z)\,V_J(0,0)\right\rangle
\label{Zammetric}\ee
on the moduli space $\modul(\{g^I\})$ of $\sigma$-model couplings $\{g^I\}$. Then
the right-hand side of (\ref{beta2}) also corresponds to the gradient of the
potential $V$ in inflationary models:
\be
\ddot\phi+3H\dot\phi=-\frac{dV}{d\phi}
\label{infl3}\ee
where $\phi$ is the inflaton field in a sufficiently homogeneous domain of the
universe.

\subsubsection{The Hartle-Hawking Wavefunction}

In \cite{ms1,emn2} it was shown, through the energy dependence of quantum
uncertainties, that some sort of stochasticity characterizes non-critical
Liouville string dynamics, implying that the analogy of eq. (\ref{beta2}) with
the equations of motion in inflationary models should be made with those
involving chaotic inflation \cite{lm3}. Let us now briefly review the
properties of these latter models. In such cases, the ground state wavefunction
of the universe may be identified as \cite{hh4}:
\be
\psi_0(a,\phi)=\exp-S_E(a,\phi)
\label{psi04}\ee
where $S_E$ is the Euclidean action for the scalar field $a(\tau)$ and the
inflaton scalar field $\phi(\tau)$ which satisfy the boundary conditions:
\be
a(0)=a~~~~~~,~~~~~~\phi(0)=\phi
\label{bc}\ee
and $\tau$ is the Euclidean time.

To understand how eq. (\ref{psi04}) comes about, we appeal to the
Hartle-Hawking interpretation \cite{hh4}.
Consider the Green's function $\langle
x,t|0,t'\rangle$ of a particle which propagates from the spacetime point
$(0,t')$ to $(x,t)$:
\be
\langle x,t|0,t'\rangle=\sum_n\psi_n^\dagger(x)\psi_n(0)~\e^{iE_n(t-t')}=\int
Dx~\e^{iS(x,|t-t'|)}
\label{green5}\ee
where $\{\psi_n\}$ is the complete set of energy eigenstates with energy
eigenvalues $E_n\geq0$ (the sum in (\ref{green5}) should be replaced by an
appropriate integration in the case of a continuous spectrum).
To obtain an expression for the ground state wavefunction, we make a Wick
rotation $t=-i\tau$, and take the limit $\tau\to-\infty$ to recover the initial
state. Then in the summation over energy eigenvalues in (\ref{green5}), only
the ground state ($n=0$) term survives if $E_0=0$. The corresponding path
integral representation becomes $\int Dx~\e^{-S_E(x)}$, and one obtains eq.
(\ref{psi04}) in the semi-classical approximation.

For inflationary models which are based on the de Sitter spaces $dS_4$ with
\be
a(\tau)=\kappa^{-1}(\phi)\cos\kappa(\phi)\tau
\label{atau}\ee
one has
\be
S_E(a,\phi)=-\frac3{16V(\phi)}
\label{SEinfl}\ee
and hence
\be
\psi_0(a,\phi)=\exp\frac3{16V(\phi)}
\label{psi05a}\ee
Thus the probability density for finding the universe in a state with
$\phi={\rm const.}$, $a=\kappa^{-1}(\phi)=\sqrt{\frac3{8\pi V(\phi)}}$ is
\be
{\cal P}=|\psi_0|^2=\e^{3/8V(\phi)}
\label{P6}\ee
The distribution function (\ref{P6}) has a sharp maximum as $V(\phi)\to0$. For
inflationary models this is a bad feature, because it diminishes the
possibility of finding the universe in a state with a large $\phi$ field and
thereby having a long stage for inflation. However, from the point of view of
Liouville string theory, the result (\ref{P6}), if indeed valid, implies that
the {\it critical} string theory (since $V\propto Q^2$ there) is a favorable
situation statistically, and hence any consideration (such as those in
\cite{ms1}) made in the neighborhood of a fixed point of the renormalization
group flow on the moduli space of running coupling constants is justified.

\subsubsection{Moduli Space Wavefunctionals}

Let us now proceed to discuss the possibility of finding a Schr\"odinger wave
equation for the D-particle wavefunction. The identification (\ref{psi04}) in
the inflationary case needs some careful verification in the case of the
topological expansion of the worldsheet $\sigma$-model (\ref{sigma1}). In
Liouville string theory, the genus expansion of the partition function may be
identified \cite{emn2} with the wavefunctional of non-critical string theory in
the moduli space of coupling constants $\{g^I\}$:
\be
\Psi(\{g^I\})=\sum_{\rm genera}\int
Dx~\e^{-S_\sigma[x;\{g^I\}]}\equiv\e^{-{\cal F}[\{g^I\}]}
\label{psig7}\ee
where
\be
{\cal F}[\{g^I\}]=\sum_{h=0}^\infty(g_s)^{h-2}\,{\cal F}_h[\{g^I\}]
\label{effgenusexp}\ee
is the effective target space action functional of
the non-critical string theory. The sum on the right-hand side of
(\ref{effgenusexp}) is over all worldsheet genera, which sums up the
one-particle irreducible connected worldsheet amplitudes ${\cal F}_h$
with $h$ handles. The
gradient flow property (\ref{gflow}) of the $\beta$-functions ensures
\cite{ms1,emn2} that the Helmholtz conditions for canonical quantization are
satisfied, which is consistent with the existence of an off-shell action ${\cal
F}[\{g^I\}]$. In that case, the effective Lagrangian on moduli space whose
equations of motion coincide with the renormalization group equations
(\ref{beta2}) is given by \cite{ms1}
\be
{\cal L}_\moduls(t)=-\beta^I\,G_{IJ}\,\beta^J
\label{effcalL}\ee
and it coincides with the Zamolodchikov $C$-function. The semi-classical
wavefunction determined by (\ref{psig7}) is thereby determined by the
action $C[\lambda]$ regarded as an effective action on the space of
two-dimensional renormalizable field theories. Thus the probability density is
${\cal
P}[\{g^I\}]=\e^{-2{\cal F}[\{g^I\}]}$, which implies that the minimization of
${\cal F}[\{g^I\}]$ yields a maximization of ${\cal P}[\{g^I\}]$, provided that
the effective action is positive-definite. This is an ideal situation, since
then the minimization of ${\cal F}[\{g^I\}]$, in the sense of solutions of the
equations $\delta{\cal F}/\delta g^I=0$, corresponds to the
conformally-invariant fixed point of the $\sigma$-model moduli space, thereby
justifying the analysis in a neighborhood of a fixed point.

However, the identification (\ref{psig7}) is {\it not} the only possibility in
non-critical string theory, as will be discussed below, in particular in
connection with the Schr\"odinger dynamics of D0-branes. The main point is that
upon taking the topological expansion in Liouville string theory, the couplings
$g^I$ become quantized in such a way so that
\be
{\sum_{\rm genera}}'\int
Dx~\e^{-S_\sigma[x;\{g^I\}]}=\int\limits_{\moduls(\{g^I\})}D\alpha^I~
\e^{-\frac1{2\Gamma^2}\alpha^IG_{IJ}\alpha^J}\,\int Dx~
\e^{-S_\sigma^{(0)}[x;\{g^I+\alpha^I\}]}
\label{pinchsum8}\ee
where the prime on the sum means that the genus expansion is truncated to a sum
over pinched annuli of infinitesimal strip size, $S_\sigma^{(0)}[x;\{g^I\}]$ is
the tree-level (disc or sphere) action for the $\sigma$-model, and $\alpha^I$
are worldsheet wormhole parameters on the moduli space $\modul(\{g^I\})$ of the
two-dimensional quantum field theory. The Gaussian spread in the $\alpha^I$ in
(\ref{pinchsum8}) can be interpreted as a probability distribution
characterizing the statistical fluctuations of the coupling constants $g^I$.
The width $\Gamma$ is proportional to the logarithmic modular divergences on
the pinched annuli, which may be identified with the short-distance infinities
$\log\Lambda$ at tree-level \cite{ms1} ($\Lambda$ is the worldsheet
ultraviolet cutoff scale). The result (\ref{pinchsum8}) suggests
that one may directly identify the genus expansion of the worldsheet
partition function as the probability density
\be
\left|\Psi(\{g^I\},t)\right|^2\equiv{\cal P}(\{g^I\},t)
\label{PsicalP}\ee
for finding non-critical strings in the moduli space configuration $\{g^I\}$ at
Liouville time $t$ (the worldsheet zero mode of the Liouville field). In this
way one has a {\it natural} explanation for the reality of eq.
(\ref{pinchsum8}) on Euclidean worldsheets. If the identification of the genera
summed partition function with the probability density holds, i.e. with the
square of the wavefunction $\Psi(\{g^I\},t)$ rather than the wavefunctional
itself, then one may obtain a temporal evolution equation for (\ref{PsicalP})
using the Wilson-Polchinski renormalization group equation on the string
worldsheet \cite{hlp6}. This will be described in details later on.

One may argue formally in favour of the above identification 
in the case of Liouville strings, within a world-sheet formalism, 
by noting~\cite{emn} that the 
conventional interpretation of the Liouville (world-sheet) correlators  
as target-space $S$-matrix elements breaks down upon the 
interpretation of the Liouville zero-mode as target time. Instead,
the only well-defined concept in such a case is the non-factorizable 
$\nd{S}$-matrix, which acts on target-space density matrices rather than 
state vectors. This in turn implies that the corresponding 
world-sheet partition function, summed over topologies, which in the case 
of critical strings would be the generating functional of such $S$-matrix
elements in target space, should be identified with the 
probability density in the moduli space of the non-critical strings
(\ref{PsicalP}). 

Below we review briefly this approach~\cite{emn} by focusing on 
those aspects of the formalism that are most relevant to our 
purposes here. 
As we shall discuss, the above identification 
follows from specific properties of the Liouville string formalism.

\begin{figure}[ht]
\centerline{\epsfxsize=3.9in\epsfbox{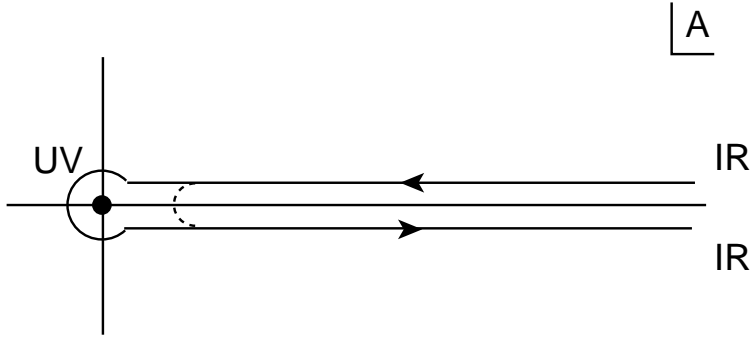}}   
\caption{Contour
of integration in the analytically-continued
(regularized) version of $\Gamma (-s)$ for $ s \in Z^+$.
The quantity A denotes the (complex) world-sheet area. 
This is known in the literature as the Saalschutz contour,
and has been used in
conventional quantum field theory to relate dimensional
regularization to the Bogoliubov-Parasiuk-Hepp-Zimmermann
renormalization method. Upon the interpretation of the 
Liouville field with target time, this curve
resembles closed-time-paths in non-equilibrium field theories.
\label{fig1}}
\end{figure}

We commence our analysis 
by considering the correlation functions among vertex operators 
in a generic Liouville theory, viewing the Liouville field
as a local renormalization-group scale on the world sheet~\cite{emn}.
Standard computations\cite{goulian} yield for an $N$-point correlation
function among world-sheet integrated
vertex operators $V_i\equiv \int d^2z V_i (z,{\bar z}) $ :
\be
A_N \equiv <V_{i_1} \dots V_{i_N} >_\mu = \Gamma (-s) \mu ^s
<(\int d^2z \sqrt{{\hat \gamma }}e^{\alpha \phi })^s {\tilde
V}_{i_1} \dots {\tilde V}_{i_N} >_{\mu =0}
\label{C12}
\ee
where the tilde denotes removal of the
Liouville  field $\phi $ zero mode, which has been
path-integrated out in (\ref{C12}).
The world-sheet scale $\mu$ is associated with cosmological
constant terms on the world sheet, which are characteristic
of the Liouville theory.
The quantity $s$ is the sum of the Liouville anomalous dimensions
of the operators $V_i$
\be
s=-\sum _{i=1}^{N} \frac{\alpha _i}{\alpha } - \frac{Q}{\alpha}
\qquad ; \qquad \alpha = -\frac{Q}{2} + \frac{1}{2}\sqrt{Q^2 + 8}
\label{C13}
\ee
The $\Gamma $ function can be regularized\cite{kogan,emn}
(for negative-integer
values of its argument) by
analytic continuation to the complex-area plane using the
the Saaschultz contour
of Fig. \ref{fig1}. Incidentally, this yields the possibility
of an increase of the running central charge
due to the induced oscillations of the dynamical
world sheet area (related to the Liouville zero mode).
This is associated with an oscillatory solution
for the Liouville central charge near the fixed point.
On the other hand, the bounce interpretation
of the infrared fixed points of the flow,
given in refs. \cite{kogan,emn},
provides an alternative picture
of the overall monotonic change
at a global level in target space-time.

To see technically why the above formalism 
leads to a breakdown in the 
interpretation of the correlator $A_N$ 
as a target-space string amplitude, which in turn leads to  
the interpretation of the world-sheet
partition function as a probability density rather than a 
wave-function in target space, 
one first expands the Liouville
field in (normalized) eigenfunctions  $\{ \phi _n \}$
of the Laplacian $\Delta $ on the world sheet
\be
 \phi (z, {\bar z}) = \sum _{n} c_n \phi _n  = c_0 \phi _0
 + \sum _{n \ne 0} \phi _n \qquad \phi _0 \propto A^{-\frac{1}{2}}
\label{C14}
\ee
with $A$ the world-sheet area,
and
\be
   \Delta \phi _n = -\epsilon_n \phi _n  \qquad n=0, 1,2, \dots,
\qquad \epsilon _0 =0
\qquad (\phi _n, \phi _m ) = \delta _{nm}
\label{C15}
\ee
The result for the correlation functions (without the Liouville
zero mode) appearing on the right-hand-side of eq. (\ref{C12})
is, then
\bea
{\tilde A}_N \propto &\int & \Pi _{n\ne0}dc_n exp(-\frac{1}{8\pi}
\sum _{n\ne 0} \epsilon _n c_n^2 - \frac{Q}{8\pi}
\sum _{n\ne 0} R_n c_n + \nn \\
~&~&\sum _{n\ne 0}\alpha _i \phi _n (z_i) c_n )(\int d^2\xi
\sqrt{{\hat \gamma }}e^{\alpha\sum _{n\ne 0}\phi _n c_n } )^s
\label{C16}
\eea
with $R_n = \int d^2\xi R^{(2)}(\xi )\phi _n $. We can compute
(\ref{C16}) if we analytically continue \cite{goulian}
$s$ to a positive integer $s \rightarrow n \in {\bf Z}^{+} $.
Denoting
\be
f(x,y) \equiv  \sum _{n,m~\ne 0} \frac{\phi _n (x) \phi _m (y)}
{\epsilon _n}
\label{fxy}
\ee
one observes that, as a result
of the lack of the zero mode,
\be
   \Delta f (x,y) = -4\pi \delta ^{(2)} (x,y) - \frac{1}{A}
\label{C17}
\ee
We may choose
the gauge condition  $\int d^2 \xi \sqrt{{\hat \gamma}}
{\tilde \phi }=0 $. This determines the conformal
properties of the function $f$ as well as its
`renormalized' local limit
\be
   f_R (x,x)=lim_{x\rightarrow y } (f(x,y) + {\rm ln}d^2(x,y))
\label{C18}
\ee
where  $d^2(x,y)$ is the geodesic distance on the world sheet.
Integrating over $c_n$ one obtains
\bea
~&& {\tilde A}_{n + N} \propto
exp[\frac{1}{2} \sum _{i,j} \alpha _i \alpha _j
f(z_i,z_j) + \nn  \\
~&&\frac{Q^2}{128\pi^2}
\int \int  R(x)R(y)f(x,y) - \sum _{i} \frac{Q}{8\pi}
\alpha _i \int \sqrt{{\hat \gamma}} R(x) f(x,z_i) ]
\label{C19}
\eea

We now consider
infinitesimal Weyl shifts of the world-sheet metric,
$\gamma (x,y) \rightarrow \gamma (x,y) ( 1 - \sigma (x, y))$,
with $x,y$ denoting world-sheet coordinates.
Under these,
the correlator $A_N$
transforms as follows~\cite{emn} 
\bea
&~&
\delta {\tilde A}_N \propto
[\sum _i h_i \sigma (z_i ) + \frac{Q^2}{16 \pi }
\int d^2x \sqrt{{\hat \gamma }} {\hat R} \sigma (x) +    \nn \\
&~&
\frac{1}{{\hat A}} \{
Qs \int d^2x \sqrt{{\hat \gamma }} \sigma (x)
       +
(s)^2 \int d^2x \sqrt{{\hat \gamma }} \sigma (x) {\hat f}_R (x,x)
+  \nn \\
&~&
Qs \int \int d^2x d^2y
\sqrt{{\hat \gamma }} R (x) \sigma (y) {\hat {\cal
 G}} (x,y) -
  s \sum _i \alpha _i
  \int d^2x
  \sqrt{{\hat \gamma }} \sigma (x) {\hat {\cal
 G}} (x, z_i) -   \nn \\
&~&
 \frac{1}{2} s \sum _i \alpha _i{\hat f}_R (z_i, z_i )
  \int d^2x \sqrt{{\hat \gamma }} \sigma (x)
-    \nn \\
&~&
 \frac{Qs}{16\pi} \int
  \int d^2x d^2y \sqrt{{\hat \gamma (x)}{\hat \gamma }(y)}
  {\hat R}(x) {\hat f}_R (x,x) \sigma (y)\} ] {\tilde A }_N
\label{dollar}
\eea
where the hat notation denotes transformed quantities,
and
the function  ${\cal G}$(x,y)
is defined as
\be
  {\cal G}(z,\omega ) \equiv
f(z,\omega ) -\frac{1}{2} (f_R (z,z) + f_R (\omega, \omega ) )
\label{C20}
\ee
and transforms simply under Weyl shifts~\cite{emn}.
We observe from (\ref{dollar}) that
if the sum of the anomalous dimensions
$s \ne 0$ (`off-shell' effect of
non-critical strings), then there are
non-covariant terms in
(\ref{dollar}), inversely proportional to the
finite-size world-sheet area $A$.
Thus the generic correlation 
function $A_N$ does not have a well-defined 
limit as $A \rightarrow 0$. 

In our approach to string time we identify~\cite{emn} 
the target time as $t=\phi_0=-{\rm log}A$, 
where $\phi_0$ is the world-sheet zero mode of the Liouville field.
The normalization follows from
a consequence 
of the canonical form of the kinetic term for the Liouville field $\phi$ 
in the Liouville $\sigma$ model~\cite{aben,emn}. 
The opposite flow of the target time, as compared to that of the 
Liouville mode, is, on the other hand, a consequence
of the `bounce' picture~\cite{kogan,emn} for Liouville flow of Fig. \ref{fig1}.
In view of this, the above-mentioned induced time (world-sheet scale 
$A$-) dependence
of the correlation functions $A_N$ implies the
breakdown of their interpretation as
well-defined $S$-matrix elements,
whenever there is a departure from criticality $s \ne 0$. 

In general, this is a feature of non-critical strings
wherever the Liouville mode is viewed as a local 
renormalization-group scale
of the world sheet~\cite{emn}. In such a case,
the central charge of the theory
flows continuously with the world-sheet scale $A$,
as a result of the Zamolodchikov
$c$-theorem \cite{zam}. In contrast, the screening operators
in conventional strings 
yield quantized values\cite{aben}.
Due to the analytic continuation curve illustrated in Fig.~\ref{fig1},
we observe that upon interpreting the Liouville field $\phi$ 
as time~\cite{emn}: $t \propto {\rm log}A$, 
the contour of Fig.~\ref{fig1} represents evolution 
in both directions of time between fixed points of the 
renormalization group: $ {\rm Infrared} ~ {\rm fixed} ~ 
{\rm point}  \rightarrow {\rm  Ultraviolet} ~ {\rm fixed}
~{\rm point} \rightarrow
 {\rm Infrared} ~ {\rm fixed} ~ {\rm point}$.

When one integrates over the Saalschultz contour in fig. \ref{fig1}, 
the integration
around the simple pole at $A=0$ yields an imaginary part~\cite{kogan,emn},
associated with the instability of the Liouville vacuum. We note, on the  
other hand, that the integral around the dashed contour
shown in Fig. \ref{fig1}, 
which does not encircle the pole at $A=0$, is well defined.
This can be interpreted as a well-defined $\nd{S}$-matrix element,
which is not, however, factorisable into a product of 
$S-$ and $S^\dagger -$matrix elements, due to the 
$t$ dependence acquired after the identification 
$t=-{\rm log}A$. 

Note that 
this formalism is similar to the 
Closed-Time-Path (CTP) formalism used in non-equilibrium 
quantum field theories~\cite{ctp}.
Such formalisms are characterized by a `doubling of degrees of 
freedom' (c.f. the two directions of the time (Liouville scale) 
curve of figure \ref{fig1}, 
in each of which one 
can define a set of dynamical fields in target space). 
As we discussed above, this prompts one 
to identify the corresponding 
Liouville correlators $A_N$ with $\nd{S}$-matrix elements 
rather than $S$-matrix elements in target space. 
Such elements act on the 
density matrices $\rho={\rm Tr}_{{\cal M}}|\Psi><\Psi|$ 
rather than wave vectors $|\Psi>$ in target space of the string:
$\rho_{out} = \nd{S} \rho_{in}$ (c.f. the analogy 
with the $S$-matrix, $|out> =S|in>$). 

This in turn implies that the world-sheet partition function ${\tilde 
{\cal 
Z}}_{\chi,L}$ 
of a Liouville string at a given world-sheet genus $\chi$,
which is connected to the generating functional 
of the Liouville correlators $A_N$, 
when 
{\it defined} over the closed Liouville (time) path (CTP) 
of figure \ref{fig1},  
can be associated 
with the {\it probability density} (diagonal element of a density matrix)
rather than the 
wavefunction in the space of couplings. Indeed, one has
\begin{equation}
{\tilde {\cal Z}}_{\chi,L}[g^I] = 
\int_{CTP} d\phi_0 {\cal Z}_{\chi,L}[\phi_0, g^I] 
\label{liouvpartfnct}
\end{equation}
where 
$\{ g^I \}$ denotes the set of couplings of the (non-conformal) deformations, 
$\phi_0 \sim {\rm ln}A$ is the Liouville zero mode, and 
$A$  is the world-sheet  area (renormalization-group scale). 
If one naively interprets 
${\cal Z}_{\chi,L}[\phi_0,g^I]$ as a wavefunctional 
in moduli space $\{ g^I \}$, $\Psi [\phi_0, g^I ]$, 
then,  
in view of the double contour of figure \ref{fig1},
over which ${\tilde {\cal Z}}_{\chi,L}$ is defined, 
one 
encounters at each slice of constant $\phi_0$ 
a product of $\Psi [\phi_0, g^I]\Psi^\dagger [\phi_0, g^I]$, 
the complex conjugate wavefunctional corresponding to the 
second branch of the contour of opposite sense to the branch 
defining $\Psi [\phi_0,g^I]$. This is analogous to the 
doubling of degrees of freedom in conventional 
thermal field theories~\cite{ctp}. 
Such products represent clearly 
probability densities ${\cal P}[t,g^I]$ 
in moduli space of the non-critical strings
upon the identification of 
the Liouville zero mode $\phi_0$ with the target time $t$~\cite{emn}. 

In the above spirit, one may then consider 
the (formal) summation over world-sheet topologies $\chi$, and 
identify the summed-up world-sheet partition function 
$\sum_{\chi} {\cal Z}_{\chi,L}[\phi_0, g^I]$ 
with the associated probability density in moduli space. 
In the case of D-particles, discussed in this work, 
the moduli space coincides with the configuration space (collective) 
coordinates of the D-particle 
soliton, and hence the corresponding probability density 
is associated with the position of the D-particle in target space. 
We stress once again that the above conclusion 
is based on the 
crucial assumption of the 
definition of the Liouville-string world-sheet partition 
function over the closed-time-path of figure \ref{fig1}. 
As we demonstrate in the main text, 
the specific D--brane example provide us  
with highly non-trivial 
consistency checks of this approach.

We would like now to give an explicit demonstration
of the above ideas for the specific (simplified) case  
of recoiling (Abelian) D-particles. We shall demonstrate below
that, upon considering the non-critical $\sigma$-model of a recoiling 
D-particle at a fixed world-sheet (Liouville) scale $\phi_0={\rm ln}A$, 
and identifying the 
Liouville mode with the target time, 
the Euclideanized 
world-sheet partition function can describe a  probability density
in moduli (collective coordinate) space. 

To this end, let us first consider the pertinent 
$\sigma$ model partition 
function for a D-particle, at tree level and 
in a {\it Minkowskian} world-sheet $\Sigma$ formalism:
\begin{equation} 
{\cal Z}_{\chi=0, L} = \int (DX^i)~ e^{-i\frac{1}{4\pi \alpha'} \int _{\Sigma} \partial X^i 
{\overline \partial} X^j \eta_{ij} - i\frac{1}{2\pi \alpha'} 
\int _{\partial \Sigma} \left(\epsilon g_i^C + 
g_i^D \frac{1}{\epsilon} \right) \partial_n X^i }
\label{dparticlepf}
\end{equation} 
where $\epsilon^{-2} \sim {\rm ln}\Lambda ^2={\rm ln}A$ 
(c.f. (\ref{epLambdaid})), 
on account of the 
logarithmic algebra~\cite{kmw12}. In our approach
$\epsilon^{-2}$ is identified with the 
target time. 
This is why in (\ref{dparticlepf}) we have not path-integrated
over $X^0$, but we consider 
an integral only over the spatial collective coordinates 
$X^i, i=1, \dots 9$ of the D-particle. The combination of $\sigma$-model 
couplings 
$ \epsilon g_i^C + g_i^D \frac{1}{\epsilon} $ may be identified with the
generalized (Abelian) position
$\epsilon Y^i$ of the recoiling D-particle (\ref{recoilY}).
Notice that, since here we have already identified the time with the 
scale $\epsilon^{-2} >0$, the step function in the recoil deformations
of the $\sigma$-model (\ref{recoilops}) 
acquires trivial meaning.
We shall come back to a discussion on how one can incorporate
a world-sheet dependence in the time coordinate later on.  

Suppose now that, following the spirit of critical strings~\cite{hlp6},
one identifies the Minkowskian world-sheet partition function 
(\ref{dparticlepf})  
with a wavefunctional $\Psi[Y^i, \phi_0=t]$.
The probability density in $Y^i$ space, 
${\cal P}[Y^i, t]=\Psi[Y^i,t]\Psi^*[Y^i,t]$,  
reads in this case:
{\small 
\begin{eqnarray} 
&& \left|{\cal Z}_{\chi=0, L}[Y^i,t]\right|^2 = \nonumber \\
&& \int DX^i \int DX'^j e^{-i\frac{1}{4\pi \alpha'} 
\int _{\Sigma} \partial X^i 
{\overline \partial} X^j \eta_{ij} + i\frac{1}{4\pi \alpha'}
\int _{\Sigma} \partial X'^i 
{\overline \partial} X'^j \eta_{ij} 
-i \frac{1}{2\pi \alpha'} 
\int _{\partial \Sigma} \epsilon Y_i(t)\partial_n (X^i - X'^i)} =
\nonumber \\
&& \left(\int DX_{-}^i e^{i\frac{1}{4\pi\alpha'}\int _{\Sigma} \partial X_{-}^i 
{\overline \partial} X_{-}^j \eta_{ij} 
-i \frac{1}{2\pi \alpha'}\int _{\partial \Sigma} Y_i(t) \partial_n X_{-}^i }\right) \otimes 
(\left(\int DX_{+}^i e^{-i\frac{1}{4\pi \alpha'} \int _{\Sigma} 
\partial X_{+}^i 
{\overline \partial} X_{+}^j \eta_{ij}}\right)~, \nonumber \\
\label{factorization}
\end{eqnarray}}
where $X_{\pm}^i=X^i \pm X'^i $. 
Upon passing to a Euclidean world-sheet formalism, 
and taking into account that the $Y_i$ independent factor 
can be absorbed in appropriate normalization of the 
$\sigma$-model correlators, one then 
proves our statement
that $\sigma$-model partition functions in non-critical strings
can be identified with moduli space probability densities. 

Notice that similar conclusions can be reached even in the case
where the time $X^0$ is included in the analysis as a full fledged 
world-sheet field and {\it is only eventually} 
identified with the Liouville mode. 
In such a case, by considering the probability density as above, 
one is confronted with path integration over $X_{\pm}^0 = X_0 \pm X'^0$ 
$\sigma$-model fields, which also appear in the arguments
of the step function operators $\Theta_\epsilon ( X_{\pm}^0 )$
in the recoil deformations (c.f. below, (\ref{recoilops})),  
that 
are non trivial in this case. 
However, upon Liouville dressing and the {\it requirement} 
that the  Liouville mode be identified with the target  time,
one is forced to restrict oneself on the hypersurface $X_{-}=0$
in the corresponding  path integral $\int DX_{+}^0 DX_{-}^0 ( \dots )$. 
As a consequence, one is then left
with a world-sheet partition 
function 
integrated only over 
the Liouville mode $X_{+}=2\phi$ (c.f. ${\tilde {\cal Z}}$ 
in (\ref{liouvpartfnct})), 
and hence the identification of 
a Liouville string partition 
function with a probability density in 
moduli space  
is still valid, upon passing onto a Euclideanized
world-sheet formalism. It can also be seen, in a straightforward manner,
that summing upon higher world-sheet topologies, 
as in \cite{ms1}, will not change this conclusion.

Notice that if one interprets the topological expansion of the worldsheet
partition function as the probability density for the non-critical string
configuration $\{g^I\}$, then the simple argument leading to eq. (\ref{psi04})
is not valid here. In such a situation the action in eq. (\ref{green5}), which
refers to the string moduli space, is {\it not} the same as the effective
target space action ${\cal F}[\{g^I\}]$, but rather something different,
corresponding to the phase of the wavefunctional $\Psi(\{g^I\},t)$ whose
probability density (\ref{PsicalP}) corresponds to the genera summed worldsheet
partition function. This is not necessarily a bad feature, as we shall see,
although in most treatments the target space effective action ${\cal
F}[\{g^I\}]$ is identified with the moduli space action upon identification of
the Liouville zero mode (i.e. the local worldsheet renormalization group scale)
with target time. For this, we observe that the statistical interpretation of
the resummed worldsheet partition function is {\it compatible} with the
interpretation in \cite{ms1} of the Gaussian wormhole parameter distribution
function in eq. (\ref{pinchsum8}) as being responsible for the quantum
uncertainties of D-branes. This follows trivially from the fact that
\be
\left|\Psi(\{g^I\},t)\right|^2=\e^{-2{\cal F}(\{g^I\},t)}
\label{PsicalF}\ee
Then, any correlation function may be written as
\bea
&&\left\langle V_{I_1}\cdots
V_{I_n}\right\rangle = \int\limits_{\moduls(\{g^I\})}Dg^I~
\left|\Psi(\{g^I\},t)\right|^2\,V_{I_1}\cdots V_{I_n} = \nonumber\\
&& \int\limits_{\moduls(\{g^I\})}Dg^I~\int\limits_{\moduls(\{g^I\})}D\alpha^I~
\e^{-\frac1{2\Gamma^2}\alpha^IG_{IJ}\alpha^J}\,\int Dx~\e^{-S_
\sigma^{(0)}[x;\{g^I+\alpha^I\}]}\,V_{I_1}\cdots V_{I_n}
\label{VIcorr}\eea
which using eq. (\ref{PsicalF}) gives the connection between the two
probability distributions.

\subsection{Matrix D-brane Dynamics}

In this section we shall briefly review the worldsheet description of
\cite{ms1} for matrix D0-brane dynamics. The partition function is given by
\cite{dorn}
\bea
{\cal Z}[A_0,Y]&=&\int
D\mu(x,\bar\xi,\xi)~\exp\left(-\frac1{4\pi\alpha'}\int\limits_\Sigma
d^2z~\eta_{\mu\nu}\,\partial x^\mu\,\bar\partial x^\nu+\frac1{2\pi\alpha'}
\oint\limits_{\partial\Sigma}d\tau~x_i(\tau)\,\partial_\sigma x^i(\tau)
\right)\nonumber\\& &\times\,{\cal
W}[x,\bar\xi,\xi]
\label{partfnD}\eea
where
\be
{\cal W}[x,\bar\xi,\xi]=\exp
ig_s\oint\limits_{\partial\Sigma}d\tau~\left(\bar\xi_a(\tau)
A_0^{ab}\xi_b(\tau)\,\partial_\tau x^0(\tau)+\frac i{2\pi\alpha'}
\,\bar\xi_a(\tau)Y_i^{ab}(x^0)\xi_b(\tau)\,\partial_\sigma x^i(\tau)\right)
\label{wilsonloop}\ee
is the deformation action of the free $\sigma$-model in (\ref{partfnD}). Here
the indices $\mu=0,1,\dots,9$ and $i=1,\dots,9$ label spacetime and spatial
directions of the target space, which we assume has a flat metric
$\eta_{\mu\nu}$. The functional
integration measure in (\ref{partfnD}) is given by
\bea
&& D\mu(x,\bar\xi,\xi)=Dx^\mu~D\bar\xi~D\xi~\exp\left[-\sum_{a=1}^N
\left(\oint\limits_{\partial\Sigma}d\tau~\bar\xi_a(\tau)\,\partial_\tau
\xi_a(\tau)+\bar\xi_a(0)\xi_a(0)\right)\right]
\times \nonumber \\
&& \sum_{a=1}^N\bar\xi_a(0)\xi_a(1)
\label{Dmeasure}\eea
The complex auxiliary fields $\bar\xi_a(\tau)$ and $\xi_a(\tau)$,
$a=1,\dots,N$,
transform in the fundamental representation of the brane gauge group, and they
live on the boundary of the worldsheet $\Sigma$ which at tree-level is a disc
whose boundary is a circle $\partial\Sigma$ with periodic longitudinal
coordinate
$\tau\in[0,1]$ and normal coordinate $\sigma\in\real$. They have the propagator
$\langle\bar\xi_a(\tau)\xi_b(\tau')\rangle=\delta_{ab}\,\Theta(\tau'-\tau)$,
where $\Theta$ denotes the usual step function. The integration over the
auxiliary fields with the measure (\ref{Dmeasure}) therefore turns
(\ref{wilsonloop}) into a path-ordered exponential functional of the fields $x$
which is the $T$-dual of the usual Wilson loop operator for the ten-dimensional
gauge field $(A^0,-\frac1{2\pi\alpha'}Y^i)$ dimensionally reduced to the
D-particle worldlines. In this picture, $A^0$ is thought of as a gauge field
living on the brane worldline, while $Y_i^{aa}$,
$a=1,\dots,N$, are the transverse coordinates of
the $N$ D-particles and $Y_i^{ab}$, $a\neq b$, of the short open string
excitations connecting them. We shall subtract out the center of mass motion of
the
assembly of $N$ D-branes and assume that $Y_i\in su(N)$. We shall also use
$SU(N)$-invariance of the theory (\ref{partfnD}) to select the temporal gauge
$A^0=0$.

The action in (\ref{partfnD}) may be formally identified with the
deformed conformal field theory (\ref{sigma1}) by taking the couplings $g^I\sim
Y_i^{ab}$ and introducing the one-parameter family of bare matrix-valued vertex
operators
\be
V_{ab}^i(x;\tau)=\frac{g_s}{2\pi\alpha'}\,\partial_\sigma
x^i(\tau)\,\bar\xi_a(\tau)\xi_b(\tau)
\label{Dvertexops}\ee
This means that there is a one-parameter family of Dirichlet boundary
conditions for the fundamental string fields $x^i$ on $\partial\Sigma$,
labeled by $\tau\in[0,1]$ and the configuration fields
\be
y_i(x^0;\tau)=\bar\xi_a(\tau)\,Y_i^{ab}(x^0(\tau))\,\xi_b(\tau)
\label{bdryfields}\ee
Instead of being forced to sit on a unique hypersurface as in the case of a
single D-brane, in the non-abelian case there is an infinite set of
hypersurfaces on which the string endpoints are situated. In this sense the
coordinates (\ref{bdryfields}) may be thought of as an ``abelianization'' of
the non-abelian D-particle coordinate fields $Y_i^{ab}$.

To describe the non-relativistic dynamics of heavy D-particles, the natural
choice is to take the couplings to correspond to the Galilean boosted
configurations $Y_i^{ab}(x^0)=Y_i^{ab}+U_i^{ab}x^0$, where $U_i$ is the
non-relativistic velocity matrix. However, logarithmic modular divergences
appear in matter field amplitudes at higher genera
when the string propagator $L_0$ is computed with
Dirichlet boundary conditions. These modular divergences are canceled by
adding logarithmic recoil operators \cite{ms1,kmw12} to the matrix
$\sigma$-model action in (\ref{partfnD}). From a physical point of view, if one
is to use low-energy probes to observe short-distance spacetime structure, such
as a generalized Heisenberg microscope, then one needs to consider the
scattering of string matter off the assembly of D-particles. For the
Galilean-boosted multiple
D-particle system, the recoil is described by taking the
deformation of the $\sigma$-model action in (\ref{partfnD}) to be of the form
\cite{ms1}
\be
Y_i^{ab}(x^0)=\sqrt{\alpha'}\,Y_i^{ab}C_\epsilon(x^0)+U_i^{ab}D_\epsilon(x^0)=
\left(\sqrt{\alpha'}\,\epsilon Y_i^{ab}+U_i^{ab}x^0\right)\Theta_\epsilon(x^0)
\label{recoilY}\ee
where
\be
C_\epsilon(x^0)=\epsilon\,\Theta_\epsilon(x^0)~~~~~~,~~~~~~
D_\epsilon(x^0)=x^0\,\Theta_\epsilon(x^0)
\label{recoilops}\ee
and
\be
\Theta_\epsilon(x^0)=\frac1{2\pi
i}\int\limits_{-\infty}^{+\infty}\frac{dq}{q-i\epsilon}~\e^{iqx^0}
\label{stepfnreg}\ee
is the regulated step function whose $\epsilon\to0^+$ limit is the usual step
function. The operators (\ref{recoilops}) have non-vanishing matrix elements
between different string states and therefore describe the appropriate change
of quantum
state of the D-brane background. They can be thought of as describing the
recoil of the assembly of D-particles in an impulse approximation, in which it
starts moving as a whole only at time $x^0=0$. The collection of constant
matrices $\{Y^i_{ab},U^j_{cd}\}$ now form the set of coupling constants
$\{g^I\}$ for the
worldsheet $\sigma$-model (\ref{partfnD}).

As discussed previously, 
the recoil operators (\ref{recoilops}) possess a very important property. They
lead to a deformation of the free $\sigma$-model action in (\ref{partfnD})
which is not conformally-invariant, but rather defines a logarithmic conformal
field theory \cite{gurarie,lcftfurther}.
Such a quantum field theory contains logarithmic scaling violations in its
correlation functions on the worldsheet, which can be seen in the present case
by computing the pair correlators of the fields (\ref{recoilops})
\cite{kmw12}
\bea
\Bigl\langle
C_\epsilon(z)\,C_\epsilon(0)\Bigr\rangle&=&0\nn\\\Bigl\langle
C_\epsilon(z)\,D_\epsilon(0)\Bigr\rangle&=&\frac
b{z^{h_\epsilon}}\nn\\\Bigl\langle
D_\epsilon(z)\,D_\epsilon(0)\Bigr\rangle&=&\frac{b\,\alpha'}
{z^{h_\epsilon}}\,\log z
\label{2ptconfalg}\eea
where
\be
h_\epsilon=-\frac{|\epsilon|^2\,\alpha'}2
\label{CDconfdim}\ee
is the conformal dimension of the recoil operators. The constant $b$ is fixed
by the leading logarithmic divergence of the conformal blocks of the theory.
Note that (\ref{CDconfdim}) vanishes as $\epsilon\to0$, so that the logarithmic
worldsheet divergences in (\ref{2ptconfalg}) cancel the modular annulus
divergences mentioned above. An essential ingredient for this cancellation is
the identification \cite{kmw12}
\be
\frac1{\epsilon^2}=-2\alpha'\log\Lambda
\label{epLambdaid}\ee
which relates the target space regularization parameter $\epsilon$ to the
worldsheet ultraviolet cutoff scale $\Lambda$.

Logarithmic conformal field theories are characterized by the fact that their
Virasoro generator $L_0$ is not diagonalizable, but rather admits a Jordan cell
structure. Here the operators (\ref{recoilops}) form the basis of a $2\times2$
Jordan block and they appear in the spectrum of the two-dimensional quantum
field theory as a consequence of the zero modes that arise from the breaking of
the target space translation symmetry by the topological defects. The mixing
between $C$ and $D$ under a conformal transformation of the worldsheet can be
seen explicitly by considering a scale transformation
\be
\Lambda\to\Lambda'=\Lambda\,\e^{-t/\sqrt{\alpha'}}
\label{Lambdatransf}\ee
Using (\ref{epLambdaid}) it follows that the operators (\ref{recoilops}) are
changed according to $D_\epsilon'=D_\epsilon+t\sqrt{\alpha'}C_\epsilon$ and
$C_\epsilon'=C_\epsilon$. Thus in order to maintain scale-invariance of the
theory (\ref{partfnD}) the coupling constants must transform under
(\ref{Lambdatransf}) as \cite{kmw12,lm11} $Y'^i=Y^i+U^it$ and $U'^i=U^i$,
which are just the Galilean transformation laws for the positions $Y^i$ and
velocities $U^i$. Thus a scale transformation of the worldsheet is equivalent
to a
Galilean transformation of the moduli space of $\sigma$-model couplings, with
the parameter $\epsilon^{-2}$ identified with the time evolution parameter
$t=-\sqrt{\alpha'}\log\Lambda$. The corresponding
$\beta$-functions for the worldsheet renormalization group flow are
\bea
\beta_{Y_i}&\equiv&\frac{dY_i}{dt}=h_\epsilon\,Y_i+\sqrt{\alpha'}\,U_i
\nn\\\beta_{U_i}&\equiv&\frac{dU_i}{dt}=h_\epsilon\,U_i
\label{betafns}\eea
and they generate the Galilean group $G(9)^{N^2}$ in nine-dimensions.

The associated Zamolodchikov metric
\be
G_{ab;cd}^{ij}=2N\Lambda^2\left\langle
V_{ab}^i(x;0)\,V_{cd}^j(x;0)\right\rangle
\label{ZamD}\ee
can be evaluated to leading orders in $\sigma$-model perturbation theory using
the logarithmic conformal algebra (\ref{2ptconfalg}) and the propagator of the
auxiliary fields to give \cite{ms1}
\bea
G_{ab;cd}^{ij}&=&\frac{4\bar g_s^2}{\alpha'}\left[\eta^{ij}\,I_N\otimes
I_N+\frac{\bar g_s^2}{36}\left\{I_N\otimes\left(\bar U^i\bar U^j+\bar U^j\bar
U^i\right)\right.\right.\nn\\& &\biggl.\left.+\,\bar U^i\otimes\bar U^j+\bar
U^j\otimes\bar U^i+\left(\bar U^i\bar U^j+\bar U^j\bar U^i\right)\otimes
I_N\right\}\biggr]_{db;ca}+ \nonumber \\ 
& & {\cal O}\left(\bar g_s^6\right)
\label{zammetricexpl}\eea
where $I_N$ is the identity operator of $SU(N)$ and we have introduced the
renormalized coupling constants
\be
\bar g_s=g_s/\sqrt{\alpha'}\epsilon~~~~~~,~~~~~~\bar
U^i=U^i/\sqrt{\alpha'}\epsilon
\label{rencc}\ee
{}From the renormalization group equations (\ref{betafns}) it follows that the
renormalized velocity operator in target space is truly marginal,
\be
\frac{d\bar U^i}{dt}=0
\label{dUdt}\ee
which ensures uniform motion of the D-branes. It can also be shown that the
renormalized string coupling $\bar g_s$ is time-independent \cite{ms1}. If we
further define the position renormalization
\be
\bar Y^i=Y^i/\sqrt{\alpha'}\epsilon
\label{renY}\ee
then the $\beta$-function equations (\ref{betafns}) coincide with the Galilean
equations of motion of the D-particles, i.e.
\be
\frac{d\bar Y^i}{dt}=\bar U^i
\label{dYdt}\ee
Note that the Zamolodchikov metric (\ref{zammetricexpl}) is a complicated
function of the D-brane dynamical parameters, and as such it represents the
appropriate effective target space geometry of the D-particles. The moduli
space Lagrangian (\ref{effcalL}) is then readily seen to coincide with the
expansion to ${\cal O}(\bar g_s^4)$ of the symmetrized form of the non-abelian
Born-Infeld action for the D-brane dynamics \cite{tseytlin},
\be
{\cal L}_{\rm NBI}=\frac1{\sqrt{2\pi\alpha'}\bar g_s}\,\tr~{\rm
Sym}\,\sqrt{\det_{\mu,\nu}\left[\eta_{\mu\nu}\,I_N+2\pi\alpha'\bar
g_s^2\,F_{\mu\nu}\right]}
\label{NBIaction}\ee
where tr denotes the trace in the fundamental representation of $SU(N)$,
\be
{\rm Sym}(M_1,\dots,M_n)=\frac1{n!}\sum_{\pi\in S_n}M_{\pi_1}\cdots
M_{\pi_n}
\label{symprod}\ee
is the symmetrized matrix product and the components of the
dimensionally reduced field strength tensor are given by
\be
F_{0i}=\frac1{2\pi\alpha'}\,\frac{d{\bar Y}_i}{dt}~~~~~~,~~~~~~
F_{ij}=\frac{\bar g_s}{(2\pi\alpha')^2}\left[\bar Y_i\,,\,\bar Y_j\right]
\label{Fred}\ee

\subsection{Evolution Equation for the Probability Distribution}

In this section we will derive the temporal evolution equation for the
probability density ${\cal P}(\{g^I\},t)$ following the identification of time
with a worldsheet renormalization group scale (i.e. the Liouville zero mode).
The basic identity is the Wilson-Polchinski equation for the case of the
worldsheet action (\ref{sigma1}) which reads \cite{hlp6}
{\small \bea
0&=&\frac{\partial{\cal Z}}{\partial\log\Lambda}\nonumber\\&=&\int
Dx^\mu~\e^{-S_\sigma[x;\{g^I\}]}\,\left\{\frac{\partial S_{\rm
int}}{\partial\log\Lambda}\right.\nonumber\\& &\left.-\,\int\limits_\Sigma
d^2z~\int\limits_\Sigma
d^2w~\left(\frac\partial{\partial\log\Lambda}G(z-w)\right)
\left[
\frac{\delta^2S_{\rm int}}{\delta x^\mu(z)\delta x_\mu(w)}+
\frac{\delta S_{\rm int}}{\delta x^\mu(z)}\frac{\delta S_{\rm int}}
{\delta x_\mu(w)}\right]\right\} 
\label{WPeq9}\eea}
and it is the requirement of conformal invariance of the quantum string theory.
Here $S_{\rm int}=S_\sigma-S_*$, $\cal Z$ is the partition function of the
$\sigma$-model, and
\be
G(z-w)=\left\langle\NO x^\mu(z)x_\mu(w)\NO\right\rangle_*
\label{2ptfn}\ee
is the two-point function computed with respect to the conformal field theory
action $S_*[x]$. The basic assumption in arriving at eq. (\ref{WPeq9}) is that
the ultra-violet cutoff $\Lambda$ on the string worldsheet appears explicitly
only in the propagator $G(z-w)$, as can always be arranged by an appropriate
regularization \cite{hlp6}.

Henceforth we shall concentrate on the specific case of interest of a system of
$N$ interacting D-particles. Then, upon summing up over pinched genera, there
are extra logarithmic divergences in the Green's function (\ref{2ptfn}) coming
from pinched annulus diagrams, which may be removed by the introduction of
logarithmic recoil operators, as explained in the previous section. Using
primes to denote the result of resumming the topological expansion over pinched
genera, we then have that
\be
\frac{\partial}{\partial\log\Lambda}G(z-w)'=
\frac\partial{\partial\log\Lambda}{\sum_{\rm genera}}'\left\langle\NO
x^\mu(z)x_\mu(w)\NO\right\rangle=\frac\partial{\partial\log\Lambda}
\left\langle\NO x^\mu(z)x_\mu(w)\NO\right\rangle_{\rm int}
\label{corrint}\ee
where the correlator $\langle\cdot\rangle_{\rm int}$ includes the disc and
recoil interaction contributions. Subtracting the disc $\Lambda$-dependence in
normal ordering, the remaining dependence on the worldsheet cutoff comes from
the two-point functions of the logarithmic recoil operators, giving terms of
the form
\be
\frac\partial{\partial\log\Lambda}\left\langle\NO
x^\mu(z)x_\mu(w)\NO\Bigl(a_{CC}C_\epsilon(z)C_\epsilon(w)+a_{CD}
C_\epsilon(z)D_\epsilon(w)+a_{DD}D_\epsilon(z)D_\epsilon(w)\Bigr)
\right\rangle_*
\label{recoilcontrtot}\ee
The leading divergence comes from the correlation function $\langle
D_\epsilon(z)D_\epsilon(w)\rangle_*\sim\log\Lambda$, which follows upon the
identification (\ref{epLambdaid}). Thus we may write
\be
\frac\partial{\partial\log\Lambda}G(z-w)\simeq
c_G\,(\alpha')^2\log|z-w|\,\sum_{i=1}^9\,\sum_{a,b=1}^N|U_{ab}^i|^2
\label{leadingGLambda}\ee
where $c_G>0$ is a numerical coefficient whose precise value is not important,
and we have used the fact that $U^i\in su(N)$.

Next, we observe that in the case of D-particles the second term in eq.
(\ref{WPeq9}) becomes
\bea
& &\int
D\mu(x,\bar\xi,\xi)~\e^{-S_\sigma}\,\oint\limits_{\partial\Sigma}d\tau~\oint
\limits_{\partial\Sigma}d\tau'~(\alpha')^2c_G\sum_{i=1}^9\sum_{a,b=1}^N
|U_{ab}^i|^2\,\log[2-2\cos(\tau-\tau')]\nonumber\\& &~~~~~~\times\,
\left[\frac{\delta^2S_{\rm int}}{\delta x^\mu(\tau)\delta x_\mu(\tau')}
+\frac{\delta S_{\rm int}}{\delta x^\mu(\tau)}
\frac{\delta S_{\rm int}}{\delta x_\mu(\tau')}\right]
\label{2ndterm}\eea
where the interaction Lagrangian is given by
\be
S_{\rm
int}=\frac{g_s}{2\pi\alpha'}\oint\limits_{\partial\Sigma}d\tau~\partial_\sigma
x^i(\tau)\,\bar\xi_a(\tau)\,Y_i^{ab}(x^0)\,\xi_b(\tau)
\label{DSint}\ee
In the case of a system of recoiling D0-branes, the $\sigma$-model couplings in
eq. (\ref{DSint}) are given by (\ref{recoilY}) with the abelianized couplings
(\ref{bdryfields}) of $Y_i^{ab}$ viewed as the boundary values for the open
string embedding fields $x^i(\tau)$ on the D-brane. This means that the fields
$x^i(\tau)$ are simply identified with $\bar\xi_a(\tau)Y_i^{ab}\xi_b(\tau)$.
All the non-trivial dependence comes from the $x^0$ field which obeys Neumann
boundary conditions and is not constant on the boundary of $\Sigma$. Then we
may write
\bea
& &\frac{\delta^2S_{\rm int}}{\delta x^\mu(\tau)\delta
x_\mu(\tau')}+\frac{\delta S_{\rm int}}{\delta x^\mu(\tau)}\frac{\delta S_{\rm
int}}{\delta x_\mu(\tau')}\nonumber\\& &~~~~=\nabla_{y_i}^2S_{\rm
int}+\left(\nabla_{y_i}S_{\rm
int}\right)^2+\left(\frac{g_s}{2\pi\alpha'}\right)^2\,U_i^{ab}U_j^{cd}
\bar\xi_a(\tau)\xi_b(\tau)\bar\xi_c(\tau')\xi_d(\tau')\nonumber
\\& &~~~~~~\times\,\partial_\sigma x^i(\tau)\partial_\sigma x^j(\tau')
\Theta_\epsilon(x^0(\tau))\Theta_\epsilon(x^0(\tau'))
\label{Sintderivs}\eea
where $y_i$ denotes the constant abelianized zero modes of $x^i(\tau)$ on
$\partial\Sigma$. Here we have used the fact that terms of the form
$x^0\delta(x^0)$ and $\Theta(x^0)\delta(x^0)$ vanish with the regularization
(\ref{stepfnreg}) \cite{ms1}. The terms involving $\partial_\sigma
x^i(\tau)\partial_\sigma x^j(\tau')$ will average out to yield terms of the
form
\be
|U_{ab}^i|^2\left\langle\Theta_\epsilon(x^0(\tau))\Theta_\epsilon(x^0(\tau'))
\right\rangle_*=\alpha'|\bar U_{ab}^i|^2\left\langle C_\epsilon(\tau)
C_\epsilon(\tau')\right\rangle_*\sim{\cal O}(\epsilon^2)
\label{Oepterms9b}\ee
where we have used the logarithmic conformal algebra. At leading orders, these
terms vanish, but we shall see the importance of such sub-leading terms later
on.

Using the Dirichlet correlator
\be
\left\langle\partial_\sigma x^i(\tau)\partial_\sigma
x_i(\tau')\right\rangle_*=-\frac{36\pi^2\alpha'}{1-\cos(\tau-\tau')}
\label{Dcorr}\ee
we find that the boundary integrations in eq. (\ref{2ndterm}) are of the form
\cite{ms1}
\be
\oint\limits_{\partial\Sigma}d\tau~\oint\limits_{\partial\Sigma}d\tau'~
\frac{\log[2-2\cos(\tau-\tau')]}{1-\cos(\tau-\tau')}\sim\log\Lambda
\label{bdryints}\ee
which has the effect of renormalizing the velocity matrix $U_{ab}^i\to\bar
U_{ab}^i$. Thus, ignoring the ${\cal O}(\epsilon^2)$ terms for the moment, we
find that the remaining terms in the Wilson-Polchinski renormalization group
equation (\ref{WPeq9}) yield a diffusion term for the probability density:
\be
\partial_t{\cal P}[Y,U;t]=c_G\sqrt{\alpha'}\sum_{j=1}^9\sum_{a,b=1}^N|\bar
U_{ab}^j|^2\,\nabla_{y_i}^2{\cal P}[Y,U;t]+{\cal O}(\epsilon^2)
\label{diffprob10}\ee
This equation is of the Fokker-Planck type,
with diffusion coefficient
\be
{\cal D}=c_G\sqrt{\alpha'}\sum_{i=1}^9\sum_{a,b=1}^N|\bar U_{ab}^i|^2
\label{diffcoeff11}\ee
coming from the quantum recoil of the assembly of D-particles. The diffusion
disappears when there is no recoil. Note that (\ref{diffcoeff11}) naturally
incorporates the short-distance quantum gravitational smearings for the open
string interactions (compare with eq. (\ref{Deltayia})), and it arises as an
abelianized velocity for the constant auxiliary field configuration
$\bar\xi_a(\tau)=\xi_a(\tau)=1~~\forall a=1,\dots,N$.

The evolution equation (\ref{diffprob10}) should be thought of as a
modification of the usual continuity equation for the probability density.
Indeed, as we will now show, the ${\cal O}(\epsilon^2)$ terms in eq.
(\ref{diffprob10}) coming from (\ref{Oepterms9b}) are of the form
$-\nabla_{y_i}{\cal J}_i$, where
\be
{\cal
J}_i=\frac{\hbar_\moduls}{2im}\left(\Psi^\dagger\,\nabla_{y_i}\Psi-
\Psi\,\nabla_{y_i}\Psi^\dagger\right)
\label{current13}\ee
is the probability current density. Here
\be
m=\frac1{\sqrt{\alpha'}\bar g_s}~~~~~~,~~~~~~\hbar_\moduls=4\bar g_s
\label{mhbardefs}\ee
are, respectively, the
BPS mass of the D-particles and the moduli space
``Planck constant''.~\footnote{\baselineskip=12pt The
identification (\ref{mhbardefs}) of
Planck's constant in the D-particle quantum mechanics on moduli space
with the string coupling constant is actually not unique
in the present context of considering only the exchange of strings
between D-particles. As discussed in \cite{ms1}, the most general
relation, compatible with the
logarithmic conformal algebra, involves an arbitrary exponent $\chi$ through
$\hbar_\moduls=4(\bar g_s)^{1 + \chi/2}$. The exponent $\chi$
arises from specific mechanisms for the cancellation of modular
divergences on pinched annular surfaces by appropriate
world-sheet short-distance
infinities at lower genera. The only restriction imposed
on $\chi$ is that it be positive definite.
As shown in \cite{ms1}, the standard kinematical properties of D-particles
are reproduced by the choice $\chi=\frac23$.
A choice of $\chi\ne0$ seems more natural from the point of view
that modular divergences should be suppressed for
weakly interacting strings. However, in the present case,
we assume for simplicity the value $\chi=0$, which yields the standard
string smearing $\sqrt{\alpha '}$ for the minimum length uncertainty.
The incorporation of an arbitrary $\chi\ge 0$ in the formalism
is straightforward and would not affect the qualitative properties of
the following results.}

For this, we note first of all that such terms should generically come in the
form
\be
-\nabla_{y_i}{\cal
J}_i=-\frac{\hbar_\moduls}m\left(\nabla_{y_i}^2\,{\rm arg}\,\Psi\right){\cal
P}-\frac{\hbar_\moduls}m\left(\nabla_{y_i}\,{\rm
arg}\,\Psi\right)\nabla_{y_i}{\cal P}
\label{genterms}\ee
The second term in eq. (\ref{genterms}), upon
identification of the probability density $\cal P$ with the genera resummed
partition function on the string worldsheet, is proportional to the worldsheet
renormalization group $\beta$-function, given the gradient flow property
(\ref{gflow}) of the string effective action \cite{mm3b}, so that
\be
\nabla_{y_i}{\cal P}=-2{\cal P}\,G_{ij}\beta^j
\label{gflowcalP}\ee
which is to be understood in terms of abelianized quantities. In the present
case the renormalization group equations are given by (\ref{dUdt}) and
(\ref{dYdt}) and, since the couplings $\bar U_i^{ab}$ are truly marginal, we
are left in (\ref{gflowcalP}) with only a Zamolodchikov metric contribution
$G_{CC}=2N\Lambda^4\langle C_\epsilon(\tau)C_\epsilon(\tau)\rangle_*$ (Note
that here one should use suitably normalized correlators
$\langle\cdot\rangle_*$
which yield the behavior (\ref{gflowcalP})). From the logarithmic conformal
algebra it therefore follows that a term with the structure of the second piece
in eq. (\ref{genterms}) is hidden in the contributions (\ref{Oepterms9b}) which
were dropped as being subleading in $\epsilon$. Furthermore, from
(\ref{zammetricexpl}), (\ref{genterms}), (\ref{leadingGLambda}),
(\ref{Sintderivs}) and (\ref{Oepterms9b}) it follows that to leading orders
\be
\nabla_{y_i}\,{\rm arg}\,\Psi=-\frac{c_G}{\sqrt{\alpha'}\bar
u_i}\left(\sum_{j=1}^9\sum_{a,b=1}^N|\bar U_{ab}^j|^2\right)^2
\label{thetaeq}\ee
where $\bar u_i=d\bar y_i/dt$ is the worldsheet zero mode of the abelianized,
renormalized velocity operator. It then follows that to leading orders we have
$\nabla_{y_i}^2\,{\rm arg}\,\Psi=0$.\footnote{\baselineskip=12pt
Noncommutative position
dependent terms arising from commutators $[Y_i,Y_j]$ appear
only at two-loop order in $\sigma$-model perturbation theory \cite{ms1}. An
interesting extension of the present analysis would be to generalize the
results to include these higher-order terms into the quantum dynamics. However,
given that the pertinent equations involve only the abelianized coordinates
(\ref{bdryfields}), we do not expect the inclusion of such terms to affect the
ensuing qualitative conclusions. The effect of the noncommutativity is to
render the quantum wave equation for the system of D-particles non-linear,
through the recoil-induced diffusion from the multi-brane interactions, as we
discuss in the subsequent sections (for a single brane one would obtain a free
wave equation governing the quantum dynamics).}

Thus, keeping the subleading terms in the target space regularization parameter
$\epsilon$ leads to the complete Fokker-Planck equation for the probability
density ${\cal P}=\Psi^\dagger\Psi$:
\be
\partial_t{\cal P}[Y,U;t]=-\nabla_{y_i}{\cal J}_i[Y,U;t]+{\cal
D}\,\nabla_{y_i}^2{\cal P}[Y,U;t]
\label{FPeq12}\ee
where ${\cal J}_i$ is the probability current density (\ref{current13}), with
$\Psi$ the wavefunctional for the system of D-branes:
\be
\Psi[Y,U;t]=\prod_{i=1}^9\exp\left[-\frac{ic_G}{\sqrt{\alpha'}}\frac{y_i}{\bar
u_i}\left(\sum_{j=1}^9\tr\,|\bar
U^j|^2\right)^2\right]\,\Bigl|\Psi[Y,U;t]\Bigr|
\label{Psi14}\ee
Such quantum diffusion is characteristic of all Liouville string theories
\cite{emn2,emn10,emn10b}. The resulting quantum dynamics, including the
quantum diffusion which arises from the D-brane recoil, is described by the
Schr\"odinger wave equation which corresponds to this Fokker-Planck equation.
This equation is analyzed in detail in the next section.

\subsection{Non-linear Schr\"odinger Wave Equations}

Given the Fokker-Planck equation (\ref{FPeq12}), there is no unique solution
 for the wavefunction $\Psi$, as we discuss below, and the resulting
Schr\"odinger wave equation is necessarily non-linear, due to the diffusion
term \cite{dg7,dg8}. Consider the quantum mechanical system with diffusion
which is described by the
Fokker-Planck equation (\ref{FPeq12}) for the probability density ${\cal
P}=\Psi^\dagger\Psi$. In \cite{dg7} it was shown that, by imposing
diffeomorphism invariance in the space $\vec y\in\modul$ and representing the
symmetry through the infinite-dimensional kinematical symmetry algebra
$C^\infty(\modul)~\semiplus{\rm Vect}(\modul)$, one may arrive
at the following {\it non-linear} Schr\"odinger wave equation:
\be
i\hbar_\moduls\,\frac{\partial\Psi}{\partial t}={\cal H}_0\Psi+iI(\Psi)\Psi
\label{nlse16a}\ee
where ${\cal H}_0$ is the linear Hamiltonian operator
\be
{\cal H}_0=-\frac{\hbar_\moduls^2}{2m}\,\nabla_{y_i}^2+V_\moduls(\vec y,\vec u;t)
\label{Ham16b}\ee
and
\be
I(\Psi)=\frac12\,\hbar_\moduls\,{\cal
D}\,\frac{\nabla_{y_i}^2(\Psi^\dagger\Psi)}{\Psi^\dagger\Psi}
\label{IPsi}\ee
Here $V_\moduls(\vec y,\vec u;t)$ is the interaction potential on moduli space
and the real continuous quantum number $\cal D$ in (\ref{diffcoeff11}) is the
classification parameter of the unitarily inequivalent diffeomorphism group
representations. Other models which have more than one type of diffusion
coefficient
can be found in \cite{dg7,dg8}.

A crucial point \cite{dg8} is that there exist non-linear phase
transformations of the wavefunction $\Psi$ (known as quantum mechanical ``gauge
transformations'') which leave invariant appropriate families of
non-linear Schr\"odinger equations, and also the probability density $\cal P$.
Such transformations do not affect any physical observables of the system. This
implies that the choice of $\Psi$ is ambiguous, once a density $\cal P$ is
found as a solution of eq. (\ref{FPeq12}) on the collective coordinate space
$\{Y_i^{ab}\}$ of the D-branes. An important ingredient in finding such
transformations is the assumption \cite{dg8,fh9} that all measurements of
quantum mechanical systems can be made so as to reduce eventually to position
and time measurements. Because of this possibility, a theory formulated in
terms of position measurements is complete enough in principle to describe all
quantum phenomena. This point of view is certainly met by the D-brane moduli
space, whereby the wavefunctional depends only on the couplings $\{g^I\}$ and
not on the conjugate momenta $p_I=-i\hbar_\moduls\,\partial/\partial g^I$.
The group of non-linear gauge transformations acts on each leaf in a foliation
of a family of non-linear Schr\"odinger equations, such that the
two-dimensional leaves of the foliation consist of sets of equivalent
quantum mechanical evolution equations.

It follows that then one can perform the following local, two-parameter
projective gauge transformation of the wavefunction \cite{dg8}:
\be
\Psi'=N_{\gamma,\lambda}(\Psi)=|\Psi|\exp(i\gamma\log|\Psi|+i\lambda\,{\rm
arg}\,\Psi)
\label{phasetransf17}\ee
under which the probability density is invariant, but the probability
current transforms as
\be
{\cal J}_i'=\lambda\,{\cal J}_i+\frac\gamma2\,\nabla_{y_i}{\cal P}
\label{probcurrenttransf}\ee
Here $\gamma(t)$ and $\lambda(t)\neq0$ are some real-valued time-dependent
functions. The collection of all non-linear transformations
$N_{\gamma,\lambda}$
obeys the multiplication law of the one-dimensional affine Lie group
$Aff(1)$. Under (\ref{phasetransf17})
there are families of non-linear Schr\"odinger equations that are
{\it closed} (in the sense of ``gauge closure''). A generic form of such a
family, to which the non-linear Schr\"odinger equation (\ref{nlse16a}) belongs,
is
\bea
i\,\frac{\partial\Psi}{\partial
t}&=&\frac1{\hbar_\moduls}\,{\cal
H}_0\Psi+i\nu_2R_2[\Psi]\,\Psi+\mu_1R_1[\Psi]\,\Psi+\left(\mu_2-
\mbox{$\frac12$}\,\nu_1\right)R_2[\Psi]\,\Psi\nn\\& &+\,(\mu_3+\nu_1)
R_3[\Psi]\,\Psi+\mu_4R_4[\Psi]\,\Psi+\left(\mu_5+\mbox{$\frac14$}\,\nu_1
\right)R_5[\Psi]\,\Psi\nn\\&=&i\sum_{i=1,2}\nu_iR_i[\Psi]\,\Psi+
\sum_{j=1}^5\mu_jR_j[\Psi]\,\Psi +\frac1{\hbar_\moduls}\,
V_\moduls(\vec y,\vec u;t)\,\Psi
\label{family18}\eea
where $\nu_i,\mu_j$ are real-valued coefficients which are related to diffusion
coefficients $\cal D$ and ${\cal D}'$ by
\bea
\nu_1&=&-\frac{\hbar_\moduls}{2m}\nn\\\nu_2&=&\frac12\,{\cal
D}\nn\\\mu_1&=&c_1{\cal D}'\nn\\\mu_2&=&-\frac{\hbar_\moduls}{4m}+c_2{\cal
D}'\nn\\\mu_3&=&\frac{\hbar_\moduls}{2m}+c_3{\cal D}'\nn\\\mu_4&=&c_4{\cal
D}'\nn\\\mu_5&=&\frac{\hbar_\moduls}{8m}+c_5{\cal D}'
\label{numu19}\eea
and $R_j[\Psi]$ are non-linear homogeneous functionals of degree 0 which are
defined by
\bea
R_1&=&\frac m{\hbar_\moduls}\,\frac{\nabla_{y_i}{\cal J}_i}{\cal
P}\nn\\R_2&=&\frac{\nabla_{y_i}^2{\cal P}}{\cal
P}\nn\\R_3&=&\frac{m^2}{\hbar_\moduls^2}\,\frac{{\cal J}_i^2}{{\cal
P}^2}\nn\\R_4&=&\frac m{\hbar_\moduls}\,\frac{{\cal J}_i\,\nabla_{y_i}{\cal
P}}{{\cal P}^2}\nn\\R_5&=&\frac{(\nabla_{y_i}{\cal P})^2}{{\cal P}^2}
\label{Rj20}\eea
In eq. (\ref{numu19}) the $c_j$ are constants, while in eq. (\ref{Rj20}) the
probability current density is given by (\ref{current13}) with ${\cal
P}=\Psi^\dagger\Psi$.

The gauge group $Aff(1)$ acts on the parameter space of the family
(\ref{family18}). Some members of this family are thereby linearizable to an
ordinary Schr\"odinger wave equation under the action of (\ref{phasetransf17}).
These are the members for which there exists a specific relation between $\cal
D$ and ${\cal D}'$ \cite{dg8}, and for which Ehrenfest's theorem of quantum
mechanics receives no dissipative corrections. The quantum mechanics of
D-particles is not of this type, given that there is definite diffusion,
dissipation and thus time irreversibility.
However, as discussed in \cite{ms1,kmw12,lm11}, one needs to also maintain
Galilean invariance, which is a property originating from the logarithmic
conformal algebra of the recoil operators. As described in \cite{dg8}, there is
a class of non-linear Schr\"odinger wave equations which is Galilean invariant
but which violates time-reversal symmetry. For this, it is useful to first
construct a parameter set of equations of the form (\ref{numu19}) which remain
{\it invariant} under the gauge transformations (\ref{phasetransf17}). We may
describe the parameter family of equations (\ref{family18}) in terms of orbits
of $Aff(1)$ by regarding $\gamma=2m\mu_1$ and $\lambda=2m\nu_1$ as the group
parameters of an $Aff(1)$ gauge transformation (\ref{phasetransf17}).
Then the remaining five parameters in (\ref{numu19}) are taken to be
the functionally-independent parameters
$\eta_j$,
$j=1,\dots,5$, which are invariant under $Aff(1)$ and are defined by
\bea
\eta_1&=&\nu_2-\frac12\,\mu_1\nn\\\eta_2&=&\nu_1\mu_2-\nu_2\mu_1\nn\\
\eta_3&=&\frac{\mu_3}{\nu_1}\nn\\\eta_4&=&\mu_4-\mu_1\,\frac{\mu_3}{\nu_1}
\nn\\\eta_5&=&\nu_1\mu_5-\nu_2\mu_4+(\nu_2)^2\,\frac{\mu_3}{\nu_1}
\label{etaj21}\eea
A detailed discussion of the corresponding physical observables is given in
\cite{dg8}. For our purposes, we simply select the following relevant property
of the non-linear Schr\"odinger equation based on the parameter set
(\ref{etaj21}).

Consider the effect of time-reversal on the non-linear Schr\"odinger wave
equation. Setting $t\to-t$ is equivalent to introducing the following new set
of coefficients:
\bea
(\nu_i)^T&=&-\nu_i~~~~i=1,2\nn\\(\mu_j)^T&=&-\mu_j~~~~j=1,\dots,5\nn\\
(V_\moduls)^T&=&-V_\moduls
\label{Tcoeffs22}\eea
where the superscript $T$ denotes the time-reversal transformation. It is
straightforward to show \cite{dg8} that, in terms of the $\eta_j$'s, there is
time-reversal invariance in the non-linear Schr\"odinger equation if
the two parameters $\eta_1$ and $\eta_4$ are both non-vanishing. On the
other hand, a straightforward calculation also shows \cite{dg8} that Galilean
invariance sets $\eta_4=0$, thereby implying that a family of
non-linear Schr\"odinger wave equations which is invariant under $G(9)$ but not
time-reversal invariant indeed {\it exists}. For a single diffusion coefficient
${\cal D}\neq0$, as in the case (\ref{diffcoeff11}) of recoiling D-branes, one
may set ${\cal D}'c_j=0$ (corresponding to the $Aff(1)$ gauge choice $\mu_1=0$)
and thereby obtain the set of gauge invariant
parameters:
\bea
\eta_1&=&\frac12\,{\cal
D}\nn\\\eta_2&=&2\alpha'\bar g_s^4\nn\\\eta_3&=&-1\nn\\\eta_4&=&0\nn\\
\eta_5&=&-\alpha'\bar g_s^4-\frac14\,{\cal D}^2
\label{etainv23}\eea
The parameter set (\ref{etainv23}) breaks time-reversal invariance, as expected
from the non-trivial entropy production and decoherence characterizing the
worldsheet renormalization group approach to target space time involving
Liouville string theory \cite{ms1,emn2,emn13}.
But it {\it does} preserve Galilean
invariance, as is required by conformal invariance of the non-relativistic,
recoiling system of D-particles.

One may therefore propose that the Fokker-Planck equation for the probability
density $\cal P$ on the moduli space of collective coordinates of a system of
interacting D-branes implies a Schr\"odinger wave equation for the pertinent
wavefunctional which is non-linear, Galilean-invariant and has a time arrow,
corresponding to entropy production, and hence explicitly broken time-reversal
invariance. The existence of a dissipation ${\cal D}\propto\tr\,|\bar U^i|^2$,
due to the quantum recoil of the D-branes, implies that the Ehrenfest relations
acquire extra dissipative terms for this family of non-linear Schr\"odinger
equations. For example, one can immediately obtain the relations \cite{dg7}
\bea
\frac
d{dt}\Bigl\langle\!\!\Bigl\langle\widehat{p}_i\Bigr\rangle\!\!\Bigr\rangle
&=&-\Bigl\langle\!\!\Bigl\langle\nabla_{y_i}
V_\moduls\Bigr\rangle\!\!\Bigr\rangle-m\int\limits_\moduls d\vec y~\Psi^\dagger
\left(\frac{{\cal J}_i^{({\cal D}=0)}}{\cal P}\right)
\left(-\frac{{\cal D}\,\nabla_{y_j}^2{\cal P}}{\cal P}\right)\Psi\nn\\&
 &+\,m\int\limits_\moduls d\vec y~\Psi^\dagger\left(-\frac{{\cal D}\,
\nabla_{y_i}{\cal P}}{\cal P}\right)\left(\frac{\nabla_{y_j}
{\cal J}_j^{({\cal D}=0)}}{\cal P}\right)\Psi\nn\\\frac d{dt}
\left\langle\!\!\left\langle\widehat{\bar
y}_i\right\rangle\!\!\right\rangle&=&\left\langle\!\!\left\langle\widehat{\bar
u}_i\right\rangle\!\!\right\rangle
\label{ehren24}\eea
where ${\cal J}_i^{({\cal D}=0)}$ is the undissipative current density
(\ref{current13}). Note that the fundamental renormalization group equations
(\ref{betafns}) receive no corrections due to the dissipation. The existence of
extra dissipation terms in (\ref{ehren24}) in the Ehrenfest relation for the
momentum operator $\widehat{p}_i=-i\hbar_\moduls\,\nabla_{y_i}$, which are
proportional to $\tr\,|\bar U^i|^2$, may now be compared to the generalized
Heisenberg uncertainty relations that were derived in \cite{ms1}. These extra
terms are determined by the total kinetic energy of the D-branes
and their open string excitations, and they show how the recoil of the
D-brane background produces quantum fluctuations of the classical spacetime
dynamics\cite{msnl}. 

Thus, it seems that
in this example, the identification of the world-sheet RG scale
with the target time of the string leads to non-linear
quantum mechanical equations for D-particle. Such equations
have caused some controversy as far as 
their physical meaning and uniqueness are concerned.
One may therefore question the above identification
of the Liouville field with target time. 
However, as we shall discuss below, supersymmetrization
of the world-sheet formalism, as required 
for the target-space {\it stability} of the D-particles,  
eliminates the 
leading ultraviolet world-sheet divergences (\ref{recoilcontrtot})
leading to the diffusion term (\ref{leadingGLambda}). 
We next proceed to discuss this issue. For pedagogical purposes
we also give the definition of the associated 
N=1 Logarithmic superconformal algebras used in the construction
of the super D-brane recoil problem.

\section{Definition and Properties of the $N=1$ Logarithmic Superconformal
  Algebra\label{SLCFTgen}}

We will start by looking at an abstract logarithmic superconformal field theory
to see what some of the general features are. Throughout we
will deal for simplicity with situations in which the two-dimensional
field theory contains only a single Jordan cell of rank~2, but our
considerations easily extend to more general situations. In this
section we shall begin by discussing how to properly incorporate the
Ramond sector of the theory.

\subsection{Operator Product Expansions}

Consider a logarithmic superconformal field theory defined on the complex plane
$\complex$ (or the Riemann sphere $\complex\cup\{\infty\}$) with
coordinate $z$. For the most part we will only write formulas
explicitly for the holomorphic sector of the two-dimensional field
theory. We will also use a superspace notation, with complex
supercoordinates $\scz=(z,\theta)$, where $\theta$ is a complex
Grassmann variable, $\theta^2=0$. The superconformal algebra is
generated by the holomorphic super energy-momentum tensor
\beq
\scT(\scz)=G(z)+\theta\,T(z)
\label{superEMtensor}\eeq
which is a chiral superfield of dimension $\frac32$. Here $T(z)$ is
the bosonic energy-momentum tensor of conformal dimension 2,
while $G(z)$ is the fermionic supercurrent of dimension $\frac32$ with the
boundary conditions
\beq
G(\e^{2\pi\ii}\,z)=\e^{\pi\ii\lambda}\,G(z) \ ,
\label{Gbcs}\eeq
where $\lambda=0$ in the NS sector of the theory (corresponding to periodic
boundary conditions on the fermion fields) and $\lambda=1$ in the R sector
(corresponding to anti-periodic boundary conditions).

The $N=1$ superconformal algebra may then be characterized by the
anomalous operator product expansion
\beq
\scT(\scz_1)\,\scT(\scz_2)=\frac{\hat c}4\,\frac1{(\scz_{12})^3}+
\frac{2\theta_{12}}{(\scz_{12})^2}\,\scT(\scz_2)+\frac12\,\frac1{\scz_{12}}
\,\deriv_{\scz_2}\scT(\scz_2)+\frac{\theta_{12}}{\scz_{12}}\,
\partial_{z_2}\scT(\scz_2)+\dots \ ,
\label{TOPE}\eeq
where in general we introduce the variables
\beq
\scz_{ij}=z_i-z_j-\theta_i\theta_j \ , ~~
\theta_{ij}=\theta_i-\theta_j
\label{zijthetaijdef}\eeq
corresponding to any set of holomorphic superspace coordinates
$\scz_i=(z_i,\theta_i)$. Here
\beq
\deriv_{\scz}=\partial_\theta+\theta\,\partial_z \ , ~~
\deriv_\scz^2=\partial_z
\label{supercovderivdef}\eeq
is the superspace covariant derivative, and $\hat
c=2c/3$ is the superconformal central charge
with $c$ the ordinary Virasoro central charge. An ellipsis will always
denote terms which are regular in the operator product expansion
as $\scz_1\to\scz_2$. By introducing the usual mode expansions
\bea
T(z)&=&\sum_{n=-\infty}^\infty L_n~z^{-n-2} \ , \non
G(z)&=&\sum_{n=-\infty}^\infty\,\frac12\,G_{n+(1-\lambda)/2}
{}~z^{-n-2+\lambda/2}
\label{TGmodes}\eea
with $L_n^\dag=L_{-n}$ and $G_r^\dag=G_{-r}$, the operator product expansion
(\ref{TOPE}) is equivalent to the usual relations of the $N=1$ supersymmetric
extension of the Virasoro algebra,
\bea
[L_m,L_n]&=&(m-n)L_{m+n}+\frac{\hat c}8\,\Bigl(m^3-m\Bigr)\,\delta_{m+n,0}
\ , \non  {~}[L_m,G_r]&=&\left(\frac m2-r\right)G_{m+r} \ ,
\non \{G_r,G_s\}&=&2L_{r+s}+\frac{\hat c}2\,\left(r^2-\frac14\right)\,
\delta_{r+s,0} \ ,
\label{SUSYViralg}\eea
where $m,n\in\zed$, and $r,s\in\zed+\frac12$ for the NS algebra while
$r,s\in\zed$ for the R algebra. In particular, the five operators
$L_0$, $L_{\pm1}$ and $G_{\pm1/2}$ generate the orthosymplectic Lie
algebra of the global superconformal group $OSp(2,1)$.

In the simplest instance, logarithmic superconformal operators of
weight $\Delta_C$ correspond to a pair of superfields
\bea
\scC(\scz)&=&C(z)+\theta\,\chi^{~}_{C}(z) \ , \non
\scD(\scz)&=&D(z)+\theta\,\chi^{~}_{D}(z)
\label{CDsuperfields}\eea
which have operator product expansions with the super
energy-momentum tensor given by~\cite{KAG,MavSz}
\bea
\scT(\scz_1)\,\scC(\scz_2)&=&\frac{\Delta_C\,\theta_{12}}{(\scz_{12})^2}\,
\scC(\scz_2)+\frac12\,\frac1{\scz_{12}}\,\deriv_{\scz_2}\scC(\scz_2)+
\frac{\theta_{12}}{\scz_{12}}\,\partial_{z_2}\scC(\scz_2)+\dots \ , \non
\scT(\scz_1)\,\scD(\scz_2)&=&\frac{\Delta_C\,\theta_{12}}{(\scz_{12})^2}\,
\scD(\scz_2)+\frac{\theta_{12}}{(\scz_{12})^2}\,\scC(\scz_2)+
\frac12\,\frac1{\scz_{12}}\,\deriv_{\scz_2}\scD(\scz_2)+
\frac{\theta_{12}}{\scz_{12}}\,\partial_{z_2}\scD(\scz_2)+\dots \ . \non &&
\label{CDsuperOPE}\eea
Note that $\scC(\scz)$ is a primary superfield of the superconformal
algebra of dimension $\Delta_C$, which is necessarily an
integer~\cite{ckt}. The appropriately normalized superfield
$\scD(\scz)$ is its quasi-primary logarithmic partner. This latter
assumption, i.e. that $[L_n,\scD(z)]=[G_r,\scD(z)]=0$ for $n,r>0$, is
not necessary, but it will simplify some of the arguments which
follow. 
The operators $C(z)$ and $D(z)$ correspond to an ordinary
logarithmic pair and their superpartners $\chi_C^{~}(z)$ and
$\chi_D^{~}(z)$ are generated through the operator products with the
fermionic supercurrent as (in the 
Neveu-Schwarz sector of the theory,
corresponding to the choice of anti-periodic boundary conditions on the
worldsheet spinor fields) 
\bea
G(z)\,C(z)&=&\frac{1/2}{z-w}\,\chi_C^{~}(w)+\dots \ , \non
G(z)\,D(z)&=&\frac{1/2}{z-w}\,\chi_D^{~}(w)+\dots \ .
\label{chiphiOPE}\eea
In particular, in the NS algebra we may write the superpartners as
$\chi_C^{~}(z)=[G_{-1/2},C(z)]$ and $\chi_D^{~}(z)=[G_{-1/2},D(z)]$.

For later use we give here the component form 
the ${\cal N}=1$ supersymmetric completion of the
logarithmic conformal algebra (\ref{TCD}) 
and the associated OPEs. It 
is as follows:  
\bea
T(z)\,G(w)&=&\frac{3/2}{(z-w)^2}\,G(w)+\frac1{z-w}\,\partial_wG(w)+\dots \ ,
\nn\\G(z)\,G(w)&=&\frac{\hat c}{(z-w)^3}+\frac2{z-w}\,T(w)+\dots \ ,
\label{TGOPE}\eea
where $\hat c=2c/3$ is the superconformal central charge. We introduce
fermionic fields $\chi^{~}_C$ and $\chi^{~}_D$ which are the worldsheet
superpartners of the operators $C$ and $D$, respectively. 
The pair $(C,\chi^{~}_C)$ satisfies
the standard algebraic relations of a primary superconformal multiplet of
dimension $\Delta$, while the additional relations for $\chi^{~}_D$ can be
obtained by differentiating those involving $\chi^{~}_C$ with the formal
identification $\chi^{~}_D=\partial\chi^{~}_C/\partial\Delta$.
The ${\cal N}=1$ logarithmic superconformal algebra is thereby characterized by
the operator product expansions (\ref{TCD}), (\ref{chiphiOPE}), and
\bea
T(z)\,\chi^{~}_C(w)&=&\frac{\Delta+1/2}{(z-w)^2}\,\chi^{~}_C(w)+\frac1{z-w}
\,\partial_w\chi^{~}_C(w)+\dots \ , \nn\\T(z)\,\chi^{~}_D(w)&=&
\frac{\Delta+1/2}{(z-w)^2}\,\chi^{~}_D(w)+\frac1{(z-w)^2}\,\chi^{~}_C(w)+
\frac1{z-w}\,\partial_w\chi^{~}_D(w)+\dots \ , \nn\\G(z)\,\chi^{~}_C(w)&=&
\frac\Delta{(z-w)^2}\,C(w)+\frac{1/2}{z-w}\,\partial_wC(w)+\dots \ ,
\nn\\G(z)\,\chi^{~}_D(w)&=&\frac\Delta{(z-w)^2}\,D(w)+
\frac1{(z-w)^2}\,C(w)+\frac{1/2}{z-w}\,\partial_wD(w)+\dots \ .
\nn \\ \label{SUSYOPE}\eea
In addition to the Green's functions (\ref{CD2pt}), the two-point functions
involving the extra fields can also be readily worked out to be
\bea
\Bigl\langle\phi(z)\,\chi^{~}_{\phi'}(w)\Bigr\rangle&=&0 \ , ~~ \phi,\phi'=
C,D \ , \nn\\\Bigl\langle\chi^{~}_C(z)\,\chi^{~}_C(w)\Bigr
\rangle&=&0 \ ,\nn\\\Bigl\langle\chi^{~}_C(z)\,\chi^{~}_D(w)
\Bigr\rangle&=&\frac{2\Delta\xi}{(z-w)^{2\Delta+1}} \ , \nn
\\\Bigl\langle\chi^{~}_D(z)\,
\chi^{~}_D(w)\Bigr\rangle&=&\frac2{(z-w)^{2\Delta+1}}\,\Bigl(-2\Delta\xi
\ln(z-w)+\xi+\Delta d\Bigr) \ .\nn \\ 
\label{SUSY2pt}\eea

Analogous results can be obtained for higher order correlators. 
It is also possible to
generalize these results to the case where
there is more than one Jordan block. Note that, under the assumption that the
logarithmic partner fields are quasi-primary, any such Jordan block implies the
existence of a Jordan cell for the identity operator, which has vanishing
scaling dimension. Thus, if there exists a Jordan block with $\Delta\neq0$,
then there are automatically at least two Jordan blocks for the logarithmic
conformal field theory.

\subsection{Highest-Weight Representations\label{HighestWeight}}

The quantum Hilbert space $\cal H$ of the superconformal field theory
decomposes into two subspaces,
\beq
{\cal H}={\cal H}_{\rm NS}\oplus{\cal H}_{\rm R} \ ,
\label{Hilbertsplit}\eeq
corresponding to the two types of boundary conditions obeyed by the fermionic
fields. They carry the representations of the NS and R algebras, respectively.
In this space, we assume that some of the highest-weight representations of the
$N=1$ superconformal algebra are
indecomposable~\cite{GabKausch1,rohsiepe}. Then a (rank 2)
highest-weight Jordan cell of energy $\Delta_C$ is generated by a pair of
appropriately normalized states $|C\rangle$, $|D\rangle$ obeying the conditions
\bea
L_0|C\rangle&=&\Delta_C|C\rangle \ , \nn \\L_0|D\rangle&=&\Delta_C|D\rangle+
|C\rangle \ , \non L_n|C\rangle&=&L_n|D\rangle~=~0 \ , ~~ n>0 \ , \non
G_r|C\rangle&=&G_r|D\rangle~=~0 \ , ~~ r>0 \ .
\label{highestwtdef}\eea
A highest-weight representation of the logarithmic superconformal algebra is
then generated by applying the raising operators $L_n$, $G_r$, $n,r<0$ to these
vectors giving rise to the descendant states of the theory. Note that
$|C\rangle$ is a highest-weight state of the irreducible sub-representation of
the superconformal algebra contained in the Jordan cell.

\subsubsection*{Neveu-Schwarz Sector}

The NS sector ${\cal H}_{\rm NS}$ of the Hilbert space contains the
normalized, $OSp(2,1)$-invariant vacuum state $|0\rangle$ which is the unique
state of lowest energy $\Delta=0$ in a unitary theory,
\beq
L_0|0\rangle=0 \ .
\label{L0vacuum}\eeq
In this sector, the states defined by (\ref{highestwtdef}) are in a one-to-one
correspondence with the logarithmic operators satisfying the operator product
expansions (\ref{CDsuperOPE}). Namely, under the usual operator-state
correspondence of local quantum field theory, the superfields $\scC(\scz)$
and $\scD(\scz)$ are associated with highest weight states of energy
$\Delta_C$ through
\bea
C(0)|0\rangle&=&|C\rangle^{~}_{\rm NS} \ , \non
\chi^{~}_C(0)|0\rangle&=&G_{-1/2}|C\rangle^{~}_{\rm NS} \ , \non
D(0)|0\rangle&=&|D\rangle^{~}_{\rm NS} \ , \non
\chi^{~}_D(0)|0\rangle&=&G_{-1/2}|D\rangle^{~}_{\rm NS} \ .
\label{NSopstate}\eea
In this way, the NS sector is formally analogous to an ordinary, bosonic
logarithmic conformal field theory. Note that the vacuum state
$|0\rangle$ itself corresponds to the identity operator $I$.

\subsubsection*{Ramond Sector}

Things are quite different in the R sector $\hil_{\rm R}$. Consider a
highest weight state $|\Delta\rangle^{~}_{\rm R}$ of energy $\Delta$,
\beq
L_0|\Delta\rangle^{~}_{\rm R}=\Delta|\Delta\rangle^{~}_{\rm R} \ .
\label{L0h}\eeq
{}From the superconformal algebra (\ref{SUSYViralg}), we see that the operators
$L_0$ and $G_0$ commute in the R sector, so that the supercurrent zero
mode $G_0$ acts on the highest weight states. As a consequence, the state
$G_0|\Delta\rangle^{~}_{\rm R}$ also has energy $\Delta$. Therefore,
the highest weight states of the R sector $\hil_{\rm R}$ come in
orthogonal pairs $|\Delta\rangle^{~}_{\rm R}$,
$G_0|\Delta\rangle^{~}_{\rm R}$ of the same energy. Under the
operator-state correspondence, the Ramond highest weight states are
created from the vacuum $|0\rangle$ by the application of spin fields
$\Sigma^\pm_\Delta(z)$~\cite{FQS} which are ordinary conformal fields
of dimension $\Delta$,
\bea
\Sigma_\Delta^+(0)|0\rangle&=&|\Delta\rangle^{~}_{\rm R} \ , \non
\Sigma_\Delta^-(0)|0\rangle&=&G_0|\Delta\rangle^{~}_{\rm R} \ .
\label{hpmspinfield}\eeq

The operator product expansions of the spin fields with the super
energy-momentum tensor may be computed from (\ref{L0h}) and
(\ref{hpmspinfield}) and are given by
\bea
T(z)\,\Sigma_\Delta^\pm(w)&=&\frac\Delta{(z-w)^2}\,\Sigma_\Delta^\pm(w)+
\frac1{z-w}\,\partial_w\Sigma_\Delta^\pm(w)+\dots \ ,
\label{TSigmaOPE}\nopg G(z)\,\Sigma_\Delta^+(w)&=&\frac12\,
\frac1{(z-w)^{3/2}}\,\Sigma_\Delta^-(w)+\dots \ ,
\label{GSigma+OPE}\nopg G(z)\,\Sigma_\Delta^-(w)&=&\frac12\,
\left(\Delta-\frac{\hat c}{16}\right)\,
\frac1{(z-w)^{3/2}}\,\Sigma_\Delta^+(w)+\dots \ ,
\label{GSigma-OPE}\eea
where we have used the super-Virasoro algebra (\ref{SUSYViralg}) to
write
\beq
G_0^2=L_0-\frac{\hat c}{16} \ .
\label{G02L0}\eeq
The operator product (\ref{TSigmaOPE}) merely states that
$\Sigma_\Delta^\pm(z)$ is a dimension $\Delta$ primary field
of the ordinary, bosonic Virasoro algebra, while (\ref{GSigma+OPE})
and (\ref{GSigma-OPE}) show that the fermionic supercurrent
$G(z)$ is double-valued with respect to the spin fields, since they are
equivalent to the monodromy conditions
\beq
G(\e^{2\pi\ii}\,z)\,\Sigma_\Delta^\pm(w)=-G(z)\,\Sigma_\Delta^\pm(w) \ .
\label{GSigmamonodromy}\eeq
It follows that Ramond
boundary conditions can be regarded as due to a branch cut in the complex plane
connecting the spin fields $\Sigma_\Delta^\pm(z)$ at $z=0$ and
$z=\infty$. The spin fields make the entire superconformal field
theory non-local, and correspond to the irreducible representations of
the Ramond algebra. Note that the ordinary superfields are block
diagonal with respect to the decomposition (\ref{Hilbertsplit}),
i.e. they are operators on $\hil_{\rm NS}\to\hil_{\rm NS}$ and
$\hil_{\rm R}\to\hil_{\rm R}$, while the spin fields
$\Sigma_\Delta^\pm:\hil_{\rm NS}\to\hil_{\rm R}$ are block
off-diagonal.

The spin fields $\Sigma_\Delta^\pm(z)$ do not affect the integer weight fields
$C(z)$ and $D(z)$, while their operator product expansions with the
fermionic partners to the logarithmic operators in the R sector are given by
\bea
\chi^{~}_C(z)\,\Sigma_\Delta^\pm(w)&=&\frac1{\sqrt{z-w}}\,
\widetilde{\Sigma}^\pm_{C,\Delta}(w)+\dots
\ , \non\chi^{~}_D(z)\,\Sigma_\Delta^\pm(w)&=&\frac1{\sqrt{z-w}}\,
\widetilde{\Sigma}^\pm_{D,\Delta}(w)+\dots \ .
\label{chiCDSigma}\eea
The relations (\ref{chiCDSigma}) define two different excited twist fields
$\widetilde{\Sigma}^\pm_{C,\Delta}(z)$ and
$\widetilde{\Sigma}^\pm_{D,\Delta}(z)$ which are conjugate to the spin fields
$\Sigma_\Delta^\pm(z)$. They are also double-valued with respect to
$\chi^{~}_C$ and $\chi^{~}_D$, respectively, and they each act within
the Ramond sector as operators on ${\cal H}_{\rm NS}\to{\cal H}_{\rm
  R}$. The relative non-locality of the operator product
expansions (\ref{chiCDSigma}) yields the global $\zed_2$-twists in the
boundary conditions required of the R sector fermionic fields.

While $\widetilde{\Sigma}_{C,\Delta}^\pm(z)$ are primary fields of
conformal dimension $\Delta_C+\Delta$, the conjugate spin fields
$\widetilde{\Sigma}^\pm_{D,\Delta}(z)$ exhibit
logarithmic mixing behavior. This can be seen explicitly by applying the
operator product expansions to both sides of (\ref{chiCDSigma}) using
(\ref{CDsuperOPE}) and (\ref{TSigmaOPE})--(\ref{GSigma-OPE}) to get
\bea
T(z)\,\widetilde{\Sigma}_{C,\Delta}^\pm(w)&=&\frac{\Delta_C+\Delta}{(z-w)^2}\,
\widetilde{\Sigma}_{C,\Delta}^\pm(w)+
\frac1{z-w}\,\partial_w\widetilde{\Sigma}_{C,\Delta}^\pm(w)+\dots \ ,
\label{TSigmaC}\nopg
T(z)\,\widetilde{\Sigma}_{D,\Delta}^\pm(w)&=&\frac{\Delta_C+\Delta}{(z-w)^2}\,
\widetilde{\Sigma}_{D,\Delta}^\pm(w)+
\frac1{(z-w)^2}\,\widetilde{\Sigma}_{C,\Delta}^\pm(w)+\frac1{z-w}\,\partial_w
\widetilde{\Sigma}_{D,\Delta}^\pm(w)+
\dots \ , \non&&\label{TSigmaD}\nopg
G(z)\,\widetilde{\Sigma}_{C,\Delta}^+(w)&=&\frac12\,\frac1{(z-w)^{3/2}}\,
\widetilde{\Sigma}_{C,\Delta}^-(w)+
\dots \ , \label{GSigmaC+}\nopg
G(z)\,\widetilde{\Sigma}_{C,\Delta}^-(w)&=&\frac12\,\left(\Delta-\frac{\hat
    c}{16}\right)\,\frac1{(z-w)^{3/2}}\,\widetilde{\Sigma}_{C,\Delta}^+(w)+
\dots \ , \label{GSigmaC-}\nopg
G(z)\,\widetilde{\Sigma}_{D,\Delta}^+(w)&=&\frac12\,\frac1{(z-w)^{3/2}}\,
\widetilde{\Sigma}_{D,\Delta}^-(w)+\dots \ , \label{GSigmaD+}\nopg
G(z)\,\widetilde{\Sigma}_{D,\Delta}^-(w)&=&\frac12\,\left(\Delta-\frac{\hat
    c}{16}\right)\,\frac1{(z-w)^{3/2}}\,
\widetilde{\Sigma}_{D,\Delta}^+(w)+\dots \ .
\label{GSigmaD-}\eea
The operator product expansions (\ref{TSigmaC}) and (\ref{TSigmaD})
yield a pair of ordinary, bosonic logarithmic conformal algebras,
while (\ref{GSigmaC+})--(\ref{GSigmaD-}) show that both
$\widetilde{\Sigma}_{C,\Delta}^\pm(z)$ and
$\widetilde{\Sigma}_{D,\Delta}^\pm(z)$ twist the fermionic supercurrent
$G(z)$ in exactly the same way that the original spin fields
$\Sigma_\Delta^\pm(z)$ do. In particular, the set of degenerate spin fields
$\widetilde{\Sigma}_{C,\Delta}^\pm(z)$,
$\widetilde{\Sigma}_{D,\Delta}^\pm(z)$ generate a pair of reducible but
indecomposable representations (\ref{highestwtdef}) of the R algebra,
of the {\it same} shifted weight $\Delta_C+\Delta$. The corresponding
excited highest-weight states $|C,\Delta\rangle^\pm_{\rm R}$,
$|D,\Delta\rangle^\pm_{\rm R}$ of the mutually
orthogonal degenerate Jordan blocks for the action of the Virasoro
operator $L_0$ on ${\cal H}_{\rm R}$ are created from the NS ground
state through the application of the logarithmic spin operators as
\bea
\widetilde{\Sigma}_{C,\Delta}^\pm(0)|0\rangle&=&|C,\Delta
\rangle^\pm_{\rm R} \ , \non\widetilde{\Sigma}_{D,\Delta}^\pm(0)
|0\rangle&=&|D,\Delta\rangle^\pm_{\rm R} \ ,
\label{SigmaCDpmdef}\eea
with
\bea
L_0|C,\Delta\rangle^\pm_{\rm R}&=&(\Delta_C+\Delta)|C,\Delta
\rangle^\pm_{\rm R} \ , \non L_0|D,\Delta\rangle^\pm_{\rm R}&=&
(\Delta_C+\Delta)|D,\Delta\rangle^\pm_{\rm R}+
|C,\Delta\rangle^\pm_{\rm R} \ .
\label{L0CDhpm}\eea

In the following we will be primarily interested in the spin fields associated
with the Ramond ground state $|\Delta\rangle^{~}_{\rm R}$ which is defined
by the condition $G_0|\Delta\rangle_{\rm R}^{~}=0$. This lifts the
degeneracy of the highest weight representation which by (\ref{G02L0})
necessarily has dimension $\Delta=\hat c/16$, corresponding to the
lowest energy in a unitary theory whereby $G_0^2\geq0$. In this case, the
Ramond state $G_0|\Delta\rangle^{~}_{\rm R}$ is a null vector and the R
sector contains a single copy of the logarithmic superconformal
algebra, as in the NS sector. We will return to the issue of
logarithmic null vectors within this context in
section~\ref{nullvectors}. The spin field
$\Sigma_{\hat c/16}^-(z)$ is then an irrelevant operator and may be
set to zero, while the other spin field will be simply denoted by
$\Sigma(z)\equiv\Sigma_{\hat c/16}^+(z)$. The spin field $\Sigma(z)$
corresponds to the unique supersymmetric ground state $|\frac{\hat
  c}{16}\rangle_{\rm R}^{~}$ of the Ramond system, with supersymmetry
generator $G_0$, in the logarithmic superconformal field
theory. Similarly, we may set $\widetilde{\Sigma}_{C,\hat
  c/16}^-(z)=\widetilde{\Sigma}_{D,\hat c/16}^-(z)=0$, and we denote the
remaining excited spin fields simply by
$\widetilde{\Sigma}_C(z)\equiv\widetilde{\Sigma}_{C,\hat
  c/16}^+(z)$ and $\widetilde{\Sigma}_D(z)\equiv\widetilde{\Sigma}^+_{D,\hat
c/16}(z)$.

\subsection{Correlation Functions\label{Correlators}}

Carrying on with an abstract logarithmic superconformal algebra, we
shall now describe the structure of logarithmic correlation functions
in both the NS and R sectors. In particular, we will determine all
two-point correlators involving the various logarithmic operators.

\subsubsection{Ward Identities and Neveu-Schwarz Correlation
  Functions\label{Ward}}

In the NS sector, we define the correlator of any periodic operator $\sf O$ as
its vacuum expectation value
\beq
\langle{\sf O}\rangle^{~}_{\rm NS}=\langle0|{\sf O}|0\rangle \ .
\label{NCcorrelator}\eeq
Such correlators of logarithmic operators, and their descendants, may be
derived as follows. Consider a collection of Jordan blocks in
the superconformal field theory of rank~2, weight $\Delta_{C_i}$, and
spanning logarithmic superfields $\scC_i(\scz)$, $\scD_i(\scz)$. Then, in
the standard way, we may deduce from the operator product expansions
(\ref{CDsuperOPE}) the superconformal Ward identities

\vbox{\bea
&&\Bigl\langle\scT(\scz)\,\scC_n(\scz_n)\cdots\scC_{n+k}(\scz_{n+k})\,
\scD_m(\scw_m)\cdots\scD_{m+l}(\scw_{m+l})\Bigr\rangle_{\rm
 NS}\non&&~~=~\left(\,\sum_{i=n}^{n+k}\left[\frac12\,\frac1{z-z_i-\theta\,
\theta_i}\,\deriv_{\scz_i}+\frac{\theta-\theta_i}{z-z_i-\theta\,\theta_i}
\,\partial_{z_i}+\frac{\Delta_{C_i}\,(\theta-\theta_i)}
{(z-z_i-\theta\,\theta_i)^2}\right]\right.\non&&~~~~~~+\left.
\sum_{i=m}^{m+l}\left[\frac12\,\frac1{z-w_i-\theta\,\zeta_i}
\,\deriv_{\scw_i}+\frac{\theta-\zeta_i}{z-w_i-\theta\,\zeta_i}
\,\partial_{w_i}+\frac{\Delta_{C_i}\,(\theta-\zeta_i)}
{(z-w_i-\theta\,\zeta_i)^2}\right]\right)
\non&&~~~~~~\times\,\Bigl\langle\scC_n(\scz_n)\cdots\scC_{n+k}(\scz_{n+k})\,
\scD_m(\scw_m)\cdots\scD_{m+l}(\scw_{m+l})\Bigr\rangle_{\rm
  NS}\non&&~~~~~~+\,\sum_{i=m}^{m+l}\frac{\theta-\zeta_i}
{(z-w_i-\theta\,\zeta_i)^2}
\,\Bigl\langle\scC_n(\scz_n)\cdots\scC_{n+k}(\scz_{n+k})\Bigr.\non&&~~~~~~
\times\Bigl.\scD_m(\scw_m)\cdots\scD_{i-1}(\scw_{i-1})\,
\scC_i(\scw_i)\,\scD_{i+1}(\scw_{i+1})\cdots
\scD_{m+l}(\scw_{m+l})\Bigr\rangle_{\rm NS},
\label{susyWardids}\eea}
\noindent
where the supercoordinates in (\ref{susyWardids}) are $\scz=(z,\theta)$,
$\scz_i=(z_i,\theta_i)$ and $\scw_i=(w_i,\zeta_i)$. These identities
can be used to derive correlation functions of descendants of the logarithmic
operators in terms of those involving the original superfields
$\scC_i$ and $\scD_i$. Notice, in particular, that the Ward identity
connects amplitudes of the descendants of $\scD_i$ with amplitudes
involving the primary superfields~$\scC_i$.

By expanding the super
energy-momentum tensor into modes using (\ref{TGmodes}) we may equate
the coefficients on both sides of (\ref{susyWardids}) corresponding to
the actions of the $OSp(2,1)$ generators $L_0$, $L_{\pm1}$ and
$G_{\pm1/2}$. By using global superconformal invariance of the vacuum
state $|0\rangle$, we then arrive at a set of superfield differential equations
\bea
0&=&\left(\,\sum_{i=n}^{n+k}\deriv_{\scz_i}+\sum_{i=m}^{m+l}\deriv_{\scw_i}
\right)\Bigl\langle\scC_n(\scz_n)\cdots\scC_{n+k}(\scz_{n+k})\,
\scD_m(\scw_m)\cdots\scD_{m+l}(\scw_{m+l})\Bigr\rangle_{\rm NS} \ , \non&&
{~~}^{~~}_{~~}\non
0&=&\left(\,\sum_{i=n}^{n+k}\Bigl[z_i\,\deriv_{\scz_i}+\theta_i\,
\partial_{\theta_i}+2\Delta_{C_i}\Bigr]+\sum_{i=m}^{m+l}\Bigl[w_i\,
\deriv_{\scw_i}+\zeta_i\,\partial_{\zeta_i}+2\Delta_{C_i}
\Bigr]\right)\non&&\times\,\Bigl\langle\scC_n(\scz_n)\cdots
\scC_{n+k}(\scz_{n+k})\,\scD_m(\scw_m)\cdots\scD_{m+l}(\scw_{m+l})
\Bigr\rangle_{\rm NS}\non&&+\,2\,\sum_{i=m}^{m+l}\Bigl\langle
\scC_n(\scz_n)\cdots\scC_{n+k}(\scz_{n+k})\Bigr.\non&&\times\Bigl.
\scD_m(\scw_m)\cdots\scD_{i-1}(\scw_{i-1})\,
\scC_i(\scw_i)\,\scD_{i+1}(\scw_{i+1})\cdots
\scD_{m+l}(\scw_{m+l})\Bigr\rangle_{\rm NS} \ ,\nonumber
\eea

\vbox{\bea
0&=&\left(\,\sum_{i=n}^{n+k}
\left[z_i^2\,\deriv_{\scz_i}+z_i\,(\theta_i\,\partial_{\theta_i}+2
\Delta_{C_i})\right]+\sum_{i=m}^{m+l}
\left[w_i^2\,\deriv_{\scw_i}+w_i\,(\zeta_i\,\partial_{\zeta_i}+2
\Delta_{C_i})\right]\right)\non&&\times\,\Bigl\langle\scC_n(\scz_n)\cdots
\scC_{n+k}(\scz_{n+k})\,\scD_m(\scw_m)\cdots\scD_{m+l}(\scw_{m+l})
\Bigr\rangle_{\rm NS}\non&&+\,2\,\sum_{i=m}^{m+l}w_i\,\Bigl\langle
\scC_n(\scz_n)\cdots\scC_{n+k}(\scz_{n+k})\Bigr.\non&&\times\Bigl.
\scD_m(\scw_m)\cdots\scD_{i-1}(\scw_{i-1})\,
\scC_i(\scw_i)\,\scD_{i+1}(\scw_{i+1})\cdots
\scD_{m+l}(\scw_{m+l})\Bigr\rangle_{\rm NS} \ .
\label{globalsusyWard}\eea}
\noindent
These equations can be used to
determine the general structure of the logarithmic correlators.

For the two-point correlation functions of the logarithmic superfields
one finds~\cite{KAG}
\bea
\Bigl\langle\scC(\scz_1)\,\scC(\scz_2)\Bigr\rangle^{~}_{\rm NS}&=&0 \
, \label{CC}\nopg
\Bigl\langle\scC(\scz_1)\,\scD(\scz_2)\Bigr\rangle^{~}_{\rm
  NS}&=&\Bigl\langle\scD(\scz_1)\,\scC(\scz_2)\Bigr\rangle^{~}_{\rm
  NS}~=~\frac b
{(\scz_{12})^{2\Delta_C}} \ , \label{CD}\nopg \Bigl\langle\scD(\scz_1)\,
\scD(\scz_2)\Bigr
\rangle^{~}_{\rm NS}&=&\frac1{(\scz_{12})^{2\Delta_C}}\,\Bigl(-2b\ln\scz_{12}+d
\Bigl) \ ,
\label{DD}\eea
where the constant $b$ is fixed by the leading logarithmic divergence of the
conformal blocks of the theory (equivalently by the normalization of
the $D$ operator), and the integration constant $d$ can be changed
by the field redefinitions
$\scD(\scz)\mapsto\scD(\scz)+\lambda\,\scC(\scz)$ which are induced by
the scale transformations $z\mapsto\e^\lambda\,z$. In
particular, the equality of two-point functions in (\ref{CD})
immediately implies that the conformal dimension $\Delta_C$ of the
logarithmic pair is necessarily an integer~\cite{ckt}. For the three-point
functions one gets~\cite{KAG}
\bea
\Bigl\langle\scC(\scz_1)\,\scC(\scz_2)\,\scC(\scz_3)\Bigr
\rangle^{~}_{\rm NS}&=&0 \ , \label{CCC}\nopg
\Bigl\langle\scC(\scz_1)\,\scC(\scz_2)\,\scD(\scz_3)\Bigr
\rangle^{~}_{\rm
  NS}&=&\frac1{(\scz_{12})^{\Delta_C}\,(\scz_{13})^{\Delta_C}
\,(\scz_{23})^{\Delta_C}}\,\Bigl(b_1+\beta_1\,\theta_{123}\Bigr) \ ,
\label{CCD}\nopg\Bigl\langle\scC(\scz_1)\,\scD(\scz_2)\,\scD(\scz_3)\Bigr
\rangle^{~}_{\rm
  NS}&=&\frac1{(\scz_{12})^{\Delta_C}\,(\scz_{13})^{\Delta_C}
\,(\scz_{23})^{\Delta_C}}\,\Bigl(b_2+\beta_2\,\theta_{123}-
2(b_1+\beta_1\,\theta_{123})\ln\scz_{23}\Bigr) \ ,\non&&
\label{CDD}\nopg\Bigl\langle\scD(\scz_1)\,\scD(\scz_2)\,\scD(\scz_3)\Bigr
\rangle^{~}_{\rm
  NS}&=&\frac1{(\scz_{12})^{\Delta_C}\,(\scz_{13})^{\Delta_C}
\,(\scz_{23})^{\Delta_C}}\,\Bigl[b_3+\beta_3\,\theta_{123}\Bigr.\non&&
-\,(b_2+\beta_2\,\theta_{123})\ln\scz_{12}\,\scz_{13}\,\scz_{23}
+(b_1+\beta_1\,\theta_{123})\Bigl(2\ln\scz_{12}\ln\scz_{13}\Bigr.\non&&+\left.
\left.2\ln\scz_{12}\ln\scz_{23}+2\ln\scz_{13}\ln\scz_{23}-\ln^2\scz_{12}-
\ln^2\scz_{13}-\ln^2\scz_{23}\right)\right] \ , \non
\label{DDD}\eea
where $b_i$ and $\beta_i$ are undetermined Grassmann even and odd
constants, respectively, and we have generally defined
\beq
\theta_{ijk}=\frac1{\sqrt{\scz_{ij}\,\scz_{jk}\,\scz_{ki}}}\,\Bigl(
\theta_i\,\scz_{jk}+\theta_j\,\scz_{ki}+\theta_k\,\scz_{ij}+
\theta_i\,\theta_j\,\theta_k\Bigr) \ .
\label{thetaijk}\eeq
The remaining three-point correlation functions can be obtained via
cyclic permutation of the superfields in (\ref{CCD}) and (\ref{CDD}). The
general form of the four-point functions may also be found in~\cite{KAG}.

\subsubsection{Ramond Correlation Functions\label{RCorrelators}}

In the R sector, we define the correlator of any operator $\sf O$ to
be its normalized expectation value in the supersymmetric Ramond ground state,
\beq
\langle{\sf O}\rangle^{~}_{\rm R}=\frac{\langle0|\Sigma(\infty)\,
{\sf O}\,\Sigma(0)|0\rangle}{\langle0|\Sigma(\infty)\,\Sigma(0)|0
\rangle} \ ,
\label{Rcorrelator}\eeq
where we have used the standard asymptotic out-state definition
\beq
\langle0|\Sigma(\infty)=\lim_{z\to\infty}\,\langle0|\Sigma(z)\,z^{\hat
  c/8}
\label{outstatedef}\eeq
and the fact that the spin field $\Sigma(z)$ is a primary field of the
ordinary Virasoro algebra of dimension $\Delta=\hat c/16$. In
particular, the two-point function of the (appropriately normalized)
spin operator is given by
\beq
\langle0|\Sigma(z)\,\Sigma(w)|0\rangle=
\frac1{(z-w)^{\hat c/8}} \ .
\label{Sigma2pt}\eeq
Since $\Sigma(z)$ does not act on the bosonic fields $C(z)$ and
$D(z)$, their R sector correlation functions coincide with those of
the NS sector, i.e. with those of an ordinary logarithmic conformal
field theory. In particular, for the two-point functions we
find~\cite{gurarie,lcftfurther,ckt}
\bea
\Bigl\langle C(z)\,C(w)\Bigr\rangle^{~}_{\rm R}&=&0 \
, \non \Bigl\langle C(z)\,D(w)\Bigr\rangle^{~}_{\rm
  R}&=&\Bigl\langle D(z)\,C(w)\Bigr\rangle^{~}_{\rm R}~=~\frac b
{(z-w)^{2\Delta_C}} \ , \non \Bigl\langle D(z)\,D(w)\Bigr
\rangle^{~}_{\rm R}&=&\frac{d-2b\ln(z-w)}{(z-w)^{2\Delta_C}} \ .
\label{DDR}\eea

For the correlation functions of the fermionic fields, we proceed as
follows. Let us introduce the function
\beq
g_C^{~}(z,w|z_1,z_2)=\frac{\langle0|\Sigma(z_1)\,
\chi_C^{~}(z)\,\chi_C^{~}(w)\,\Sigma(z_2)|0\rangle}{\langle0|
\Sigma(z_1)\,\Sigma(z_2)|0\rangle} \ .
\label{g1zw}\eeq
All fields appearing in (\ref{g1zw}) behave as ordinary
primary fields under the action of the Virasoro algebra. The Green's
function (\ref{g1zw}) can therefore be evaluated using standard
conformal field theoretic methods~\cite{DFMS}. It obeys the asymptotic
conditions

\vbox{\bea
g_C^{~}(z,w|z_1,z_2)&\simeq&0+\dots ~~~~ {\rm as}~~z\to w \ ,
\label{g1ztow}\nopg
&\simeq&\frac{(z_1-z_2)^{\hat c/8}}{\sqrt{z-z_1}}\,
\langle0|\widetilde{\Sigma}_C(z_1)\,\chi_C^{~}(w)\,\Sigma(z_2)|0\rangle
+\dots ~~~~ {\rm as}~~z\to z_1 \ , \non&&\label{g1ztoz1}\nopg
&\simeq&\frac{(z_1-z_2)^{\hat c/8}}{\sqrt{z-z_2}}\,
\langle0|\Sigma(z_1)\,\chi^{~}_C(w)\,\widetilde{\Sigma}_C(z_2)|0\rangle
+\dots ~~~~ {\rm as}~~z\to z_2 \ .\non&&
\label{g1ztoz2}\eea}
\noindent
The first condition (\ref{g1ztow}) arises from the fact that the short
distance behavior of the quantum field theory is independent of the
global boundary conditions, so that in the limit $z\to w$ the function
(\ref{g1zw}) should coincide with the corresponding Neveu-Schwarz
two-point function determined in (\ref{CC}), i.e. $\langle
\chi_C^{~}(z)\,\chi_C^{~}(w)\rangle^{~}_{\rm NS}=0$. The local
monodromy conditions (\ref{g1ztoz1}) and (\ref{g1ztoz2}) follow from
the operator product expansions (\ref{chiCDSigma}). In addition, by
Fermi statistics the Green's function (\ref{g1zw}) must be
antisymmetric under the exchange of its arguments $z$ and $w$,
\beq
g^{~}_C(z,w|z_1,z_2)=-g_C^{~}(w,z|z_1,z_2) \ .
\label{gCantisym}\eeq

By translation invariance, the conditions (\ref{g1ztow}) and
(\ref{gCantisym}) are solved by any
odd analytic function $f$ of $z-w$. Since the correlators appearing in
(\ref{g1ztoz1}) and (\ref{g1ztoz2}) involve only ordinary, primary conformal
fields, global conformal invariance dictates that the function
$f(z-w)$ must multiply a quantity which is a
function only of the $SL(2,\complex)$-invariant
anharmonic ratio $x$ of the four points of
$g^{~}_C(z,w|z_1,z_2)$ given by
\beq
x=\frac{(z-z_1)(w-z_2)}{(z-z_2)(w-z_1)} \ .
\label{crossratio}\eeq
By conformal invariance, the odd analytic function $f(z-w)$
is therefore identically $0$, and hence
\beq
g^{~}_C(z,w|z_1,z_2)=0 \ .
\label{gC0}\eeq

Using this result we can determine a number of correlation
functions. Setting $z_1=\infty$ and $z_2=0$ gives the Ramond correlator
\beq
\Bigl\langle\chi^{~}_C(z)\,\chi^{~}_C(w)\Bigr\rangle^{~}_{\rm R}=0 \ .
\label{chiCCR}\eeq
{}From (\ref{g1ztoz2}) and (\ref{gC0}) we obtain in
addition the vanishing mixed correlator
\beq
\langle0|\Sigma(z_1)\,\chi^{~}_C(z_2)\,\widetilde{\Sigma}_C(z_3)|0\rangle=0 \ .
\label{mixedNScorr0}\eeq
Fusing together the fields $\Sigma(z_1)$ and $\chi^{~}_C(z_2)$ in
(\ref{mixedNScorr0}) using (\ref{chiCDSigma}) then gives the conjugate
spin-spin correlator
\beq
\langle0|\widetilde{\Sigma}_C(z)\,\widetilde{\Sigma}_C(w)|0\rangle=0 \ .
\label{PiPi}\eeq
The vanishing of the $\widetilde{\Sigma}_C\widetilde{\Sigma}_C$
correlation function is consistent with
the fact that the excited spin field $\widetilde{\Sigma}_C(z)$ obeys
the logarithmic conformal algebra
(\ref{TSigmaC},\ref{TSigmaD})~\cite{gurarie,lcftfurther,ckt}.

Next, let us consider the function
\beq
g_D^{~}(z,w|z_1,z_2)=\frac{\langle0|\Sigma(z_1)\,
\chi_C^{~}(z)\,\chi_D^{~}(w)\,\Sigma(z_2)|0\rangle}{\langle0|
\Sigma(z_1)\,\Sigma(z_2)|0\rangle} \ .
\label{gDzw}\eeq
The action of the Virasoro algebra in (\ref{gDzw}) does not
produce any additional terms from the logarithmic mixing of the
fermionic field $\chi^{~}_D(w)$, because of the vanishing property
(\ref{gC0}). Therefore, this function can also be evaluated as if the
theory were an ordinary conformal field theory~\cite{DFMS}. Using
(\ref{chiCDSigma}) and (\ref{CD}) the asymptotic conditions
(\ref{g1ztow})--(\ref{g1ztoz2}) are now replaced with
\bea
g_D^{~}(z,w|z_1,z_2)&\simeq&\frac{2\Delta_C\,b}{(z-w)^{2\Delta_C+1}}+\dots
{}~~~~ {\rm as}~~z\to w \ , \label{gDztow}\nopg
&\simeq&\frac{(z_1-z_2)^{\hat c/8}}{\sqrt{z-z_1}}\,
\langle0|\widetilde{\Sigma}_C(z_1)\,\chi_D^{~}(w)\,\Sigma(z_2)|0\rangle
+\dots ~~~~ {\rm as}~~z\to z_1 \ , \non&&\label{gDztoz1}\nopg
&\simeq&\frac{(z_1-z_2)^{\hat c/8}}{\sqrt{z-z_2}}\,
\langle0|\Sigma(z_1)\,\chi^{~}_D(w)\,\widetilde{\Sigma}_C(z_2)|0\rangle
+\dots ~~~~ {\rm as}~~z\to z_2 \ , \non&&\label{gDztoz2}\nopg
&\simeq&\frac{(z_1-z_2)^{\hat c/8}}{\sqrt{w-z_1}}\,
\langle0|\widetilde{\Sigma}_D(z_1)\,\chi_C^{~}(z)\,\Sigma(z_2)|0\rangle
+\dots ~~~~ {\rm as}~~w\to z_1 \ , \non&&\label{gDwtoz1}\nopg
&\simeq&\frac{(z_1-z_2)^{\hat c/8}}{\sqrt{w-z_2}}\,
\langle0|\Sigma(z_1)\,\chi^{~}_C(z)\,\widetilde{\Sigma}_D(z_2)|0\rangle
+\dots ~~~~ {\rm as}~~w\to z_2 \ .\non&&
\label{gDwtoz2}\eea

Again, from (\ref{mixedNScorr0}) it follows that the correlators in
(\ref{gDztoz1})--(\ref{gDwtoz2}) involving a single logarithmic operator
can be treated as an ordinary conformal correlator for primary
fields. In particular, we can treat (\ref{gDzw}) as a correlator for
two identical conformal fermion fields of dimension
$\Delta_C+\frac12$ and require it to be antisymmetric under exchange
of $z$ and $w$, as in (\ref{gCantisym}). This property follows from
the fact that the local NS correlator (\ref{gDztow}) is antisymmetric
in $z$ and $w$ and this feature should extend globally in the quantum
field theory. Again, by $SL(2,\complex)$-invariance the quantity
$(z-w)^{-2\Delta_C-1}\,g^{~}_D(z,w|z_1,z_2)$ is a function only of the
anharmonic ratio (\ref{crossratio}). The precise dependence on $x$ is
uniquely determined by the boundary conditions
(\ref{gDztow})--(\ref{gDwtoz2}) and the antisymmetry of $g^{~}_D$, and
we find\footnote{\baselineskip=12pt To show explicitly that
  (\ref{gDfinal}) is the unique function of $z$ and $w$ with the
  desired properties, we write it as
$$
g^{~}_D(z,w|z_1,z_2)=\frac1{\sqrt{(z-z_1)(z-z_2)(w-z_1)(w-z_2)}}\,
\frac{\Delta_C\,b}{(z-w)^{2\Delta_C+1}}\,\Bigl((z-z_1)(w-z_2)+
(z-z_2)(w-z_1)\Bigr) \ .
$$
The first factor here gives the correct behavior for $g^{~}_D$ as
$z,w\to z_1,z_2$, while the second factor is the required pole at $z=w$ of
order $2\Delta_C+1$. The third factor is then chosen so that the
residue of the pole is $2\Delta_C\,b$ and such that it cancels the
lower order poles arising from the first factor in the limit $z\to w$,
and by further requiring that the overall combination be antisymmetric in $z$
and $w$.}
\begin{eqnarray} \nonumber 
g^{~}_D(z,w|z_1,z_2)=\frac{\Delta_C\,b}{(z-w)^{2\Delta_C+1}}\,
\left(\,\sqrt{\frac{(z-z_1)(w-z_2)}{(z-z_2)(w-z_1)}}+
\sqrt{\frac{(z-z_2)(w-z_1)}{(z-z_1)(w-z_2)}}~\right) \ . \\
\label{gDfinal}\end{eqnarray}

By taking various limits of (\ref{gDfinal}) we can generate another set
of correlation functions for Ramond sector operators. In the
simultaneous limit $z_1\to\infty$ and $z_2\to0$, the function
(\ref{gDfinal}) yields the Ramond two-point correlators
\beq
\Bigl\langle\chi^{~}_C(z)\,\chi^{~}_D(w)\Bigr\rangle^{~}_{\rm R}=
-\,\Bigl\langle\chi^{~}_D(z)\,\chi^{~}_C(w)\Bigr\rangle^{~}_{\rm R}=
\frac{\Delta_C\,b}{(z-w)^{2\Delta_C+1}}\,\left(\,\sqrt{\frac zw}+
\sqrt{\frac wz}~\right) \ .
\label{chiCDR}\eeq
Note that the term in parentheses has branch cuts at $z=0,\infty$ and
$w=0,\infty$, yielding the antiperiodic boundary conditions on the
spinor fields as they circle around the origin in the complex plane
and across the cut
connecting the spin operators $\Sigma(0)$ and $\Sigma(\infty)$ in
(\ref{Rcorrelator}). Taking the limits $z\to z_1,z_2$ and $w\to
z_1,z_2$ in (\ref{gDfinal}) and comparing with
(\ref{gDztoz1})--(\ref{gDwtoz2}) yields the correlation functions
\bea
\langle0|\widetilde{\Sigma}_C(z_1)\,\chi^{~}_D(z_2)\,\Sigma(z_3)|0\rangle
&=&\frac{\ii\Delta_C\,b}{(z_1-z_2)^{2\Delta_C+1/2}\,(z_1-z_3)^{\hat
    c/8-1/2}\,\sqrt{z_2-z_3}}
\non&=&-\,\langle0|\widetilde{\Sigma}_D(z_1)\,
\chi^{~}_C(z_2)\,\Sigma(z_3)|0\rangle \ .
\label{SigmaCchiDS}\eeq
Fusing $\Sigma(z_3)$ with $\chi^{~}_D(z_2)$ and $\chi^{~}_C(z_2)$ in
(\ref{SigmaCchiDS}) using (\ref{chiCDSigma}) then yields the spin-spin
correlators
\beq
\langle0|\widetilde{\Sigma}_C(z)\,\widetilde{\Sigma}_D(w)|0
\rangle=-\,\langle0|\widetilde{\Sigma}_D(z)\,\widetilde{\Sigma}_C(w)|0
\rangle=\frac{\ii\Delta_C\,b}{(z-w)^{2\Delta_C+\hat c/8}} \ .
\label{spinspinCD}\eeq
Note that the logarithmic pair
$\widetilde{\Sigma}_C,\widetilde{\Sigma}_D$ does not have the
canonical two-point functions of a logarithmic conformal field theory
(see~(\ref{DDR})). This is because the excited spin fields of the
theory are not bosonic fields, but are rather given by non-local
operators which interpolate between different sectors of the quantum
Hilbert space and which satisfy, in addition to the logarithmic
algebra, a supersymmetry algebra. In fact, their correlators are
almost identical in form to the correlation functions of the
logarithmic superpartners~$\chi^{~}_C,\chi^{~}_D$~\cite{MavSz}.

Finally, we need to compute the $DD$ type correlators. The above
techniques do not directly apply because Green's functions with two or
more logarithmic operator insertions will not transform covariantly
under the action of the Virasoro algebra. However, we may obtain the
$DD$ type correlators from the mixed $CD$ type ones above by the
following trick~\cite{KAG,MavSz,gezel}. We regard $\Delta_C$ as a continuous
weight and note that the logarithmic superconformal algebra can be
simply obtained by writing down the standard conformal operator
product expansions for the $C$ type operators, and then
differentiating them with respect to $\Delta_C$ to obtain the $D$ type
ones with the formal identifications $D=\partial C/\partial\Delta_C$,
$\chi^{~}_D=\partial\chi^{~}_C/\partial\Delta_C$ and
$\widetilde{\Sigma}_D=\partial\widetilde{\Sigma}_C/\partial\Delta_C$.
Since the basic spin fields $\Sigma(z)$ do not depend on the conformal
dimension $\Delta_C$, we can differentiate the correlation functions
(\ref{chiCDR})--(\ref{spinspinCD}) to get the
desired Green's functions. In doing so we regard the parameter $b$ as
an analytic function of the weight $\Delta_C$ and define $d=\partial
b/\partial\Delta_C$. In this way we arrive at the correlators
\bea
\Bigl\langle\chi^{~}_D(z)\,\chi^{~}_D(w)\Bigr\rangle^{~}_{\rm R}&=&
\frac{b+\Delta_C\Bigl(d-2b\ln(z-w)\Bigr)}{(z-w)^{2\Delta_C+1}}\,
\left(\,\sqrt{\frac zw}+\sqrt{\frac wz}~\right) \ , \non
\langle0|\widetilde{\Sigma}_D(z_1)\,\chi^{~}_D(z_2)\,\Sigma(z_3)|0\rangle
&=&-\frac{b+\Delta_C\Bigl(d-2b\ln(z_1-z_2)\Bigr)}
{\ii(z_1-z_2)^{2\Delta_C+1/2}\,(z_1-z_3)^{\hat
    c/8-1/2}\,\sqrt{z_2-z_3}} \ , \non
\langle0|\widetilde{\Sigma}_D(z)\,\widetilde{\Sigma}_D(w)|0
\rangle&=&-\frac{b+\Delta_C\Bigl(d-2b\ln(z-w)\Bigr)}
{\ii(z-w)^{2\Delta_C+\hat c/8}} \ .
\label{DDR3corrs}\eea

In a completely analogous way, we may easily determine the vanishing
two-point correlation functions
\beq
\Bigl\langle\phi(z)\,\chi^{~}_{\phi'}(w)\Bigr\rangle^{~}_{\rm R}~=~0~=~
\langle0|\widetilde{\Sigma}_{\phi'}(z_1)\,\phi(z_2)\,\Sigma(z_3)
|0\rangle \ ,
\label{othervanish}\eeq
where $\phi$ and $\phi'$ label either of the two fields $C$ or $D$.
The present technique unfortunately does not directly determine higher order
correlation functions of the fields. As they will not be required in
what follows, we will not pursue this issue in this paper.

\subsection{Null Vectors, Hidden Symmetries and Spin Models\label{nullvectors}}

It has been suggested~\cite{MavSz} that, in the limit $\Delta_C=0$,
the fermionic field $\chi^{~}_C(z)$ in (\ref{CDsuperfields}) may be a
null field, since its two-point correlation functions with all other
logarithmic fields vanish for zero conformal dimension. Furthermore,
the logarithmic scaling violations in the fermionic two-point
functions involving the field $\chi^{~}_D(z)$ disappear in this
limit. While this latter property is certainly true for all Green's
functions of the conformal field theory, a quick examination of the
three-point correlators (\ref{CCC}) and (\ref{CCD}) shows that
$\chi^{~}_C(z)$ is {\it not} a null field if $\beta_1\neq0$. The
situation is completely analogous to what happens generically to its
superpartner $C(z)$. Since the primary field $C(z)$ creates a
zero-norm state, and since $\Delta_C\in\zed$, there is a new hidden
continuous symmetry in the theory~\cite{ckt} generated by the conserved
holomorphic current $C(z)$, which is a symmetric tensor of rank
$\Delta_C$. For $\Delta_C=0$, the extra couplings of the $\chi^{~}_C$
field for $\beta_1\neq0$ show that it corresponds to a non-trivial,
dynamical fermionic symmetry of the logarithmic superconformal field
theory. In fact, in the R~sector the structure of these continuous
symmetries is even richer, given that the excited spin field
$\widetilde{\Sigma}_C(z)$ also creates a zero-norm state in the
logarithmic superconformal field theory, and that it has vanishing
two-point functions for $\Delta_C=0$. In $c\neq0$ theories where the
bosonic energy-momentum tensor $T(z)$ has a logarithmic partner, the
identity operator $I$ generates a Jordan cell with
$\Delta_C=0$~\cite{KogNich1} and the zero-norm state is the vacuum,
$\langle0|0\rangle=0$. In this case, of course, the fermion field
$\chi^{~}_I(z)=0$ is trivially a null field, and its
partner $\chi^{~}_D(z)$ is an ordinary, non-logarithmic primary
field of the Virasoro algebra of conformal dimension
$\frac12$. Similarly, in this case $\widetilde{\Sigma}_C(z)=0$, while
$\widetilde{\Sigma}_D(z)$ is an ordinary, non-logarithmic twist field
of weight $\hat c/16$.

In the Ramond sector, there are natural ways to generate null
states for any $\Delta_C$. One way is to build the representation of
the Ramond algebra from the supersymmetric ground state $|\frac{\hat
  c}{16}\rangle^{~}_{\rm R}$ as described at the end of
section~\ref{HighestWeight}. Another way is to introduce the fermion
parity operator $\Gamma=(-1)^F$, where $F$ is the fermion number operator
of the superconformal field theory. The operator $\Gamma$ commutes
with integer spin fields and anticommutes with half-integer spin
fields. It defines an inner automorphism $\pi^{~}_\Gamma:{\cal
  C}\to{\cal C}$ of the maximally extended chiral symmetry algebra
$\cal C$ of the superconformal field theory, such that there is an
exact sequence of vector spaces
\beq
0~\longrightarrow~{\cal C}^+~\longrightarrow~{\cal C}~\longrightarrow~
{\cal C}^-~\longrightarrow~0 \ , ~~
\pi^{~}_\Gamma({\cal C}^\pm)=\pm\,{\cal C}^\pm \ .
\label{calCGamma}\eeq
Under the operator-state correspondence, this determines a fermion
parity grading of the Hilbert space of states as
\beq
{\cal H}={\cal H}^+\oplus{\cal H}^- \ , ~~ \Gamma\,{\cal H}^\pm=\pm\,
{\cal H}^\pm \ .
\label{calHGammagrading}\eeq
Since $G_0$ reverses chirality, the paired Ramond ground states have
opposite chirality,
\beq
\Gamma\,\Sigma_\Delta^\pm(0)|0\rangle=\pm\,\Sigma_\Delta^\pm(0)|0\rangle \ .
\label{oppchirality}\eeq
The opposite chirality spin fields $\Sigma_\Delta^\pm(z)$ are non-local
with respect to each other
(c.f.~(\ref{GSigma+OPE}) and (\ref{GSigma-OPE})). In a unitary theory,
whereby $G_0^2\geq0$, all $\Delta=\hat c/16$ states are chirally asymmetric
highest-weight states, since the state $G_0|\Delta\rangle^{~}_{\rm R}$ is
then a null vector in the Hilbert space. On the other hand, the
orthogonal projection $\frac12\,(1+\Gamma):{\cal H}\to{\cal H}^+$
onto states of even fermion parity $\Gamma=1$ eliminates
the spin field $\Sigma_\Delta^-(z)$ and gives a local field theory which is
customarily referred to as a ``spin model''~\cite{FQS}. The fields of
the spin model live in the local chiral algebra ${\cal C}^+$. This
projection eliminates $G(z)$ and the other half-integer weight
fields. When combined with the projection onto $G_0=0$ it gives the
``GSO projection'' which will be important in the D-brane applications
of the next section.

The main significance of the chiral subalgebra restriction
$\frac12\,(1+\pi^{~}_\Gamma):{\cal C}\to{\cal C}^+$
is that the fermionic fields of the superconformal field theory can be
reconstructed from the $\Gamma=1$ spin fields $\Sigma(z)$, at least in
the examples that we consider in this paper. In an analogous way, the
logarithmic superpartners $\chi^{~}_C(z)$ and $\chi^{~}_D(z)$ can be
reconstructed from the $\Gamma=1$ excited spin fields
$\widetilde{\Sigma}_C(z)$ and $\widetilde{\Sigma}_D(z)$. By
supersymmetry, this yields the bosonic partners $C(z)$ and $D(z)$, and
so in this way the spin model determines the entire logarithmic
superconformal field theory. In fact, the spin field $\Sigma(z)$ can
be uniquely constructed from the underlying chiral current algebra
generated by currents which are formed by the primary fermionic fields
of the theory~\cite{FMS1}. The fermionic current algebra will thereby
completely determine the entire logarithmic superconformal field theory.

\section{The Recoil Problem in Superstring Theory\label{RecoilProb}}

In the remainder of this paper we will consider a concrete model
to illustrate the above formalism explicitly. This example
will also serve to describe some of the basic constructions of
logarithmic spin operators and will illustrate the applicability of
the superconformal logarithmic formalism. In this section we will
discuss the logarithmic superconformal field theory that describes the
recoil of a D-particle in string theory~\cite{kmw,ms1,MavSz}. This is
the simplest example which serves to illustrate the formalism, but 
also captures the essential
features of the general theory of the previous section in a
very simple setting. Moreover, for our purposes here, we 
shall use it as a concrete case of demonstration 
of the consistency within the 
superstring formalism of the identification of time 
with a world-sheet 
renormalization group (Liouville) scale~\cite{kogan,emn}.

\subsection{Supersymmetric Impulse Operators}

We will now derive the ${\cal N}=1$ supersymmetric completion of the impulse
operator (\ref{vertexD},\ref{Yirecoil}). For this, we introduce $2\times2$
Dirac matrices $\rho^\alpha$, $\alpha=1,2$, and real two-component Majorana
fermion fields $\psi^\mu$ which are the worldsheet superpartners of the string
embedding fields $x^\mu$. A convenient basis for the worldsheet spinors is
given by
\beq
\rho^1=\begin{pmatrix} 0&-i\cr i&0\cr \end{pmatrix} \ , ~~ \rho^2=
\begin{pmatrix} 0& i\cr i&0\cr \end{pmatrix}\ ,
\label{rhobasis}\eeq
in which the fermion fields decompose as
\beq
\psi^\mu=\begin{pmatrix} \psi^\mu_-\cr\psi^\mu_+\cr \end{pmatrix} \ .
\label{WSspinordecomp}\eeq
The fields (\ref{WSspinordecomp}) obey the boundary conditions
$\psi_+^\mu|_{\partial\Sigma}=\pm\,\psi_-^\mu|_{\partial\Sigma}$, where the
sign depends on whether they belong to the Ramond or Neveu-Schwarz
sector of the worldsheet theory. The global worldsheet supersymmetry
is determined by the supercharge $\cal Q$ which generates the infinitesimal
${\cal N}=1$ supersymmetry transformations
\bea
\Bigl[{\cal Q}\,,\,x^\mu\Bigr]&=&\psi^\mu \ , \non\Bigl\{{\cal Q}\,,\,
\psi^\mu\Bigr\}&=&-i\,\rho^\alpha\,\partial_\alpha x^\mu \ .
\label{WSsusytransfs}\eea

The fermionic fields $\psi_+(z)$ have conformal dimension $\frac12$, and from
(\ref{WSsusytransfs}) it follows that the superpartner of the tachyon vertex
operator $\e^{i\omega x^0}$ is $\sqrt{\alpha'}\,\omega\,\psi_+^0\,\e^{i\omega
x^0}$, so that in the Neveu-Schwarz sector we may write
\beq
G(z)~\e^{i\omega x^0(w)}=\frac{\sqrt{\alpha'}\,\omega/2}{z-w}\,
\psi_+^0(w)~\e^{i\omega x^0(w)}+\dots \ .
\label{tachyonsusy}\eeq
In what follows it will be important to note the factor of
$\sqrt{\alpha'}\,\omega$ that appears in the supersymmetry transformation
(\ref{tachyonsusy}). Because of it, and the fact that the tachyon vertex
operator has conformal dimension $\alpha'\omega^2/2$, the inverse
transformation is given by
\beq
G(z)\,\psi_+^0(w)~\e^{i\omega x^0(w)}=\frac{\sqrt{\alpha'}\,\omega/2}{(z-w)^2}~
\e^{i\omega x^0(w)}+\frac{i/2\sqrt{\alpha'}}{z-w}\,
\Bigl(\partial_wx^0(w)\Bigl)~\e^{i\omega x^0(w)}+\dots \ .
\label{tachyonsusyinv}\eeq

To compute the superpartners of the logarithmic pair (\ref{CDops}), we use
(\ref{ThetaFourier}) and (\ref{tachyonsusy}) to write
\beq
G(z)\,C_\epsilon(w)=\frac{\epsilon\,(\alpha')^{3/2}/4\pi i}{z-w}\,
\psi_+^0(w)\,\int\limits_{-\infty}^\infty\frac{d\omega}{\omega-i\epsilon}~
\Bigl[(\omega-i\epsilon)+i\epsilon\Bigr]~\e^{i\omega x^0(w)}+\dots \ .
\label{GCepsilon}\eeq
In the first term of the integrand in (\ref{GCepsilon}) there is no pole and so
after contour integration it vanishes. Formally it is a delta functional
$\delta(x^0(w))$ which we neglect since we are interested here in only the
asymptotic time-dependence of string solitons. Then, only the second term
contributes, and comparing with (\ref{chiphiOPE}) we find
\beq
\chi^{~}_{C_\epsilon}(x^0,\psi^0)=i\,\epsilon\,C_\epsilon(x^0)\,\psi_+^0 \ .
\label{chiCepsilon}\eeq
Similarly, we have
\beq
G(z)\,D_\epsilon(w)=-\frac{\sqrt{\alpha'}/4\pi}{z-w}\,\psi_+^0(w)\,
\int\limits_{-\infty}^\infty\frac{d\omega}{(\omega-i\epsilon)^2}~
\Bigl[(\omega-i\epsilon)+i\epsilon\Bigr]~\e^{i\omega x^0(w)}+\dots \ ,
\label{GDepsilon}\eeq
which using (\ref{chiphiOPE}) gives
\beq
\chi^{~}_{D_\epsilon}(x^0,\psi^0)=i\,\left(\epsilon\,D_\epsilon(x^0)-\frac1
{\epsilon\,\alpha'}\,C_\epsilon(x^0)\right)\,\psi_+^0 \ .
\label{chiDepsilon}\eeq
The operators (\ref{chiCepsilon}) and (\ref{chiDepsilon}) have conformal
dimension $\Delta_\epsilon+\frac12$.

It is straightforward to now check that the remaining relations of the ${\cal
N}=1$ logarithmic superconformal algebra are satisfied. By using (\ref{TCD}),
(\ref{Deltavarep}), (\ref{chiCepsilon}) and (\ref{chiDepsilon}), it is easy to
verify the first two operator product expansions of (\ref{SUSYOPE}) in this
case. For the operator products with the fermionic supercurrent, we use in
addition the Fourier integral (\ref{ThetaFourier}) along with
(\ref{tachyonsusyinv}) to get
\bea
G(z)\,\chi^{~}_{C_\epsilon}(w)&=&-\frac{\epsilon^2(\alpha')^{3/2}/4\pi i}
{(z-w)^2}\,\int\limits_{-\infty}^\infty\frac{d\omega}{\omega-i\epsilon}~
\Bigl[(\omega-i\epsilon)+i\epsilon\Bigr]\e^{i\omega x^0(w)}\nn\\&&
+\,\frac1{z-w}\,\partial_w\chi^{~}_{C_\epsilon}(w)+\dots\nn\\
&=&-\frac{\sqrt{\alpha'}\,\epsilon^2/2}{(z-w)^2}\,C_\epsilon(w)+\frac1{z-w}\,
\partial_w\chi^{~}_{C_\epsilon}(w)+\dots \ , \\&&~~~~~\nn\\
G(z)\,\chi^{~}_{D_\epsilon}(w)&=&
-\frac{\sqrt{\alpha'}/4\pi}{(z-w)^2}\,\int\limits_{-\infty}^\infty
\frac{d\omega}{\omega-i\epsilon}\,\left[\Bigl((\omega-i\epsilon)+i\epsilon
\Bigr)+\frac{i\epsilon}{\omega-i\epsilon}\,\Bigl((\omega-i\epsilon)+i
\epsilon\Bigr)\right]\nn\\&&\times\,\e^{i\omega x^0(w)}+\frac1{z-w}\,
\partial_w\chi^{~}_{D_\epsilon}(w)+\dots\nn\\&=&\frac{1/\sqrt{\alpha'}}
{(z-w)^2}\,\left(C_\epsilon(w)-\frac{\alpha'\epsilon^2}2\,D_\epsilon(w)\right)
+\frac1{z-w}\,\partial_w\chi^{~}_{D_\epsilon}(w)+\dots \ ,
\label{Gchiepsilon}\eeq
which also agree with (\ref{SUSYOPE}) in this case.

For the two-point correlation functions (\ref{SUSY2pt}), we use the fermionic
Green's function in the upper half-plane,
\beq
\Bigl\langle\psi_+^0(z)\,\psi_+^0(w)\Bigr\rangle=\frac1{z-w} \ ,
\label{fermGF}\eeq
and the fact that bosonic and fermionic field correlators factorize from each
other in the free superconformal $\sigma$-model on $\Sigma$. The first set of
relations in (\ref{SUSY2pt}) are then satisfied in this case because
$\langle\psi_+^0(z)\rangle=0$. The second relation holds to order $\epsilon^4$
since $\Delta_\epsilon\propto\epsilon^2$ and $\langle
C_\epsilon(z)C_\epsilon(w)\rangle=0$ to order $\epsilon^2$. For the remaining
correlators, we use (\ref{CD2pt}), (\ref{chiCepsilon}), (\ref{chiDepsilon}),
(\ref{fermGF}) and factorization to compute
\bea
\Bigl\langle\chi^{~}_{C_\epsilon}(z)\,\chi^{~}_{D_\epsilon}(w)\Bigr\rangle&=&
-\frac{\epsilon^2\xi}{(z-w)^{2\Delta_\epsilon+1}} \ , \nn\\
\Bigl\langle\chi^{~}_{D_\epsilon}(z)\,\chi^{~}_{D_\epsilon}(w)\Bigr\rangle&=&
\frac1{(z-w)^{2\Delta_\epsilon+1}}\,\left[\frac{2\xi}{\alpha'}-
\epsilon^2\Bigl(-2\xi\ln(z-w)+d_\epsilon\Bigr)\right] \ ,
\label{chiepsilon2pt}\eea
which upon using (\ref{Deltavarep}) are also seen to agree with
(\ref{SUSY2pt}). Therefore, the supersymmetric extensions (\ref{chiCepsilon})
and (\ref{chiDepsilon}) of the impulse operators (\ref{CDops}) give precisely
the right combinations of operators that generate the full algebraic structure
of a logarithmic superconformal field theory. This yields a non-trivial
realization of the supersymmetric completion of the previous section, and
illustrates the overall consistency of the impulse operators describing the
dynamics of D-branes in closed string scattering states.

To recapitulate: we considered 
above the superconformal field theory defined by the classical worldsheet
action
\begin{eqnarray} \nonumber 
S_{\rm D0}=\frac1{2\pi}\,\int\dd z~\dd\overline{z}~\dd\theta~\dd
\overline{\theta}~~\overline{{\cal D}_\scz}\,\scx^\mu\,\deriv_\scz
\scx_\mu-\frac1\pi\,\oint\dd\tau~\dd\vartheta~\Bigl(y_i\,
\scC_\epsilon+u_i\,\scD_\epsilon\Bigr)\,\deriv^{~}_{\!\perp}\scx^i \ ,\\
\label{SD0}\end{eqnarray} 
where $\scx^\mu(\scz,\overline{\scz}\,)=\scx^\mu(\scz)+
\scx^\mu(\overline{\scz}\,)$ with $\scx^\mu(\scz)$ the
chiral scalar superfield
\beq
\scx^\mu(\scz)=x^\mu(z)+\theta\,\psi^\mu(z) \ ,
\label{scxsuperfield}\eeq
whose Neveu-Schwarz two-point functions are given by
\beq
\Bigl\langle\scx^\mu(\scz_1)\,\scx^\nu(\scz_2)\Bigr\rangle^{~}_{\rm
  NS}=-\delta^{\mu\nu}\,\ln\scz_{12} \ .
\label{scx2ptfn}\eeq
Here $x^\mu$, $\mu=1,\dots,10$ are maps from the upper complex
half-plane $\complex_+$ into
ten dimensional {\it Euclidean} space $\real^{10}$, and $\psi^\mu$ are
their spin $\frac12$ fermionic superpartners that transform in the vector
representation of $SO(10)$ and each of which is a Majorana-Weyl spinor
in two-dimensions. We will identify the coordinate $x^{10}$ as the
Euclidean time (obtained from our previously 
described $x^0$ by analytic continuation), 
while $x^i$, $i=1,\dots,9$ lie along the spatial
directions in the target space of the open strings. As in the previous
section, we concentrate on the chiral sector of the worldsheet field
theory with superfields (\ref{scxsuperfield}). The chiral super
energy-momentum tensor is given by
\beq
\scT(\scz)=-\frac12\,\deriv_\scz\scx^\mu(\scz)\,\partial_z
\scx_\mu(\scz) \ .
\label{chiralscT}\eeq

The reasons for working with Euclidean spacetime signature are
technical. First of all, it is easier to deal with spinor
representations of the Euclidean group $SO(10)$ than with those of the
Lorentz group $SO(9,1)$. In the former case all of the $\psi^\mu$ are
treated on equal footing and one is free from the possible
complications arising from the time-like nature of $x^0$, which would
otherwise imply a special role for its superpartner
$\psi^0$~\cite{gsw}. Secondly, for the recoil problem, Euclidean target
spaces are necessary to ensure convergence of worldsheet correlation
functions among the logarithmic operators~\cite{kmw}. For calculational
definiteness and convenience of the worldsheet path integrals, we
shall therefore adopt a Euclidean signature convention in the
following.

The second term in the action (\ref{SD0}) is a marginal deformation of the free
$\hat c=10$ superconformal field theory by the vertex operator describing the
recoil, within an impulse approximation, of a non-relativistic D-brane
in target space due to its interaction with closed string scattering
states~\cite{km},\cite{ms1}. It is the appropriate operator to use
when regarding the branes as string solitons. The coordinate $\tau$
parametrizes the boundary of the upper half-plane, and $\vartheta$ is
a real Grassmann coordinate. The fields in this part of the action are
understood to be restricted to the worldsheet boundary. The coupling
constants $y_i$ and $u_i$ are interpreted as the initial position and
constant velocity of the D-particle, respectively, and the subscript
$\perp$ denotes differentiation in the direction normal to the boundary of
$\complex_+$. The recoil operators are given by chiral superfields
$\scC_\epsilon(\scz)$ and $\scD_\epsilon(\scz)$ whose components are defined
in terms of superpositions over tachyon vertex operators
$\e^{\ii qx^{10}(z)}$ in the time direction as~\cite{MavSz}
\bea
C_\epsilon(z)&=&\frac\epsilon{4\pi\ii}\,\int
\limits_{-\infty}^\infty\frac{\dd q}{q-\ii\epsilon}~
\e^{\ii qx^{10}(z)} \ , \non \chi^{~}_{C_\epsilon}(z)&=&
\ii\epsilon\,C_\epsilon(z)\otimes\psi^{10}(z) \ , \non
D_\epsilon(z)&=&-\frac1{2\pi}\,\int\limits_{-\infty}^\infty\frac{\dd
  q}{(q-\ii\epsilon)^2}~\e^{\ii qx^{10}(z)} \ , \non
\chi^{~}_{D_\epsilon}(z)&=&\ii\left(\epsilon\,D_\epsilon(z)-
\frac2\epsilon\,C_\epsilon(z)\right)\otimes\psi^{10}(z) \ .
\label{susyrecoilops}\eea
Here and in the following, singular operator products taken at
coincident points are always understood to be normal ordered according
to the prescription
\beq
O(z)\,O'(z)\equiv\oint\limits_{w=z}\frac{\dd w}{2\pi\ii}~
\frac{O(w)\,O'(z)}{w-z} \ .
\label{normalordering}\eeq

The target space regularization parameter $\epsilon\to0^+$ is related
to the worldsheet ultraviolet cutoff $\Lambda\to0^+$ by
\beq
\frac1{\epsilon^2}=-\ln\Lambda \ .
\label{epsilonLambda}\eeq
In this limit, careful computations~\cite{kmw,MavSz} establish that,
to leading orders in $\epsilon$, the superfield recoil operators
(\ref{susyrecoilops}) satisfy the relations (\ref{CDsuperOPE}) and
(\ref{CC})--(\ref{DD}) of the $N=1$ logarithmic superconformal algebra
in the NS sector of the worldsheet field theory, with
\bea
\Delta_{C_\epsilon}&=&-\frac{\epsilon^2}4 \ , \non
b&=&\frac{\pi^{3/2}}4 \ , \non d&=&\frac{\pi^{3/2}}{2\epsilon^2} \ .
\label{SLCFTrecoilconsts}\eea
In the following we will describe how to properly incorporate the
Ramond sector of this system.

\subsection{Superspace Formalism}

We will now derive the explicit form of the supersymmetric extension of the
impulse vertex operator (\ref{vertexD},\ref{Yirecoil}). For this, we consider
the Wilson loop operator
\beq
W[A]=\exp i\oint\limits_{\partial\Sigma}A_\mu(x)~dx^\mu=\exp i\int\limits_0^1
d\tau~\dot x^\mu(\tau)\,A_\mu\Bigl(x(\tau)\Bigr) \ ,
\label{Wilsonloop}\eeq
where $A_\mu$ is a $U(1)$ gauge field in ten dimensions, and $\dot
x^\mu(\tau)=dx^\mu(\tau)/d\tau$. T-duality maps the operator (\ref{Wilsonloop})
onto the vertex operator (\ref{vertexD}) for a moving D-brane by the rule
$\partial_\alpha x^i\mapsto
i\,\eta^{\beta\gamma}\,\epsilon_{\alpha\beta}\,\partial_\gamma x^i$ and the
resulting replacement of Neumann boundary conditions for $x^i$ with Dirichlet
ones. The spatial components of the Chan-Paton gauge field map
onto the brane trajectory as $A_i=Y_i/2\pi\alpha'$, while the temporal
component $A_0$ becomes a $U(1)$ gauge field on the D-particle worldline.

The minimal ${\cal N}=1$ worldsheet supersymmetric extension of the operator
(\ref{Wilsonloop}) is given by
\beq
{\cal W}[A,\psi]=W[A]\,\exp\left(-\frac12\,\int\limits_0^1d\tau~
F_{\mu\nu}\,\overline{\psi}^{\,\mu}\,\rho^1\,\psi^\nu\right) \ ,
\label{WSsusyWilson}\eeq
where $F_{\mu\nu}$ is the corresponding gauge field strength tensor. For the
recoil trajectory (\ref{Yirecoil}), an elementary computation using the contour
integration techniques outlined in the previous section gives $F_{ij}=0$ and
\beq
F_{0i}(x^0)=\frac{\delta A_i(x^0)}{\delta x^0}=\frac i{2\pi\alpha'}\,\left[
y_i\,\epsilon\,C_\epsilon(x^0)+u_i\left(\epsilon\,D_\epsilon(x^0)-
\frac1{\epsilon\,\alpha'}\,C_\epsilon(x^0)\right)\right]. \nn \\
\label{F0i}\eeq
This shows that, in the T-dual Neumann picture, the canonical supersymmetric
extension of the $U(1)$ Wilson loop operator (\ref{WSsusyWilson}) yields {\it
precisely} the couplings to the operators $\chi^{~}_{C_\epsilon}$ and
$\chi^{~}_{D_\epsilon}$ that were computed in the previous section from the
supersymmetric completion of the worldsheet logarithmic conformal algebra.

T-duality acts on the worldsheet fermion fields (\ref{WSspinordecomp}) by
reversing the sign of their right-moving components $\psi_-^\mu$. By using
(\ref{WSsusyWilson},\ref{F0i}) we may thereby write down the supersymmetric
extension of the impulse operator for moving D0-branes,
\bea
V_{\rm D}^{\rm susy}&=&\exp\left(-\frac1{2\pi\alpha'}\,\int\limits_0^1d\tau~
\left\{\left[y_i\,C_\epsilon\Bigl(x^0(\tau)\Bigr)+u_i\,D_\epsilon
\Bigl(x^0(\tau)\Bigr)\right]\,\partial^{~}_{\!\perp} x^i(\tau)\right.\right.
\nn\\&&+\Biggl.\left.\left[
y_i\,\chi_{C_\epsilon}^{~}\Bigl(x^0(\tau)\,,\,\psi^0(\tau)\Bigr)+u_i\,
\chi^{~}_{D_\epsilon}\Bigl(x^0(\tau)\,,\,\psi^0(\tau)\Bigr)\right]\,
\psi^i(\tau)\right\}\Biggr) \ ,\nn \\
\label{vertexDsusy}\eea
where we have dropped the $\pm$ subscripts on the fermion fields in
(\ref{WSspinordecomp}), and the logarithmic superconformal operators in
(\ref{vertexDsusy}) are given by (\ref{CDops}), (\ref{chiCepsilon}) and
(\ref{chiDepsilon}). The vertex operator (\ref{vertexDsusy}) can be expressed
in a more compact form which makes its supersymmetry manifest. For this, we
extend the disc $\Sigma$ to an ${\cal N}=1$ super-Riemann surface $\hat\Sigma$
with coordinates $(Z,\bar
Z)=(z,\theta,\bar z,\bar\theta\,)$, where $\theta$ is a complex Grassmann
variable, and with corresponding superspace covariant derivatives ${\cal
D}_Z=\partial_\theta+\theta\,\partial_z$. Given a bosonic field $\phi(z)$ with
superpartner $\chi^{~}_\phi(z)$, we introduce the chiral worldsheet superfields
\beq
\Phi_\phi(z,\theta)=\phi(z)+\theta\,\chi^{~}_\phi(z) \ ,
\label{Phisuper}\eeq
and correspondingly we make the embedding space of the superstring an ${\cal
N}=1$ superspace with chiral scalar superfields
$X^i(z,\theta)=x^i(z)+\theta\,\psi^i(z)$. Then the impulse operator
(\ref{vertexDsusy}) can be written in a manifestly supersymmetric form in terms
of superspace quantities as
\beq
V_{\rm D}^{\rm susy}=\exp\left[-\frac1{2\pi\alpha'}\,\oint
\limits_{\partial\hat\Sigma}d\tau~d\theta~
\Bigl(y_i\,\Phi_{C_\epsilon}(\tau,\theta)+u_i\,\Phi_{D_\epsilon}(\tau,\theta)
\Bigr)\,{\cal D}^{~}_{\!\perp}X^i(\tau,\theta)\right] \ ,\nn \\
\label{VDsusysuperspace}\eeq
where in (\ref{VDsusysuperspace}) the Grassmann coordinate $\theta$ is real.

In fact, the algebraic relations of the logarithmic superconformal algebra can
be most elegantly expressed in superspace notation. For this, we introduce the
super-stress tensor ${\cal T}(Z)=G(z)+\theta\,T(z)$, and define the quantities
$Z_{12}=z_1-z_2-\theta_1\theta_2$ and $\theta_{12}=\theta_1-\theta_2$
corresponding to a pair of holomorphic superspace coordinates
$Z_1=(z_1,\theta_1)$ and $Z_2=(z_2,\theta_2)$. Then the operator product
expansions (\ref{TOPE}), (\ref{TCD}) and (\ref{TGOPE})--(\ref{SUSYOPE}) can
also be written in terms of superspace quantities as
\bea
{\cal T}(Z_1)\,{\cal T}(Z_2)&=&\frac{\hat c/4}{(Z_{12})^3}+\frac{3\theta_{12}
/2}{(Z_{12})^2}\,{\cal T}(Z_2)+\frac{1/2}{Z_{12}}\,{\cal D}_{Z_2}{\cal T}(Z_2)
+\frac{\theta_{12}}{Z_{12}}\,\partial_{z_2}{\cal T}(Z_2)+\dots \ , \nn\\
{\cal T}(Z_1)\,\Phi_C(Z_2)&=&\frac{\theta_{12}\,\Delta/2}{(Z_{12})^2}\,
\Phi_C(Z_2)+\frac{1/2}{Z_{12}}\,{\cal D}_{Z_2}\Phi_C(Z_2)+\frac{\theta_{12}}
{Z_{12}}\,\partial_{z_2}\Phi_C(Z_2)+\dots \ , \nn\\
{\cal T}(Z_1)\,\Phi_D(Z_2)&=&\frac{\theta_{12}\,\Delta/2}{(Z_{12})^2}\,
\Phi_D(Z_2)+\frac{\theta_{12}/2}{(Z_{12})^2}\,\Phi_C(Z_2)\nn\\&&
+\,\frac{1/2}{Z_{12}}\,{\cal D}_{Z_2}\Phi_D(Z_2)+\frac{\theta_{12}}
{Z_{12}}\,\partial_{z_2}\Phi_D(Z_2)+\dots \ ,
\label{superspaceOPE}\eea
while the two-point functions (\ref{SUSY2pt}) may be expressed as
\bea
\Bigl\langle\Phi_C(Z_1)\,\Phi_C(Z_2)\Bigr\rangle&=&0 \ , \nn\\
\Bigl\langle\Phi_C(Z_1)\,\Phi_D(Z_2)\Bigr\rangle&=&\frac\xi{(Z_{12})^{2
\Delta}} \ , \nn\\\Bigl\langle\Phi_D(Z_1)\,\Phi_D(Z_2)\Bigr\rangle&=&
\frac1{(Z_{12})^{2\Delta}}\,\Bigl(-2\xi\ln Z_{12}+d\Bigr) \ .
\label{superspace2pt}\eea
This superspace formalism also generalizes to the construction of higher-order
correlation functions which are built from appropriate coordinate invariants of
the supergroup $OSp(1|2)$. It emphasizes how the impulse
operator (\ref{VDsusysuperspace}), and the ensuing logarithmic algebra
(\ref{superspaceOPE},\ref{superspace2pt}), is the natural supersymmetrization
of the recoil operators for D-branes.

A remark is in order here concerning the behavior of the 
superconformal partners of the recoil operators 
under the changes of the scale $\epsilon^2$. 
By using (\ref{chiCepsilon}), (\ref{chiDepsilon}),
(\ref{epsilonscale}) and (\ref{CDscale}), we see that the superconformal
partners of the logarithmic operators are scale-invariant to order
$\epsilon^2$,
\beq
\chi^{~}_{C_{\epsilon'}}=\chi^{~}_{C_\epsilon} \ , ~~ \chi^{~}_{D_{\epsilon'}}=
\chi^{~}_{D_\epsilon} \ .
\label{chiCDscale}\eeq
The invariance property (\ref{chiCDscale}) can also be deduced from the scale
independence to order $\epsilon^2$ of the two-point correlators
(\ref{SUSY2pt}), in which the scale dependent constant $d_\epsilon$ appears
only in the invariant combination $\Delta_\epsilon d_\epsilon\sim
O(\epsilon^0)$. This means that the operator (\ref{vertexDsusy}) describes the
evolution of the D0-brane in target space with respect to only the ordinary,
bosonic Galilean group. In other words, if we introduce a superspace and
worldsheet superfields as in (\ref{Phisuper}), then a worldsheet scale
transformation in the present case acts only on the bosonic part of the
superspace. This property is of course very particular to the explicit scale
dependence of the recoil
superpartners (\ref{chiDepsilon}) in the logarithmic superconformal algebra.

The fact that the super-Galilean group is not represented in the
non-relativistic dynamics of D-branes is merely a reflection of the fact that
the motion of the brane explicitly breaks target space supersymmetry. Indeed,
while the deformed $\sigma$-model that we have been working with possesses
${\cal N}=1$ {\it worldsheet} supersymmetry, it is only after the appropriate
sum over worldsheet spin structures and the GSO projection that it has the
possibility of possessing spacetime supersymmetry. To understand better the
breaking of target space supersymmetry within the present formalism, we now
appeal to an explicit spacetime supersymmetrization of the Wilson loop operator
(\ref{Wilsonloop}). This will produce a Green-Schwarz representation of the
spacetime
supersymmetric impulse operator in the dual Neumann picture, and also yield a
physical interpretation of the supersymmetric vertex operator
(\ref{vertexDsusy}).

For this, we regard the Chan-Paton gauge field $A_\mu$ as the first component
of the ten-dimensional ${\cal N}=1$ Maxwell supermultiplet. Its superpartner is
therefore a Majorana-Weyl fermion field $\lambda$ with 32 real components. We
introduce Dirac matrices $\Gamma_\mu$ in 1+9 dimensions, and define
$\Gamma_{\mu\nu}=\frac12\,[\Gamma_\mu,\Gamma_\nu]$. The loop parametrization
$x^\mu(\tau)$ has superpartner $\vartheta(\tau)$ which couples to the photino
field $\lambda$. Then, the spacetime supersymmetric extension of
(\ref{Wilsonloop}) is given by the finite supersymmetry transformation
\beq
{\sf W}[A,\lambda]=\exp\left(\,\int\limits_0^1d\tau~
\overline{\vartheta}(\tau)\,{\sf Q}
\right)~W[A]~\exp\left(-\int\limits_0^1d\tau~\overline{\vartheta}(\tau)
\,{\sf Q}\right),
\label{susyWilsonloop}\eeq
where the supercharge $\sf Q$ generates the infinitesimal ${\cal N}=1$
supersymmetry transformations
\bea
\Bigl[{\sf Q}\,,\,A_\mu\Bigr]&=&\frac i2\,\Gamma_\mu\,\lambda \ , \non
\Bigl\{{\sf Q}\,,\,\lambda\Bigr\}&=&-\frac14\,
\Gamma_{\mu\nu}\,F^{\mu\nu} \ , \nonumber\\ \Bigl[{\sf Q}\,,\,x^\mu\Bigr]
&=&\frac i4\,\Gamma^\mu\,\vartheta \ , \non \Bigl\{{\sf Q}
\,,\,\vartheta\Bigr\}&=&4 \ .
\label{susyspacetime}\eeq
By using the Baker-Campbell-Hausdorff formula, the supersymmetric Wilson loop
(\ref{susyWilsonloop}) thereby admits an expansion
\bea
{\sf W}[A,\lambda]&=&\exp i\int\limits_0^1d\tau~\left(\dot x^\mu\,A_\mu
+\frac i4\,A_\mu\,\overline{\vartheta}\,\Gamma^\mu\,\dot\vartheta\right.\non&&
+\left.\frac i2\,\dot x^\mu\,\overline{\vartheta}\,\Gamma_\mu\,\lambda
+\frac i{16}\,\dot x^\mu F^{\nu\lambda}\,\overline{\vartheta}\,\Gamma_\mu\,
\Gamma_{\nu\lambda}\,\vartheta+\dots\right) \ ,
\label{susyWilsonexp}\eea
where the ellipsis in (\ref{susyWilsonexp}) denotes contributions from
higher-order fluctuation modes of the fields.

To identify the ten-dimensional supermultiplet which is T-dual to the
worldsheet recoil supermultiplet of (\ref{vertexDsusy}), we use the
supersymmetry algebra (\ref{susyspacetime}) to get
$\Gamma_i\,\lambda=\frac12\,F_{0i}(x^0)\,\Gamma^0\,\vartheta$, with
$F_{0i}(x^0)$ given by (\ref{F0i}). We then find that the target space
supermultiplet describing the recoil of a D0-brane is given by the
dimensionally reduced supersymmetric Yang-Mills fields
\bea
A_i(x^0)&=&\frac1{2\pi\alpha'}\,\Bigl(y_i\,C_\epsilon(x^0)+u_i\,D_\epsilon
(x^0)\Bigr) \ , \nn\\\lambda(x^0,\vartheta)&=&\frac1{36\pi\alpha'}\,\Gamma^i\,
\Bigl(y_i\,\chi^{~}_{C_\epsilon}(x^0,\Gamma^0\,\vartheta)+u_i\,
\chi^{~}_{D_\epsilon}(x^0,\Gamma^0\,\vartheta)\Bigr) \ .
\label{recoilsupermult}\eea
Therefore, the logarithmic superconformal partners to the basic recoil
operators also arise naturally in the T-dual Green-Schwarz formalism.

By using (\ref{susyWilsonexp}) and (\ref{recoilsupermult}) we can now lend a
physical interpretation to the supersymmetric impulse operator. For simplicity,
we shall neglect the stringy fluctuations in the center of mass coordinates of
the D-brane and take $y_i=0$. We consider only the long-time dynamics of the
string soliton, i.e. we take $x^0>0$ and effectively set the Heaviside function
$\Theta_\epsilon(x^0)$ to unity everywhere. We will also choose the gauge
$A_0(x)=0$. The bosonic part of the Maxwell supermultiplet of course describes
the free, non-relativistic geodesic motion of the D0-brane in flat space. To
see what sort of particle kinematics is represented by the full supermultiplet,
we substitute $A_i=u_i\,x^0/2\pi\alpha'$ and
$\lambda=u\slash\,\Gamma^0\,\vartheta/36\pi\alpha'$ into (\ref{susyWilsonexp}),
where $u\slash=u_i\,\Gamma^i$, and we have again ignored stringy $O(\epsilon)$
uncertainties in position and velocity. Note that, generally, the fermionic
operator (\ref{chiDepsilon}) also induces a velocity-dependent stringy
contribution to the phase space uncertainty principle in the sense described
in~\cite{kmw}. This is reminiscent of the energy-dependent smearings that were
found in~\cite{ms1}. Heuristically, this identical stringy smearing of position
and velocity is responsible for the violation of super-Galilean invariance in
(\ref{chiCDscale}).

With these substitutions we find ${\sf W}[A,\lambda]=\e^{i\,{\sf
S}/2\pi\alpha'}$, where
\bea
{\sf S}&=&\int\limits_0^1d\tau~\left(\dot x^i\,u_i\,x^0+\frac i4\,x^0\,
\overline{\vartheta}\,u\slash\,\dot\vartheta+\frac i{36}\,\dot x^0\,
\overline{\vartheta}\,u\slash\,\vartheta-\frac i4\,\dot x^i\,u_i\,
\overline{\vartheta}\,\Gamma^0\,\vartheta\right.\nn\\&&+\left.
\frac i{32}\,\dot x^0\,\overline{\vartheta}\,\left[\Gamma^0\,,\,u\slash
\right]\,\vartheta+\frac i{32}\,\dot{x\slash}\,\overline{\vartheta}
\left[\Gamma^0\,,\,u\slash\right]\,\vartheta+\dots\right)
\label{superpartaction}\eea
can be interpreted as the action of a certain kind of superparticle in the
${\cal N}=1$ superspace spanned by the coordinates
$(x^i,\vartheta,\overline{\vartheta}\,)$ and with worldline parametrized by the
loop coordinate $\tau$. To identify the superparticle type, we will first
simplify the last four terms in (\ref{superpartaction}). For this, we note that
in ten spacetime dimensions the Dirac matrices are taken in a Majorana
representation, so that $\Gamma^0$ is antisymmetric while $\Gamma^i$,
$i=1,\dots,9$, are symmetric matrices. We also treat
$\vartheta,\dot\vartheta$ as an anticommuting pair of variables in the action
$\sf S$. Then, it is easy to check that the third term in
(\ref{superpartaction}) vanishes, because via an integration by parts it can be
written as
\beq
-\frac i{36}\,\int\limits_0^1d\tau~x^0\,\left(\dot{\vartheta}^\top\,\Gamma^0\,
u\slash\,\vartheta+\vartheta^\top\,\Gamma^0\,u\slash\,\dot\vartheta\right)
=0 \ ,
\label{3rdterm0}\eeq
where we have used the Dirac algebra to write
$\Gamma^0\,u\slash=-u\slash\,\Gamma^0$. In a similar way one readily checks
that the fourth and fifth terms in
(\ref{superpartaction}) are zero. By the same techniques one finds that the
last term is non-vanishing, and after some algebra it can be expressed in the
form $\frac
i4\,\int_0^1d\tau~\vartheta^\top\,x^j\,u^i\,\Gamma_{ij}\,\dot\vartheta$. The
action (\ref{superpartaction}) can therefore be written as
\beq
{\sf S}=\int\limits_0^1d\tau~\left[\,p_i\left(\dot x^i+i\,
\overline{\vartheta}\,\Gamma^i\,\dot\vartheta\right)-i\,\ell^\top\,\dot
\vartheta+\dots\right] \ ,
\label{superactionfinal}\eeq
where
\beq
p_i=u_i\,x^0 \ , ~~ \ell=x^i\,u^j\,\Gamma_{ij}\,\vartheta \ ,
\label{pietadef}\eeq
and we have rescaled the worldline spinor fields $\vartheta\mapsto2\vartheta$.

The action (\ref{superactionfinal}) is, modulo mass-shell constraints, that of
a twisted superparticle\cite{MSsuper}, which admits a manifestly covariant
quantization. The first term is the standard non-relativistic superparticle
action, while the inclusion of the fermionic field $\ell$ modifies the
canonically conjugate momentum to $\vartheta$ as
$\pi_\vartheta=p\slash\vartheta-\ell$. Note that the quantity $p_i$ in
(\ref{pietadef}) is the expected momentum of the uniformly moving D-particle,
while $\ell$ is proportional to its angular momentum. In the present case
$p_\mu\,p^\mu\neq0$, so that the supersymmetric impulse operator describes a
{\it massive}, non-relativistic twisted superparticle. The twist in fermionic
momentum
$\pi_\vartheta$ vanishes if there is no angular momentum, for instance if the
D-particle recoils in the direction of scattering. The equations of motion
which follow from the action (\ref{superactionfinal},\ref{pietadef}) can be
written as
\beq
\dot x^0=u_i\,\dot x^i=u\slash\,\dot\vartheta=0 \ ,
\label{finaleqsmotion}\eeq
which imply that $x^0$ and the components of $x^i$ and $\vartheta$ along the
direction of motion are independent of the proper time $\tau$. In general the
remaining components of $x^\mu$ and $\vartheta$ are $\tau$-dependent. These
classical configurations agree with the interpretation of the worldsheet zero
mode of the field $x^0$ as the target space time and also of the uniform motion
of the D-particles. In particular, the Galilean trajectory
$x^i(\tau)=y^i(\tau)+u^i\,x^0$, appropriate for the kinematics of a heavy
D0-brane, solves (\ref{finaleqsmotion}) provided that the component of the
vector $y^i(\tau)$ along the direction of recoil is independent of the
worldline coordinate $\tau$.

There are some important differences in the present case from the standard
superparticle kinematics. The action (\ref{superactionfinal}) generically
possesses a fermionic $\kappa$-symmetry defined by the transformations
\bea
\delta_\kappa\vartheta&=&p\slash\,\kappa \ , \nn\\\delta_\kappa
\ell&=&2\,p_i\,p^i\,\kappa \ , \nn\\\delta_\kappa x^i&=&
i\,\kappa\,p\slash\,\Gamma^i\,\vartheta \ ,
\label{kappasym}\eea
where $\kappa$ is an infinitesimal Grassmann spinor parameter. It is also
generically invariant under a twisted ${\cal N}=2$ super-Poincar\'e
symmetry~\cite{MSsuper}. However, the choices (\ref{pietadef}) break these
supersymmetries, which is expected because the D-brane motion induces a
non-trivial vacuum energy. The configurations (\ref{pietadef}) of course arise
from the geodesic bosonic paths in the non-relativistic limit $u_i\ll1$, or
equivalently in the limit of heavy BPS mass for the D-particles, which is the
appropriate limit to describe the tree-level dynamics here. The Galilean
solutions of (\ref{finaleqsmotion})
described above explicitly break the $\kappa$-symmetry (\ref{kappasym}).

Therefore, we see that the supersymmetric completion
of the impulse operator (for weakly-coupled strings) describes the dynamics of
a twisted supersymmetric D-particle in the non-relativistic limit, with a
gauge-fixing that breaks its target space supersymmetries. In turn, this broken
supersymmetry implies that the vertex operator (\ref{vertexDsusy}) does not
generate the action of the super-Poincar\'e group on the brane, and
consequently the super-D-particle does not evolve in target space according to
super-Galilean
transformations\cite{MavSz}. The structure of the worldsheet logarithmic
superconformal algebra is such that these spacetime properties of D-brane
dynamics are enforced by the impulse operators.

\subsection{Spin Fields\label{logspinrecoil}}

We will now construct the operators $\Sigma$ which create cuts in the
fields $\psi^{10}$ appearing in the superpartners of the recoil operators
(\ref{susyrecoilops}) and are thereby responsible for
changing their boundary conditions as one circumnavigates the cut~\cite{gsw}.
In fact, one needs $\Sigma(z)$ in the neighborhood of the fields $\psi^{10}$
but this is readily done in bosonized form~\cite{fermspin}, as we
shall now discuss, by means of a boson translation operator which
relates $\Sigma (z)$ to $\Sigma (0)$. Bosonization of the free fermion
system defined by (\ref{SD0}) allows us to express in a local-looking
form the non-local effects of the spin operators. In what follows we
shall only require the bosonization of the spinor field appearing in
(\ref{susyrecoilops}).

In the Euclidean version of the target space theory there are ten
fermion fields $\psi^\mu$ which we can treat on equal footing. Given
the pair of right-moving NSR fermion fields $\psi^9$, $\psi^{10}$
corresponding to the light-cone of the recoiling D0-brane system, we
may form complex Dirac fermion fields
\beq
\psi^\pm(z)=\psi^9(z)\pm\ii\psi^{10}(z) \ .
\label{psipmdef}\eeq
The worldsheet kinetic energy in (\ref{SD0}) associated to this pair is
of the form
\beq
\int\dd^2z~\left(\psi^9\,\overline{\partial_z}\,\psi^9+\psi^{10}\,
\overline{\partial_z}\,\psi^{10}
\right)=\int\dd^2z~\psi^+\,\overline{\partial_z}\,\psi^- \ .
\label{nsraction}\eeq
{}From the corresponding equations of motion and (\ref{scx2ptfn}) it
follows that the field
\beq
j(z)=\psi^+(z)\,\psi^-(z)
\label{Jpsi}\eeq
is a conserved $U(1)$ fermion number current which is a primary field
of the Virasoro algebra of dimension 1 and which generates a $U(1)$
current algebra at level 1. Its presence allows the
introduction of spin fields, and hence twisted sectors in the quantum
Hilbert space, through the bosonization formulas
\bea
j(z)&=&2\ii\,\partial_z\phi(z) \ , \non
\psi^\pm(z)&=&\sqrt2\,\e^{\pm\ii\phi(z)} \ ,
\label{bosonformulas}\eea
where $\phi(z)$ is a free, real, compact chiral scalar field,
i.e. its two-point function is
\beq
\langle0|\phi(z)\,\phi(w)|0\rangle=-\ln(z-w) \ .
\label{freebos2ptfn}\eeq
In this representation all fields are taken to act in the NS
sector.

The holomorphic part of the Sugawara energy-momentum tensor
corresponding to the worldsheet action (\ref{nsraction}) is given in
bosonized form by
\beq
T_\kappa(z)=-\frac12\,\partial_z\phi(z)\,\partial_z\phi(z)+\frac{\ii\kappa}2\,
\partial_z^2\phi(z) \ ,
\label{stressboson}\eeq
where the constant $\kappa$ is arbitrary because the second term in
(\ref{stressboson}) is identically conserved for all $\kappa$. This
energy-momentum tensor derives from the Coulomb gas model defined by
the Liouville action
\beq
S_\kappa=\frac1{4\pi}\,\int\dd z~\dd\overline{z}~\sqrt g\,\left(
\partial_z\phi\,\overline{\partial_z}\,\phi+\frac{\ii\kappa}2\,
R^{(2)}\,\phi\right) \ ,
\label{Skappa}\eeq
where $g(z,\overline{z}\,)$ and $R^{(2)}(z,\overline{z}\,)$ are the
metric and curvature of the worldsheet. The topological curvature term
in (\ref{Skappa}) provides a deficit term to the central charge
$c_\kappa$ of the free boson field $\phi(z)$,
\beq
c_\kappa=1-3\kappa^2 \ ,
\label{ckappa}\eeq
and it also induces a vacuum charge at infinity (the singular point of
the metric on the Riemann sphere). In particular, the primary
field $\e^{\ii q\phi(z)}$ has dimension
\beq
\Delta_{q,\kappa}=\frac q2\,\Bigl(q-\kappa\Bigr) \ .
\label{hqkappa}\eeq
What fixes $\kappa$ here, and thereby lifts the ambiguity, is the
charge conjugation symmetry $\psi^{10}(z)\mapsto-\psi^{10}(z)$ of the
NSR model (\ref{nsraction}), which interchanges the two Dirac fields
$\psi^\pm(z)$ and hence acts on the free boson field as
$\phi(z)\mapsto-\phi(z)$. This symmetry implies that $\kappa=0$ in
(\ref{stressboson}).

Let us now consider the tachyon vertex operators corresponding to the
free boson,
\begin{equation}
           \Sigma _q(z)=\e^{\ii q\phi(z)} \ ,
\label{cutoper}\end{equation}
which have conformal dimension $\Delta_{q,0}=q^2/2$.
In bosonized language the pair of Dirac
fermion fields corresponds to the operators (\ref{cutoper}) at $q=\pm
1$, $\psi^\pm(z)=\sqrt2\,\Sigma_{\pm 1}(z)$. On the other hand, the operators
(\ref{cutoper}) at $q=\pm\,\frac12$ introduce a branch cut in the
field $\psi^{10}(z)$. To see this, we note the standard free
field formula for multi-point correlators of tachyon vertex operators,
\bea
\langle0|\Sigma_{q_1}(z_1)\cdots\Sigma_{q_n}(z_n)|0\rangle&=&
\prod_{k=1}^n\,\prod_{l=1}^n\e^{-q_kq_l\langle0|\phi(z_k)\,\phi(z_l)
|0\rangle/2}\non&=&\Lambda^{\bigl(\sum_lq_l\bigr)^2/2}\,
\prod_{k<l}(z_k-z_l)^{q_kq_l} \ ,
\label{tachyoncorr}\eeq
where we have regulated the coincidence limit of the two-point
function (\ref{freebos2ptfn}) by the short-distance cutoff
$\Lambda\to0^+$. In particular, the correlator (\ref{tachyoncorr})
vanishes unless
\beq
\sum_{l=1}^nq_l=0 \ ,
\label{selectionrule}\eeq
which is a consequence of the
continuous $U(1)$ symmetry generated by the current (\ref{Jpsi}) which
acts by global translations of the fields $\phi(z_l)$. From the
general result (\ref{tachyoncorr}) we may infer the three-point
correlation functions
\beq
\langle0|\Sigma_{\pm1/2}(z_1)\,\Sigma_{\pm1/2}(z_2)\,\Sigma_{\mp1}(z_3)
|0\rangle=\frac{(z_1 - z_2)^{1/4}}{\sqrt{(z_1 - z_3)(z_2 - z_3)}} \ .
\label{cuteq}\eeq
The correlator (\ref{cuteq}) has square root branch points at
$z_3=z_1$ and $z_3=z_2$. This implies that the elementary fermion fields
$\psi^\pm(z_3)$ are double-valued in the fields of the operators
$\Sigma_{\mp1/2}(z_1)$, respectively.

It follows that the spin operators for the recoil problem are given by
\beq
\Sigma^+_{1/8}(z)=\sqrt2\,\cos\frac{\phi(z)}2
\label{spinrecoilpm}\eeq
and they have weight $\Delta=\Delta_{\pm1/2}=\frac18$. They create
branch cuts in the fermionic fields
\beq
\psi^{10}(z)=\sqrt2\,\sin\phi(z) \ .
\label{psi10sinphi}\eeq
Note that the spin operators need only be inserted at the origin
$z=0$, because it is there that they are required to change the
boundary conditions on the fermion fields. These operators are all
understood as acting on the NS vacuum state $|0\rangle$, thereby
creating highest weight states in the Ramond sector. The spin fields
$\Sigma^\pm_{1/8}(0)$ may be extended to operators
$\Sigma^\pm_{1/8}(z)$ in the neighborhood of $\psi^{10}(z)$ via application
of the boson translation operator $\e^{z\,\partial_z}=\e^{z\,L_{-1}}$.

Using the operator product expansions
\beq
\Sigma_q(z)\,\Sigma_{q'}(w)=(z-w)^{qq'}\,\Sigma_{q+q'}(w)\Bigl(1+
\ii q\,(z-w)\,\partial_w\phi(w)\Bigr)+\dots \ , ~~ qq'<-1
\nn \\ \label{SigmaqqprimeOPE}\eeq
and (\ref{chiralscT}), it is straightforward to check that the term of
order $(z-w)^{-3/2}$ in the operator product
$G(z)\,\Sigma_{1/8}^+(w)$ vanishes, and hence that
\beq
\Sigma_{1/8}^-(z)=0 \ .
\label{Sigma18minus0}\eeq
This means that the spin field $\Sigma(z)=\Sigma_{1/8}^+(z)$
corresponds to the supersymmetric ground state
$|\frac18\rangle^{~}_{\rm R}$ in the Ramond sector of the system,
associated with superconformal central charge $\hat c=2$. By using
the selection rule (\ref{selectionrule}) and the factorization of
bosonic and fermionic correlation functions in the free superconformal
field theory determined by (\ref{SD0}), it is straightforward to
verify both the NS two-point functions (\ref{CC})--(\ref{DD}) and the
spin-spin two-point function as normalized in (\ref{Sigma2pt}). The
central charge $\hat c=2$ is the one pertinent to the recoil operators
because in the bosonized representation they only refer to two of the
ten superconformal fields present in the total action (\ref{SD0}).

Using (\ref{SigmaqqprimeOPE}) one can also easily derive
the excited logarithmic spin operators of dimension
$\Delta_{C_\epsilon}+\frac18$, which along with (\ref{chiCDSigma}) and
(\ref{susyrecoilops}) yields
\bea
\widetilde{\Sigma}_{C_\epsilon}(z)&=&\ii\epsilon\,C_\epsilon(z)\otimes
\sin\frac{\phi(z)}2 \ , \non\widetilde{\Sigma}_{D_\epsilon}(z)&=&
\ii\left(\epsilon\,D_\epsilon(z)-
\frac2\epsilon\,C_\epsilon(z)\right)\otimes\sin\frac{\phi(z)}2 \ .
\label{Sigmaexcitedrecoil}\eea
The corresponding logarithmic operator product expansions
(\ref{TSigmaC}) and (\ref{TSigmaD}) are straightforward consequences
of the factorization of the bosonic and fermionic sectors in the recoil
problem. Because of this same factorization property, all of the two-point
correlation functions of section~\ref{RCorrelators} may be easily
derived. The basic identities are given by (\ref{cuteq}) and the
four-point function
\bea
\langle0|\Sigma(z_1)\,\psi^{10}(z)\,\psi^{10}(w)\,\Sigma(z_2)|0\rangle&=&
\frac1{2(z_1-z_2)^{1/4}\,(z-w)}\non&&\times\,
\left(\,\sqrt{\frac{(z_1-z)(w-z_2)}{(z_1-w)(z-z_2)}}+
\sqrt{\frac{(z_1-w)(z-z_2)}{(z_1-z)(w-z_2)}}~\right) \ , \non&&
\label{4ptfnrecoil}\eea
where we have again used the selection rule
(\ref{selectionrule}). Thus, by using bosonization techniques it is
straightforward to describe the $N=1$ supersymmetric extension of the
logarithmic operators of the recoil problem in both the NS and R
sectors of the worldsheet superconformal field theory.

\subsection{Fermionic Vertex Operators for the Recoil Problem}

As a simple application of the above formalism, we will now construct
the appropriate spacetime vertex operators which create recoil states of the
D-branes. The crucial point is that one can now build states
that are consistent with the target space supersymmetry of Type~II
superstring theory, which thereby completes the program of
constructing recoil operators in string theory. Spacetime
supersymmetry necessitates vertex operators which describe the
excitations of fermionic states in target space. Such supersymmetric
operators were constructed in~\cite{MavSz} from a target space
perspective. Here we shall construct fermionic states for the recoil
problem from a worldsheet perspective by using appropriate
combinations of the spin operators (\ref{cutoper}). We have already
seen how this arises above, in that the Ramond state
$G_0|\frac18\rangle^{~}_{\rm R}$ is a null vector and
one recovers a single logarithmic superconformal algebra among the
physical states, as in the NS sector. This construction relies heavily
on the Euclidean signature of the spacetime, and yields states that
transform in an appropriate spinor representation of the Euclidean
group.

The recoil operators (\ref{susyrecoilops}) are all built as
appropriate superpositions of the off-shell tachyon vertex operators
$\e^{\ii qx^{10}(z)}$. It is well-known how to construct the boson and
fermion emission operators which create corresponding tachyon ground
states from the NS vacuum state $|0\rangle$~\cite{gsw}. In the bosonic
sector the vertex operator is $[G_r,\e^{\ii
  qx^{10}(z)}]=q~\e^{\ii qx^{10}(z)}\otimes\psi^{10}(z)$, where the fermion
field $\psi^{10}(z)$ has the periodic mode expansion
\beq
\psi^{10}(z)=\frac1{\sqrt2}~\sum_{n=-\infty}^\infty\psi_{n+1/2}^{10}
{}~z^{-n-1}
\label{psi10permode}\eeq
appropriate to the NS sector, with $(\psi^{10}_r)^\dag=\psi^{10}_{-r}$,
$\{\psi_r^{10},\psi_s^{10}\}=\delta_{r+s,0}$, and
$\psi^{10}_{n+1/2}|0\rangle=0~~\forall n\geq0$. By construction, the
corresponding recoil operators are of course just the fermionic operators
$\chi^{~}_{C_\epsilon}(z)$ and $\chi^{~}_{D_\epsilon}(z)$ in
(\ref{susyrecoilops}). The emission of a fermion by a spinor
$u_\alpha$ is described by the vertex operator
$\e^{\ii qx^{10}(z)}\otimes\overline{u}^{\,\alpha}(q)\,\Sigma_\alpha(z)$,
where $\alpha=\pm\,\frac12$ are regarded as spinor indices of the
two-dimensional Euclidean group $SO(2)$ and $u(q)$ is a two-component
off-shell Dirac spinor.

The recoil emission vertex
operators are therefore given by the chiral superfields
\bea
\scV_{C_\epsilon}(\scz)&=&\Xi_{C_\epsilon}(z)+\theta\,
\chi^{~}_{C_\epsilon}(z) \ , \non \scV_{D_\epsilon}(\scz)&=&
\Xi_{D_\epsilon}(z)+\theta\,\chi^{~}_{D_\epsilon}(z) \ ,
\label{recoilemission}\eea
where the boson emission operators are
\bea
\chi^{~}_{C_\epsilon}(z)&=&\frac{\epsilon^2}{4\pi}\,\int
\limits_{-\infty}^\infty\frac{\dd q}{q-\ii\epsilon}~\e^{\ii
  qx^{10}(z)}
\otimes\psi^{10}(z) \ , \non\chi^{~}_{D_\epsilon}(z)&=&-\frac1{2\pi}\,
\int\limits_{-\infty}^\infty\frac{\dd q~q}{(q-\ii\epsilon)^2}~
\e^{\ii qx^{10}(z)}\otimes\psi^{10}(z) \ ,
\label{bosemission}\eea
while the emission operators for the fermionic recoil states
are\footnote{\baselineskip=12pt Strictly speaking, the spin operators
  in these relations should include a non-trivial cocycle~\cite{KLLSW} for the
  lattice of charges $\alpha$ in the exponentials of the bosonized
  representation, which also depend on the fields
  $\Sigma_\alpha(z)$. The cocycle factor is defined on the weight
  lattice of the spinor representation of the Euclidean group, and it
  ensures that the vertex has the correct spinor transformation
  properties. Its inclusion becomes especially important in the generalization
  of these results to higher-dimensional branes. To avoid clutter in
  the formulas, we do not write these extra factors explicitly.}
\bea
\Xi_{C_\epsilon}(z)&=&\frac\epsilon{4\pi\ii}\,\int
\limits_{-\infty}^\infty\frac{\dd q}{q-\ii\epsilon}~
\e^{\ii qx^{10}(z)}\otimes\mu(z)\otimes
\overline{u}^{\,\alpha}(q)\,\Sigma_\alpha(z) \ , \non
\Xi_{D_\epsilon}(z)&=&-\frac1{2\pi}\,\int\limits_{-\infty}^\infty\frac{\dd
  q}{(q-\ii\epsilon)^2}~\e^{\ii qx^{10}(z)}\otimes\mu(z)\otimes
\overline{u}^{\,\alpha}(q)\,\Sigma_{\alpha}(z) \ .
\label{fermemission}\eea
Here $\mu(z)$ is an appropriate auxiliary ghost spin operator of
conformal dimension $-\frac18$~\cite{FMS1}. For example, it can be taken to
be a plane wave $\mu(z)=\e^{\ii k_ix^i(z)}$ in the
directions $x^i$ transverse to the $(x^9,x^{10})$ light cone, with
$k^2=-\frac14$. In the physical conformal limit $\epsilon\to0^+$, the
superfields (\ref{recoilemission}) then have vanishing superconformal
dimension.

The spinor $u(q)$ in (\ref{fermemission}) is not constrained by any
on-shell equations such as the Dirac equation which would normally
guarantee that the corresponding states respect spacetime
supersymmetry. It can be partially
restricted by implementing the GSO truncation of the superstring
spectrum. The fermion chirality operator $\Gamma$ acts on the
operators (\ref{cutoper}) as
\beq
\Gamma\,\Sigma_{q+(1-\lambda)/2}(z)\,\Gamma^{-1}=(-1)^{q-\lambda+1}
\,\Sigma_{q+(1-\lambda)/2}(z)
\label{GammaSigmaq}\eeq
for $q\in\zed$. This is only consistent with the operator product
expansions in the combined superconformal field theory including ghost
fields, because the action of $\Gamma$ on the fields of (\ref{SD0})
alone is not an automorphism of the local algebra of spin
fields~\cite{FMS1}. Then the action of the chirality operator can be
extended to the spin fields with the $\Gamma=1$ projection giving a
local field theory. The chiral $\Gamma=1$ projection requires that
$u_\alpha(q)$ be a right-handed Dirac spinor, after which the operators
(\ref{fermemission}) become local fermionic fields. Then the vertex operators
(\ref{recoilemission})--(\ref{fermemission}) describe the appropriate
supersymmetric states for the recoil problem. The relevant spacetime
supersymmetry generator $Q_\alpha$ is given by the contour integral of the
fermionic vertex corresponding to the basic tachyon operator $\e^{\ii
  qx^{10}(z)}$ at zero momentum,
\beq
Q_\alpha=\oint\limits_{z=0}\frac{\dd z}{2\pi}~\partial_zx^{10}(z)
\otimes z^{1/4}\,\mu\left(\frac1z\right)
\otimes\varepsilon_\alpha^{~\beta}\,\Sigma_\beta(z)\, \ .
\label{Qalphadef}\eeq
The integrand of (\ref{Qalphadef}), which involves the adjoint ghost
field $\mu^\dag(z)$, is a BRST invariant conformal field
of dimension~1. From the various operator product expansions above it
follows that the supercharge (\ref{Qalphadef}) relates the two vertices
(\ref{bosemission}) and (\ref{fermemission}) through the anticommutators
\beq
\Bigl\{Q_\alpha\,,\,\e^{\ii qx^{10}(z)}\otimes\mu(z)\otimes\Sigma_\beta(z)
\Bigr\}=-\ii\delta_{\alpha\beta}~\e^{\ii qx^{10}(z)}\otimes\psi^{10}(z) \ .
\label{QalphaSUSY}\eeq
Notice, however, that the target space supersymmetry alluded to here
refers only to the fields which live on the worldline of the
D-particle, or more precisely on the corresponding light-cone. The
full target space supersymmetry is of course broken by the motion of
the D-brane~\cite{MavSz}.

\subsection{Modular Behavior}

As we have seen above, in the case of recoiling bosonic 
string solitons (D-branes), the non-trivial mixing between the logarithmic
$C_\epsilon$ and $D_\epsilon$ operators leads to logarithmic modular
divergences in bosonic annulus amplitudes, and it is associated with the lack
of unitarity of the low-energy effective theory in which quantum D-brane
excitations are neglected~\cite{msnl}. We now examine how these features
are modified in the presence of the logarithmic ${\cal N}=1$ superconformal
pair. For this, we consider the open superstring propagator between two
scattering states $|{\cal E}_\alpha\rangle$ and $|{\cal E}_\beta\rangle$,
\beq
\triangle_{\alpha\beta}=\langle{\cal E}_\alpha|\frac1{L_0-1/2}
|{\cal E}_\beta\rangle=-\int\limits_{\cal F}
\frac{dq}q~\langle{\cal E}_\alpha|q^{L_0-1/2}|{\cal E}_\beta\rangle \ ,
\label{stringpropgen}\eeq
where the Virasoro operator $L_0$ is defined through the Laurent expansion of
the energy-momentum tensor $T(z)=\sum_nL_n\,z^{-n-2}$, and the factor of
$\frac12$ is the normal ordering intercept in the Neveu-Schwarz sector. Here
$q=\e^{2\pi i\tau}$, with $\tau$ the modular parameter of the worldsheet strip
separating the two states $|{\cal E}_\alpha\rangle$ and $|{\cal
E}_\beta\rangle$, and $\cal F$ is a fundamental modular domain of the complex
plane. We shall
ignore the superconformal ghosts, whose contributions would not affect the
qualitative results which follow.

For the purely bosonic string, divergent contributions to the modular integral
would come from a discrete subspace of string states of vanishing conformal
dimension corresponding to the spectrum of linearized fluctuations in the
soliton background~\cite{msnl}. Since in the
present case these are precisely the states associated with the logarithmic
recoil operators, we should analyze carefully their contributions to the
propagators (\ref{stringpropgen}). We introduce the highest weight states
$|\phi\rangle=\phi(0)|0\rangle$,
$\phi=C_\epsilon,D_\epsilon,\chi^{~}_{C_\epsilon},\chi^{~}_{D_\epsilon}$, with
the understanding that the $\partial^{~}_{\!\perp} x^i$ and $\psi^i$ parts of
the vertex operator (\ref{vertexDsusy}) are included. This has the overall
effect of replacing $\Delta_\epsilon$ in the bosonic parts of the operator
product expansions everywhere by the anomalous dimension
$h_\epsilon=1+\Delta_\epsilon$ of the impulse operator, while in the fermionic
parts $\Delta_\epsilon+\frac12$ is replaced everywhere by $h_\epsilon$. Using
(\ref{TCD}) and (\ref{SUSYOPE}), the $2\times2$ Jordan cell decompositions of
the bosonic and fermionic Virasoro generators are then given by
\beq
\new{\begin{array}{rrlrrl}
L^{\rm b}_0|C_\epsilon\rangle&=&h_\epsilon|C_\epsilon\rangle \ , ~~&
L^{\rm b}_0|D_\epsilon\rangle&=&h_\epsilon|D_\epsilon\rangle+|C_\epsilon
\rangle \ , \\L^{\rm f}_0|\chi^{~}_{C_\epsilon}\rangle&=&h_\epsilon|
\chi^{~}_{C_\epsilon}\rangle \ , ~~&L^{\rm f}_0|\chi^{~}_{D_\epsilon}
\rangle&=&h_\epsilon|\chi^{~}_{D_\epsilon}\rangle+|\chi^{~}_{C_\epsilon}
\rangle \ , \end{array}}
\label{L0Jordan}\eeq
where $L_0=L_0^{\rm b}+L_0^{\rm f}$. Using the factorization of bosonic and
fermionic states, in the Jordan blocks spanned by the logarithmic operators we
have~\cite{MSsuper,MavSz}
\beq
q^{L_0}\,|C_\epsilon,D_\epsilon\rangle\otimes|\chi^{~}_{C_\epsilon},
\chi^{~}_{D_\epsilon}\rangle=q^{h_\epsilon}\begin{pmatrix} 1&0\cr\ln q&1\cr
\end{pmatrix}
|C_\epsilon,D_\epsilon\rangle\otimes q^{h_\epsilon}
\begin{pmatrix} 1&0\cr\ln q&1\cr \end{pmatrix}
|\chi^{~}_{C_\epsilon},\chi^{~}_{D_\epsilon}
\rangle \ .\nn \\
\label{qL0Jordan}\eeq
The corresponding expectation value (\ref{stringpropgen}) in such a state is
then given by
\beq
\triangle_{CD}=-\int\limits_{\cal F}dq~q^{2\Delta_\epsilon+1/2}\,\langle
C_\epsilon,D_\epsilon|\begin{pmatrix} 1&0\cr\ln q&1\cr \end{pmatrix}
|C_\epsilon,D_\epsilon\rangle\,
\langle\chi^{~}_{C_\epsilon},\chi^{~}_{D_\epsilon}|
\begin{pmatrix} 1&0\cr\ln q&1\cr \end{pmatrix}
|\chi^{~}_{C_\epsilon},\chi^{~}_{D_\epsilon}\rangle \ .\nn \\
\label{propCD}\eeq
The dangerous region of moduli space is ${\rm Im}\,\tau\to+\infty$, in which
$q\sim\delta\to0^+$. Using $\Delta_\epsilon=0$ as $\epsilon\to0^+$, we can
easily check that the contributions to the modular integration in
(\ref{propCD}) from this region {\it vanish}. For instance, the worst behavior
comes from the term in the integrand involving $\sqrt q\,(\ln q)^2$, which upon
integration over a small strip ${\cal F}_\delta$ of width $\delta$ produces a
factor
\beq
\int\limits_{{\cal F}_\delta}
dq~\sqrt q\,(\ln q)^2\simeq\frac23\,\delta^{3/2}\,\left(
(\ln\delta)^2-\frac43\,\ln\delta+\frac89\right) \ ,
\label{stripint}\eeq
which vanishes in the limit $\delta\to0^+$. Therefore, in quantities involving
matrix elements of the string propagator in logarithmic states, the
incorporation of worldsheet superconformal partners cancels the modular
divergences that are present in the purely bosonic case. It is also
straightforward to arrive at this conclusion in the Ramond sector of the
superstring theory. Notice that although the explicit calculation above is
carried out with respect to the chosen basis (\ref{L0Jordan}) within the Jordan
cell, the same qualitative conclusion is arrived at under any change of basis
$|C_\epsilon,D_\epsilon\rangle\to|aC_\epsilon+bD_\epsilon,cC_\epsilon+d
D_\epsilon\rangle$. This is because the strip integral (\ref{stripint}) is the
worst behaved one and any change of basis will simply mix it with better
behaved modular integrals. Furthermore, physical string scattering amplitudes
will involve the superstring propagator with sums over complete sets in an
invariant, basis-independent form. Its effect on such physical quantities is
therefore independent of the chosen base.

This cancellation of infinities has dramatic consequences for the behavior of
higher genus amplitudes. As we have seen in section 3, 
in the purely bosonic case, where the modular
divergences persist, the logarithmic states yield non-trivial contributions to
the sum over string states and imply that, to leading order, the genus
expansion is dominated by contributions from degenerate Riemann surfaces whose
strip sizes become infinitely thin~\cite{km,msnl}. Such amplitudes can be
described in terms of bi-local worldsheet operators and the truncated
topological series can be summed to produce a non-trivial probability
distribution on the moduli space of running coupling constants of the slightly
marginal $\sigma$-model~\cite{ms1}. The functional Gaussian distribution has
width proportional to $\sqrt{\ln\delta}$, and the string loop divergences are
canceled by a version of the Fischler-Susskind mechanism. However, we see here
that this structure
disappears completely when one considers the full superstring theory. This
means that in the supersymmetric case one has to contend with the full genus
expansion of string theory which is not even a Borel summable series. The
dominance of pinched annular surfaces, as well as the loss of unitarity due to
the logarithmic mixing, can now be understood as merely an artifact of the
tachyonic instability of the bosonic string. Once the appropriate
superconformal partners to the logarithmic operators are incorporated, the
theory is free from divergences, at least at the level of string loop
amplitudes. Heuristically, this feature can be understood from the form of the
fermionic two-point functions (\ref{SUSY2pt}), which for $\Delta=0$ reduce to
conventional fermionic correlators with no logarithmic scaling violations on
the worldsheet. The zero dimension fermion fields, after incorporating the
worldsheet superconformal ghost fields, thereby have the usual effect of
removing instabilities from the theory.

\subsection{The Zamolodchikov Metric and Linearity in Liouville Evolution}

Another way to understand the effect of the fermionic fields in the recoil
problem is through the Zamolodchikov metric in the sector corresponding to the
logarithmic states. It is defined by the short-distance two-point functions
\beq
{\cal G}_{\phi\phi'}=\Lambda^{2h}\,\lim_{z\to w}\,\Bigl\langle\phi(z)\,
\phi'(w)\Bigr\rangle \ , ~~ \phi,\phi'=C,D,\chi^{~}_C,\chi^{~}_D \ ,
\label{Zamdef}\eeq
and by using (\ref{CD2pt}) and (\ref{SUSY2pt}) it can be represented as the
$4\times4$ matrix
\beq
{\cal G}=\begin{pmatrix} 0&\xi&0&0\cr\xi&d-2\xi\ln\Lambda&0&0\cr
0&0&0&2\Delta\xi\cr0&0&2\Delta\xi&2(\xi+\Delta d-2\Delta\xi
\ln\Lambda)\cr \end{pmatrix} \ .
\label{Zam4by4}\eeq
In the upper left bosonic $2\times2$ block we find a logarithmically divergent
term, which may be associated to the logarithmic modular divergences that are
present in the bosonic case. On the other hand, in the lower right fermionic
$2\times2$ block we find that the logarithmic divergence generically appears
only through the term which is proportional to $\Delta\ln\Lambda$. For the
recoil problem, in which the conformal dimension of the operators is correlated
with the worldsheet ultraviolet scale through the relations (\ref{Deltavarep})
and (\ref{epsilonLambdarel}), this term is a finite constant. Thus, in contrast
to its bosonic part, the fermionic part of the Zamolodchikov metric is
scale-invariant. This is just another reflection of the fact that the fermionic
logarithmic operators do not themselves lead to any logarithmic divergences and
act to cure the bosonic string theory of its instabilities. In fact, this
property on its own is motivation for the identification
(\ref{epsilonLambdarel}) of worldsheet and target space regularization
parameters which was used to arrive at the logarithmic conformal algebra. In
turn, this correlation is then also consistent with the Galilean non-invariance
(\ref{chiCDscale}) which derives from the twisted superparticle interpretation
of the previous section. Nevertheless, the vanishing correlation functions in
(\ref{CD2pt}) and (\ref{SUSY2pt}) indicate the existence of a hidden
supersymmetry in the dynamics of moving D-branes. For instance, it is
straightforward to check that the fermionic Noether supercurrents associated
with spatial translations induce the same logarithmic scaling violations that
the bosonic ones do~\cite{km}.

It is curious to note that the Zamolodchikov metric (\ref{Zam4by4}) becomes
degenerate in the conformal limit $\Delta\to0$, which corresponds to the
infrared fixed point of the worldsheet field theory (in the sense that the size
of the worldsheet is infinite in units of the ultraviolet cutoff). In this
limit all two-point correlation functions involving the fermionic field
$\chi^{~}_C$ vanish. Whether or not this implies that $\chi^{~}_C$ completely
decouples from the theory requires knowledge of higher order correlators of
the theory. Furthermore, in that case there are no logarithmic scaling
violations, since $\langle\chi^{~}_D(z)\,\chi^{~}_D(w)\rangle=2\xi/(z-w)$ in
the limit $\Delta\to0$. Generally, the vanishing of two-point functions in a
logarithmic
conformal field theory implies some special properties of the model. In the
purely bosonic cases, it is known that such a vanishing property is associated
with the existence of hidden symmetries corresponding to some conserved
current~\cite{lcftfurther}. 
A similar situation may occur in the supersymmetric case,
indicating the presence of some new fermionic symmetries. For this to be case,
there must be some other field to which the field $\chi^{~}_C$ couples. While
the extra hidden symmetry may be related to the fact that $\chi_C^{~}$ is a
null field in the subspace of primary fields, it should not be a true null
field. This interesting issue deserves further investigation. Notice however
that in the recoil problem, the pertinent correlation functions are
non-vanishing in the slightly-marginal case where $\epsilon\neq0$.

Notice also that the
degeneracy of the metric (\ref{Zam4by4}) in the limit $\Delta\to0$ may not be a
true singularity of the moduli space of coupling constants. To further
elaborate on this point requires computation of the corresponding curvature
tensor, and its associated invariants which, being invariant under changes of
renormalization group scheme, contain the true physical information of the
theory. However, this again requires knowledge of the three-point and
four-point correlation functions among the pertinent vertex operators, which at
present
are not available.

The Zamolodchikov metric is also a very important ingredient in the
construction of the effective target space action of the theory. In the bosonic
case such a moduli space action reproduces the Born-Infeld action for the
D-brane dynamics in the Neumann representation~\cite{ms1}. The
supersymmetrization of the worldsheet theory along the lines 
discussed previously 
will
produce the same
Born-Infeld action, with the only effect that the tachyonic instabilities are
again removed and no renormalization of the coupling constants are
required. This is immediate due to the form of (\ref{Zam4by4}).
On the other hand, the target space supersymmetrization of the Born-Infeld
action in ten
dimensions is known. The photino field $\lambda$ corresponds to
the Goldstino particle of the super-Poincar\'e symmetry which is spontaneously
broken by the presence of the D-brane. The resulting action does however
possess local spacetime $\kappa$-symmetry. We may then expect an appropriate
version of this action to emerge within the target space formalism of the
previous section, with corresponding breaking of the fermionic
$\kappa$-symmetry.

The most important consequence, however, of the 
form of the Zamolodchikov metric (\ref{Zam4by4}) 
for our purposes in this work, 
is that it eliminates 
in the world-sheet supersymmetric case 
the 
leading ultraviolet world-sheet divergences (\ref{recoilcontrtot})
leading to the diffusion term (\ref{leadingGLambda}) upon the 
identification of time with the world-sheet scale $\epsilon^{-2}$ 
(c.f. (\ref{epsilonLambda})),
and hence the Liouville zero mode.

Thus, upon including world-sheet supersymmetry, which appears essential
for a proper definition of D-particles, guaranteeing their target-space
stability, one obtains a diffusionless probability equation
from the RG equation of the Liouville-dressed partition function
in the supersymmetric case of the twisted super D-particle, 
and hence
an ordinary Schr\"odinger equation according to our 
discussion above. This may imply a potentially interesting link between 
supersymmetry (of some sort) and linearity of quantum mechanics.

\section{Conclusions} 

In this review/tribute to the memory of I. Kogan, 
I discussed the r\^ole of superconformal logarithmic algebras 
on the physics of recoiling membranes in string theory.
Although the formalism has been developed for D-particles,
extension to higher-dimensionality branes is straightforward, albeit 
technically more involved. 

We have also seen how LCFT and their extensions enter the discussion
of the recoil problem in curved (almost conformal) 
backgrounds corresponding to late times Robertson-Walker Cosmology. 

In all cases, the relevant pairs of LCFT describing recoil
were not marginal operators, but rather slightly relevant 
world-sheet deformations, with anomalous dimension proportional to 
$\epsilon^{-2} \sim {\rm ln}\Lambda$, 
thereby varying linearly with the RG scale.
This implied the necessity for Liouville dressing. 

In this latter context, 
we have discussed some ``pathologies'' of the bosonic string formalism,
associated with non-linearities in the quantum mechanical 
evolution of the D-branes under 
the identification of the Liouville mode with target time.
Such non-linearities were associated with leading ultraviolet
divergences in the world sheet, arising from pinched world-sheet genera.
Upon supersymmetrisation, however, such divergences disappear in a 
non-trivial way, dictated by the logarithmic superconformal
algebras, thereby rendering the above identification of Liouville mode
with time consistent with a linear quantum mechanical evolution
of super D-branes. It remains to be seen whether this curious link
between supersymmetry and linearity of quantum mechanics bears
any more general consequences.

It is my firm belief that the precise nature of time 
holds the key for a complete understanding of quantum gravity.
In this sense, therefore, the above role of super-LCFT in 
rendering the Liouville evolution linear, and completely quantum mechanical,
may be of importance. 
For instance, recently~\cite{westmuck} 
some models of supersymmetric space-time foam 
involving such supersymmetric D-particles
have been constructed as consistent ground states of 
brane world models. 
Time, and further work will show whether these 
speculations are right...

\section*{Acknowledgements}

The author wishes to thank M. Shifman and J. Wheater for inviting him 
to 
contribute to the Memorial Volume for I. Kogan. He also acknowledges 
discussions with H.-D. Doebner and G.A. Goldin. Finally he  
thanks the 
Department of  Theoretical Physics of Valencia University (Spain) 
for the hospitality during the final stages of this work. 
The work of N.E.M. is partially supported by the European Union
(contract HPRN-CT-2000-00152).

\end{document}